\documentclass[12pt]{article}
\usepackage[utf8]{inputenc}
\usepackage[top = 1.2 in, bottom = 1.2 in, outer = 0.95in, inner = 0.95in]{geometry}
\usepackage{graphicx}
\usepackage{amsmath,amssymb}
\usepackage{mathrsfs}
\usepackage{multirow}
\usepackage{cases}
\usepackage[authordate, backend=biber]{biblatex-chicago}
\usepackage{url}

\usepackage{setspace}
\newcommand{\indep}{\mathop{\perp\!\!\!\!\!\!\perp}}

\newcommand{\logistic}{\mathrm{logit}^{-1}}
\newcommand{\tr}{\mathsf{T}}

\newcommand{\argmax}{\mathop{\rm arg \, max}\limits}
\newcommand{\argmin}{\mathop{\rm arg \, min}\limits}

\newtheorem{assum}{Assumption}

\newtheorem{proposition}{Proposition}

\title{\vspace{-30pt}
\Large
\textbf{Navigated Weighting \\ to Improve Inverse Probability Weighting \\ for Missing Data Problems and Causal Inference}\thanks{I thank Naoki Egami, Adam Glynn, Kosuke Imai, Tomoya Sasaki, Yiqing Xu, Teppei Yamamoto, Soichiro Yamauchi and participants at Polmeth XXXVI and the 3rd winter meeting of the JSQPS for their helpful comments and suggestions. This work was supported by JSPS Grant-in-Aid for JSPS Research Fellow Grant Number JP17J03508.}}
\author{Hiroto Katsumata\thanks{Project Lecturer, Graduate School of Arts and Sciences, The University of Tokyo. Email:~\texttt{hrt.katsumata@gmail.com} \hspace{10pt} URL:~\texttt{https://sites.google.com/site/hirotokatsumata/}}
}
\date{This draft: August 4, 2020 \\
The first draft: May 22, 2020
}

\begin{document}

\setlength{\bibitemsep}{6pt}
\setlength{\abovedisplayskip}{4pt}
\setlength{\belowdisplayskip}{4pt}

\maketitle\thispagestyle{empty}

\vspace{-10pt}
\begin{abstract}
\begin{onehalfspace}
The inverse probability weighting (IPW) is broadly utilized to address missing data problems including causal inference but may suffer from large variances and biases due to propensity score model misspecification. To solve these problems, I propose an estimation method called the navigated weighting (NAWT), which utilizes estimating equations suitable for a specific pre-specified parameter of interest (e.g., the average treatment effects on the treated). Since these pre-specified parameters determine the relative importance of each unit as a function of propensity scores, the NAWT prioritizes important units in the propensity score estimation to improve efficiency and robustness to model misspecification. I investigate its large-sample properties and demonstrate its finite sample improvements through simulation studies and an empirical example. An R package \texttt{nawtilus} 
which implements the NAWT is developed and available from the Comprehensive R Archive Network (http://cran.r-project.org/package=nawtilus).
\end{onehalfspace}
\end{abstract}

\vspace{-20pt}
\noindent%
\vspace{20pt}
\vfill

\newpage
\setcounter{page}{1}
\section{Introduction}

Missing data problems, where outcomes of some observations are not observed, are common problems in social sciences and public health \parencites{Little2019}, including the fundamental problem in causal inference: we can only observe one of the potential outcomes with or without treatment and the other one is missing for each unit. In dealing with these missing data problems, the inverse probability weighting (IPW) method is broadly utilized, which uses the propensity score defined as the probability of missing conditional on observed covariates to construct weights \parencites{Rosenbaum1987}. The resulting inverse probability weights eliminate the dependence of missingness on these covariates under the conditional ignorability assumption without relying on the correct specification of outcome models. In observational studies where the treatment assignment is not randomized, we estimate propensity scores for treatment first and then use it for the IPW. Even in experimental studies where the treatment is randomly assigned, this randomization of treatment assignment alone cannot address some selection bias problems without the IPW in such cases as the mediation analysis, panel attrition problems, and the generalization of experimental results.

Although the IPW has broad applicability and desirable theoretical properties \parencites{Hirano2003, Lunceford2004}, there are problems due to propensity score estimation. Existing research has found that it suffers from an excessively large variance due to extreme estimated weights and is highly vulnerable to propensity score model misspecification \parencites{Imai2014, Kang2007}. 

To address the problem of propensity score estimation, I propose an estimation method called the \underline{na}vigated \underline{w}eigh\underline{t}ing (NAWT) method, which improves efficiency and reduces bias due to propensity score model misspecification by tailoring propensity score estimation for a specific pre-specified quantity of interest. Among many types of quantities of interest, this study focuses on weighted average treatment effects (WATE), which include the average treatment effects (ATE), average treatment effects on the treated (ATT), average treatment effects on the controlled (ATC), and average treatment effects for overlap population (ATO) as special cases \parencites{Li2018}. I also study the estimation of the average outcome (AO) where I can observe covariates of all units but cannot observe some of their outcomes due to missingness.

These pre-specified parameters of interest determine the relative importance of each unit as a function of propensity scores, that is, for which unit propensity scores should be precisely estimated. In other words, the estimand navigates how to improve propensity score estimation. However, applied researchers almost always estimate propensity scores via the standard logistic, or maybe the probit, regression irrespective of their quantities of interest (called the standard IPW in this study), which ignores differences in the relative importance among units. The non-parametric propensity score estimation, though it is proved to be asymptotically efficient and has no bias due to model misspecification, is rarely utilized in applied studies because of its poor finite sample performance and difficulty in parameter tuning.

The NAWT, in contrast, uses a parametric estimator but approximates the non-parametric estimates for specific important units depending on the quantities of interest. It is therefore as easy to implement as, but also more robust and efficient than the standard IPW. Importantly, the NAWT improves the standard IPW especially when the IPW suffers from large variances due to excessively large inverse probability weights.

The NAWT has following attractive characteristics in addition to its improvement in robustness and efficiency over the standard parametric IPW. First, it can incorporate various estimators for well-known quantities of interest (e.g., the ATT or ATE), such as the weighted difference-in-means, Horvitz–Thompson, doubly robust IPW, and weighted least squares estimators \parencites{Imai2014}. Second, its workflow is quite simple. It adds only one step before the standard IPW procedure: we first \textit{define the quantity of interest}, then estimate propensity scores using a modified estimator based on this user-specified estimand.

The key idea of the NAWT is to improve the estimation by tweaking the propensity score estimation depending on the pre-specified quantity of interest while being totally agnostic about outcome values nor models. This is different from the methods somehow relying on outcome models such as methods proposed in the framework of empirical likelihoods \parencites{Graham2012, Tan2010}, the covariate balancing propensity score (CBPS) method \parencites{Imai2014, Zhao2019}, and the targeted maximum likelihood estimation (TMLE) \parencites{Schuler2017, van2010}. The NAWT is also different from methods that re-define parameters of interest to avoid large inverse probability weights such as the weight trimming technique \parencites{Crump2009, Yang2018} and the method estimating the ATO \parencites{Li2018}.

The rest of the paper is organized as follows. First, I propose the navigated weighting (NAWT) and investigate its large-sample properties in Section~2. In Section~3, I extend it by incorporating covariate balancing conditions for further improvement. In Section~4, I conduct simulation studies to demonstrate that the NAWT improves the standard IPW in efficiency and robustness to model misspecification and it also outperforms the CBPS in terms of the bias when the propensity score model is misspecified. Section~5 compares the performance of the NAWT with the standard IPW and the CBPS in a canonical empirical example. The final section concludes and discusses future research directions.

Finally, an open-source software package \texttt{nawtilus} is available from the Comprehensive R Archive Network for implementing the proposed methodology \parencites{nawtilus2020}.

\section{Proposed methodology} \label{sec_method}

Suppose we have a random sample of $n$ units ($i = 1, 2, \ldots, n$) from a population. For each unit $i$, we observe a $k$-dimensional vector of pretreatment covariates $\mathbf{x}_i \in \mathbb{R}^k$. We consider the causal inference case in the main text, and the missing outcome problem is explained in Appendix~\ref{sec_missing}. In the causal inference case, each unit has two potential outcomes $y_i(0), y_i(1) \in \mathbb{R}$, only one of which is observed depending on the value of the binary treatment $t_i \in \{0, 1\}$ the unit gets. The quantity of interest here is the weighted average treatment effects (WATE):
\begin{equation}
\tau_\mathrm{WATE} \equiv \frac{\int \mathbb{E}[y_i(1) - y_i(0) \mid \mathbf{x}_i = \mathbf{x}] h(\mathbf{x}) d F(\mathbf{x})}{\int h(\mathbf{x}) d F(\mathbf{x})} ,
\end{equation}
where $h(\cdot)$ is a known function of covariates and $F(\cdot)$ is a cumulative distribution function. If we take $h(\mathbf{x}) = \Pr(t_i = 1 \mid \mathbf{x}_i = \mathbf{x})$, we can get the ATT:
\begin{equation}
\tau_\mathrm{ATT} \equiv \mathbb{E}[y_i(1) - y_i(0) \mid t_i = 1] \ \ \ \mathrm{with} \ h(\mathbf{x}) = \Pr(t_i = 1 \mid \mathbf{x}_i = \mathbf{x}).
\end{equation}
The WATE also includes the ATE,
$\tau_\mathrm{ATE} \equiv \mathbb{E}[y_i(1) - y_i(0)]$ with $h(\mathbf{x}) = 1$, and
the ATC,
$\tau_\mathrm{ATC} \equiv \mathbb{E}[y_i(1) - y_i(0) \mid t_i = 0]$ with $h(\mathbf{x}) = \Pr(t_i = 0 \mid \mathbf{x}_i = \mathbf{x})$,
as special cases depending on the choice of $h(\mathbf{x})$.

For identification, I make following two assumptions. The first one is the conditional ignorability of treatment assumption that the treatment is independent of the potential outcomes conditional on the observed covariates, which implies that units with treatment have the same expected potential outcomes as units without treatment conditional on the covariates.
\begin{assum}[Conditional ignorability of treatment]\label{as_conditionalignorabilityt}
\begin{equation} \label{eq_conditionalignorabilityt}
t_i \ \indep \ \{y_i(1), y_i(0)\} \mid \mathbf{x}_i = \mathbf{x} .
\end{equation}
\end{assum}

I introduce the propensity score for treatment, which is the probability of being treated given covariates $\pi(\mathbf{x}) \equiv \Pr(t_i = 1 \mid \mathbf{x}_i = \mathbf{x})$. This needs to be bounded away from 0 and 1.
\begin{assum}[Positivity of the treatment probability]\label{as_positivityt}
\begin{equation}
0 < \pi(\mathbf{x}) < 1.
\end{equation}
\end{assum}

When the propensity score is unknown, it must be estimated, and even when it is known, using estimated propensity scores improves estimation of the quantities of interest \parencites{Hahn1998, Hirano2003}. For the propensity score estimation, researchers typically utilize parametric model $\pi_\beta(\mathbf{x_i})$, especially the logistic model,
\begin{equation}
\Pr(t_i = 1 \mid \mathbf{x}_i) \equiv \pi_\beta(\mathbf{x}_i)
= \frac{\exp(\mathbf{x}_i^\tr \beta)}{1 + \exp(\mathbf{x}_i^\tr \beta)}, \label{eq_pslogistic}
\end{equation}
where $\beta \in B$ is a $k$-dimensional vector of unknown parameters. This standard IPW estimates parameters $\beta$ by the maximum likelihood estimation (MLE), where the binomial log-likelihood function is maximized:
\begin{equation}
\hat{\beta}_\mathrm{MLE} = \argmax_{\beta \in B} \sum_{i = 1}^n t_i \log(\pi_\beta(\mathbf{x}_i)) + (1 - t_i) \log(1 - \pi_\beta(\mathbf{x}_i)) .
\end{equation}
This leads to the following first order condition:
\begin{align}
\frac{1}{n} \sum_{i = 1}^n s_\mathrm{MLE}(\beta, t_i, \mathbf{x}_i) &= 0 \label{eq_standardscore} \\
s_\mathrm{MLE}(\beta, t_i, \mathbf{x}_i) &= \left( \frac{t_i}{\pi_\beta(\mathbf{x}_i)} - \frac{1 - t_i}{1 -  \pi_\beta(\mathbf{x}_i)} \right) \pi'_\beta(\mathbf{x}_i),
\end{align}
where $\pi'_\beta(\mathbf{x}_i) = \partial \pi_\beta(\mathbf{x}_i) / \partial \beta^\tr$.

Using the estimated propensity score $\hat{\pi}_\beta(\mathbf{x}_i)$, the quantity of interest can be estimated with inverse probability weights. For the ATT estimation, weights for the control units are $\hat{\pi}_\beta(\mathbf{x}_i) / (1 - \hat{\pi}_\beta(\mathbf{x}_i))$ and those for the treatment units are $1$. For example, this study considers the weighted difference-in-means estimator (the H{\'a}jek estimator): \begin{align} \label{eq_dimatt}
\hat{\tau}_\mathrm{ATT} =& \frac{\sum_{i = 1}^n t_i y_i}{n_1} -
\sum_{i = 1}^n \frac{(1 - t_i) \hat{\pi}_\beta(\mathbf{x}_i) y_i}{1 - \hat{\pi}_\beta(\mathbf{x}_i)} \bigg/ \sum_{i = 1}^n \frac{(1 - t_i) \hat{\pi}_\beta(\mathbf{x}_i)}{1 - \hat{\pi}_\beta(\mathbf{x}_i)} ,
\end{align}
where $n_1$ is the number of treated units. 

These weights determine the relative impact of propensity score estimation for each unit on the parameters of interest estimation and the relative importance of units with specific estimated propensity scores ($I(\pi)$), which is given by taking conditional expectation on $\hat{\pi}_i = \pi'$ of the partial derivative of $\hat{\tau}$ with respect to $\hat{\pi}_i$ with $\mu_0 \equiv \mathbb{E}[y_i(0)]$:
\begin{equation}\label{eq_relimpact}
I(\pi') \equiv \mathbb{E}\left[ \frac{\partial \hat{\tau}}{\partial \hat{\pi}_i} \middle| \hat{\pi}_i = \pi' \right] = -\mathbb{E}\left[ \frac{(1 - t_i) (y_i(0) - \mu_0)}{(1 - \hat{\pi}_i)^2} \middle| \hat{\pi}_i = \pi' \right] .
\end{equation}

This relative impact suggests that units with larger estimated propensity scores, and units with outcomes largely deviated from the mean, have more impact on the ATT estimation. This implies that the larger estimated propensity scores units have, the more important those units are in the propensity score estimation because little differences in the estimated propensity scores for the former have more impact on the ATT estimation than the latter.

Since the MLE for propensity scores does not account for these differences in the importance among units, it is not the most efficient parametric estimator of the propensity score for the ATT estimation. Although the MLE is the best unbiased estimation for nuisance parameters $\beta$ when the propensity score model is correctly specified, it may not be true for the ATT estimation. This phenomenon, though somewhat counterintuitive, is akin to the well-known result that estimated propensity scores produce a more efficient estimate of the target quantity of interest than true propensity scores \parencites{Hahn1998, Hirano2003}.

\subsection{Propensity score estimation in the navigated weighting}

To account for the importance among units, the proposed method (NAWT) weights the score function for the propensity score estimation by a function of propensity score $\omega(\pi_\beta(\mathbf{x}_i))$. The NAWT replaces the MLE score condition in \eqref{eq_standardscore} with the following weighted score condition:
\begin{equation}
\sum_{i = 1}^n s_{\mathrm{ATT}}(\beta, t_i, \mathbf{x}_i) 
\equiv \sum_{i = 1}^n \left( \frac{t_i}{\pi_\beta(\mathbf{x}_i)} - \frac{1 - t_i}{1 -  \pi_\beta(\mathbf{x}_i)} \right) \omega(\pi_\beta(\mathbf{x}_i)) \pi'_\beta(\mathbf{x}_i) = 0 .
\end{equation}
When the propensity score is specified as in Equation~\eqref{eq_pslogistic}, this is further simplified as
\begin{equation}
\sum_{i = 1}^n \left( t_i - \pi_\beta(\mathbf{x}_i) \right) \omega(\pi_\beta(\mathbf{x}_i)) \mathbf{x}_i = 0. \label{eq_nawtatt}
\end{equation}
Integrating the score function with respect to $\beta$ gives the corresponding pseudo-log-likelihood. For example, the pseudo-log-likelihood of the NAWT with $\omega(\pi_\beta(\mathbf{x}_i)) = \pi_\beta(\mathbf{x}_i)^\alpha$ is the following:
\begin{align}
l_{\mathrm{ATT}}(\beta, \mathbf{t}, \mathbf{X}) &\equiv 
\int \sum_{i = 1}^n s_{\mathrm{ATT}}(\beta, t_i, \mathbf{x}_i) d \beta \\
&= \sum_{i = 1}^n \pi_\beta(\mathbf{x}_i)^\alpha \left( \frac{t_i}{\alpha} - \frac{(1 - t_i) \pi_\beta(\mathbf{x}_i) {_2F_1}(1, 1 + \alpha, 2 + \alpha, \pi_\beta(\mathbf{x}_i))}{1 + \alpha} \right) ,
\end{align}
where $_2F_1(a, b, c, z)$ is a hyper-geometric function, $\mathbf{t}$ is a vector of the missing indicator $t_i$, and $\mathbf{X}$ is a matrix of the covariates $\mathbf{x}_i$. The choice of the weighting function is discussed in Section~\ref{subsec_efficiency} and \ref{sec_extension}. We can estimate $\beta$ via the M-estimation by maximizing the pseudo-log-likelihood $l_{\mathrm{ATT}}(\beta, \mathbf{t}, \mathbf{X})$:
\begin{equation}
\hat{\beta}_\mathrm{ATT} = \argmax_{\beta \in B} l_{\mathrm{ATT}}(\beta, \mathbf{t}, \mathbf{X}).
\end{equation}
The numerical illustration on the pseudo-log-likelihood is in Appendix~\ref{sec_llnawt}.

As the properties of the M-estimator, $\hat{\beta}_\mathrm{ATT}$ is consistently estimated:
\begin{equation}
\hat{\beta}_\mathrm{ATT} \xrightarrow{p} \beta ,
\end{equation}
and the asymptotic distribution of $\hat{\beta}_\mathrm{ATT}$ is:
\begin{align}
\begin{split}
\sqrt{n} (\hat{\beta}_\mathrm{ATT} - \beta) &\xrightarrow[d]{} \mathcal{N}(0, \mathbf{H}_{\beta\beta}^{-1} \mathbf{\Sigma}_{\beta\beta} \mathbf{H}_{\beta\beta}^{-\tr}) , \\
\mathbf{H}_{\beta\beta} &\equiv \mathbb{E} \left[ \frac{\partial s_{\mathrm{ATT}}(\beta, \mathbf{t}, \mathbf{X})}{\partial \beta^\tr} \right] 
, \\
\mathbf{\Sigma}_{\beta\beta} &\equiv \mathbb{E} \left[ s_{\mathrm{ATT}}(\beta, t_i, \mathbf{x_i}) s_{\mathrm{ATT}}(\beta, t_i, \mathbf{x_i})^\tr \right] 
.
\end{split}
\end{align}

\subsection{Parameters of interest estimation in the navigated weighting} \label{subsec_ipw}

Using the weighted difference-in-means estimator, we can estimate the quantities of interest via the M-estimation. The joint estimating equations for the ATT are:
\begin{align}
\sum_{i = 1}^n s_{\mathrm{ATT}}(\beta, t_i, \mathbf{x}_i) &= 0 \\
\sum_{i = 1}^n q_{\mathrm{ATT}}(\tau_\mathrm{ATT}, \beta, t_i, \mathbf{x}_i) &= 0 ,
\end{align}
where
\begin{align}
q_{\mathrm{ATT}}(\tau_\mathrm{ATT}, \beta, t_i, \mathbf{x}_i) 
\equiv \frac{n}{n_1} t_i y_i - n \left( \sum_{i = 1}^n \frac{(1 - t_i) \pi_\beta(\mathbf{x}_i)}{1 - \pi_\beta(\mathbf{x}_i)} \right)^{-1} \frac{(1 - t_i) \pi_\beta(\mathbf{x}_i) y_i}{1 - \pi_\beta(\mathbf{x}_i)} - \tau_\mathrm{ATT} . 
\end{align}
Note that the NAWT can incorporate various other types of estimators for the parameters of interest, such as the Horvitz–Thompson estimator, doubly robust IPW estimator, and weighted least squares estimator.

Proposition~\ref{pro_ipw} shows that the estimated ATT is consistent and asymptotically normal under some regularity conditions.
\begin{proposition}[Consistency and asymptotic normality of the NAWT] \label{pro_ipw}
The NAWT estimates of the quantity of interest (e.g., the ATT) as well as $\beta$ is $\sqrt{n}$-consistent and converges to a normal distribution.
\begin{equation}\label{eq_varcov}
\sqrt{n} \left( \begin{array}{l}
\hat{\beta}_\mathrm{ATT} - \beta \\
\hat{\tau}_\mathrm{ATT} - \tau_\mathrm{ATT}
\end{array}
\right) \xrightarrow[d]{} \mathcal{N}(0, \mathbf{H}^{-1} \mathbf{\Sigma} \mathbf{H}^{-\tr}) .
\end{equation}
\end{proposition}
See Appendix~\ref{sec_asymptoticdist} for the details and Appendix~\ref{sec_asymptoticdistate} for the details of the ATE estimation.

The NAWT is simple, flexible, and has several attractive characteristics. First, it includes the standard IPW, which uses the standard MLE for the propensity score estimation, as the special case where $\omega(\pi_\beta(\mathbf{x}_i)) = 1$. Second, in a broad range of settings, it is asymptotically more efficient than the standard IPW as explained in Section~\ref{subsec_efficiency}. Third, it includes the covariate balancing propensity score (CBPS) as the special case as explained in Section \ref{sec_related}. Note that unlike weighting the score function as the NAWT, weighting the likelihood or log-likelihood function by a function of propensity scores introduces some biases.

\subsection{Efficiency results and weighting function specification} \label{subsec_efficiency}

As demonstrated in Equation~\eqref{eq_relimpact}, units with large estimated propensity scores have large impacts on the ATT estimation. Thus, the NAWT with non-decreasing weighting functions (and decreasing functions for the ATC estimation) with respect to propensity scores improves the ATT estimation because it places much importance on these units in propensity score estimation. For example, a power function of propensity scores $\pi_\beta(\mathbf{x}_i)^{\alpha}$, where $\alpha \geq 0$ is non-decreasing and suitable for the ATT estimation. 

Equation~\eqref{eq_relimpact} also suggests that units with outcomes largely deviated from the mean, as well as units with large estimated propensity scores, are important on the estimation. To demonstrate how these two factors affect the performance of the NAWT with a large sample, I conduct a thorough simulation in Appendix~\ref{sec_simeff}. The results confirm that the NAWT performs better than the standard IPW and the CBPS in a broad range of settings. Importantly, the NAWT improves the standard IPW especially when the IPW suffers from large variances due to excessively large inverse probability weights for units whose outcomes are also deviated from the mean. Note also that the NAWT includes the standard IPW as a special case and always performs at least as well as it with an appropriate weighting function.

With a finite sample, we should also mitigate unstable propensity score estimation due to putting too much weight on only a small portion of observations. Practically, utilizing $\omega(\pi_\beta(\mathbf{x}_i)) = \pi_\beta(\mathbf{x}_i)^2$ performs well in balancing weighting important units and avoiding unstable propensity score estimation as shown in Appendix~\ref{sec_simefffinite} and \ref{sec_alpha}.

We can also easily validate whether the NAWT with a specific weighting function performs better than the standard IPW in a sample by comparing the sample version of Equation~\eqref{eq_varcov}. The simulation results in Appendix~\ref{sec_simefffinite} demonstrates that the adaptive method, where the weighting function for the NAWT is selected to minimize the sample variance, performs well irrespective of the data generating process and propensity score model (mis)specification. Further, we may specify the most efficient weighting function without assuming any outcome models, which is discussed later as a future research direction in Section~\ref{subsec_adaptive}.

\subsection{Estimation of the ATE} \label{subsec_ate}

The estimation of the ATE is not as simple as the estimation of the ATT because the ATE estimation includes two different missing data problems. One of them is the average potential outcomes of the treated without treatment $\mathbb{E}[y_i(0) \mid t_i = 1]$ and the other is the average potential outcomes of the controlled with treatment $\mathbb{E}[y_i(1) \mid t_i = 0]$, neither of which can be observed. This naturally leads to the separate propensity score estimation for the potential outcomes with and without treatment. For example, for estimating propensity scores to estimate the average potential outcomes without treatment by (inversely) weighting the controlled units, the NAWT utilizes $\omega(\pi_\beta(\mathbf{x}_i)) = \pi_\beta(\mathbf{x}_i)^\alpha$, whereas it utilizes $\omega(\pi_\beta(\mathbf{x}_i)) = (1 - \pi_\beta(\mathbf{x}_i))^\alpha$ for estimating propensity scores to estimate the average potential outcomes with treatment by (inversely) weighting the treated units. This separate estimation produces two estimated propensity scores for each combination of covariates $\mathbf{x}$, one of which is for estimating average potential outcomes with treatment $\hat{\pi}_\beta^1 (\mathbf{x}_i)$ and the other is for estimating those without treatment $\hat{\pi}_\beta^0 (\mathbf{x}_i)$. In general, these two estimated propensity scores are not equal $\hat{\pi}_\beta^1 (\mathbf{x}_i) \neq \hat{\pi}_\beta^0 (\mathbf{x}_i)$ except for the standard IPW case. Although this requires a little caution to interpreting estimated propensity scores and coefficients for them, the NAWT with the separate propensity score estimation for the ATE estimation is efficient as it is shown in Appendix~\ref{sec_simeff} and \ref{sec_combined}.

Alternatively, for the ease of interpretation, we can combine the two weighted score functions and estimate one propensity score for each combination of covariates $\mathbf{x}$, for example, using $\omega(\pi_\beta(\mathbf{x}_i)) = \pi_\beta(\mathbf{x}_i)^\alpha + (1 - \pi_\beta(\mathbf{x}_i))^\alpha$. This combined estimation has an advantage in interpretation of estimated propensity scores, but it is not efficient as shown in Appendix~\ref{sec_combined}. This combined propensity score estimation is utilized in the CBPS for the ATE estimation $1 / (\pi_\beta(\mathbf{x}_i)(1 - \pi_\beta(\mathbf{x}_i))) = 1 / \pi_\beta(\mathbf{x}_i) + 1 / (1 - \pi_\beta(\mathbf{x}_i))$, which implies that it balances covariates between the treated and controlled but not between the treated and combined nor between the controlled and combined \parencites{Chan2016}.

\subsection{Intuition behind the navigated weighting}
Intuitively, the NAWT utilizes a parametric model and approximates non-parametric estimates for specific important units depending on the quantities of interest. This leads to efficiency and robustness to model misspecification because the non-parametric propensity score estimation is asymptotically efficient for the quantity of interest estimation and free from bias due to propensity score model misspecification.

\begin{figure}[t]
\centering
\includegraphics[width=0.47\textwidth]{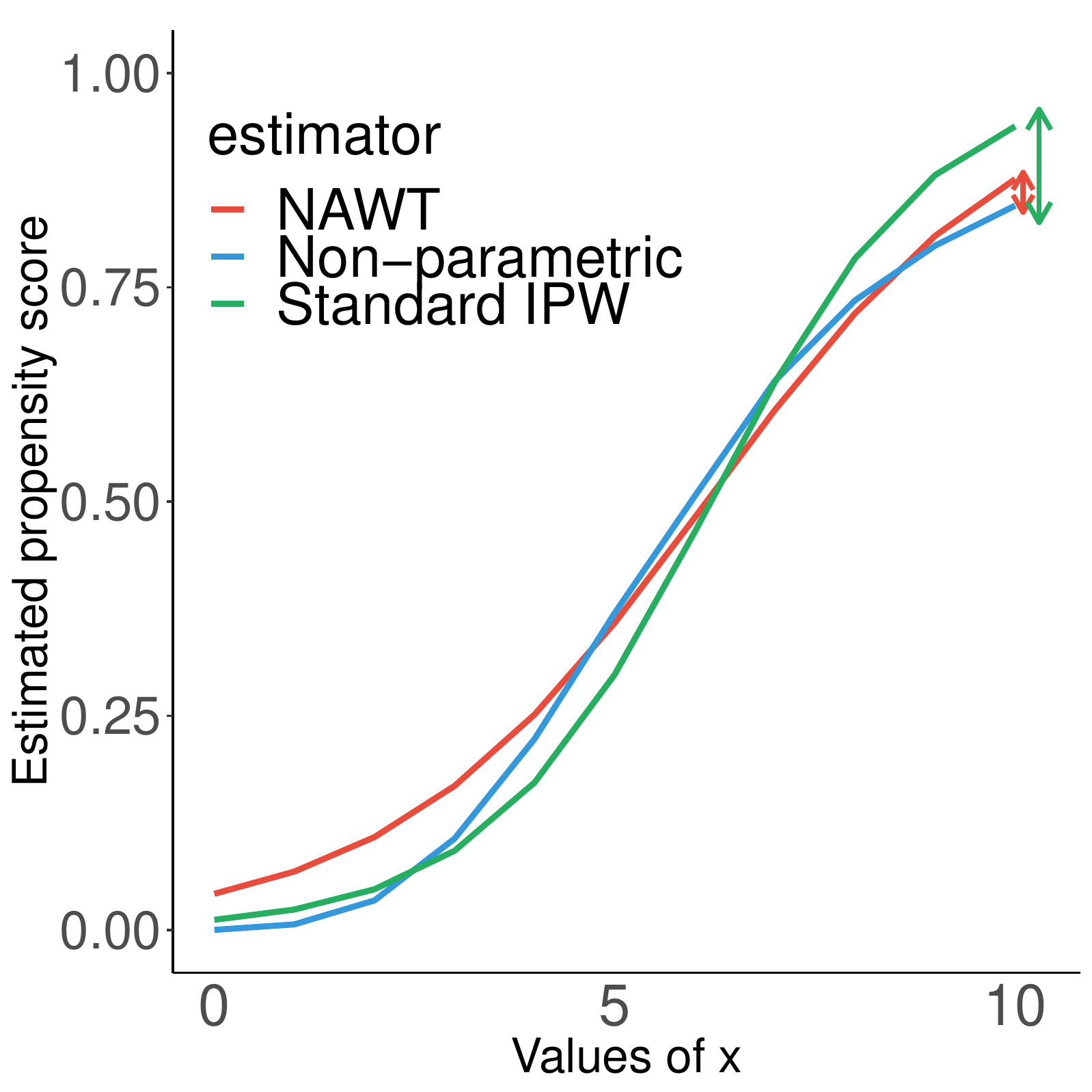}
\hspace{0.045\textwidth}
\includegraphics[width=0.47\textwidth]{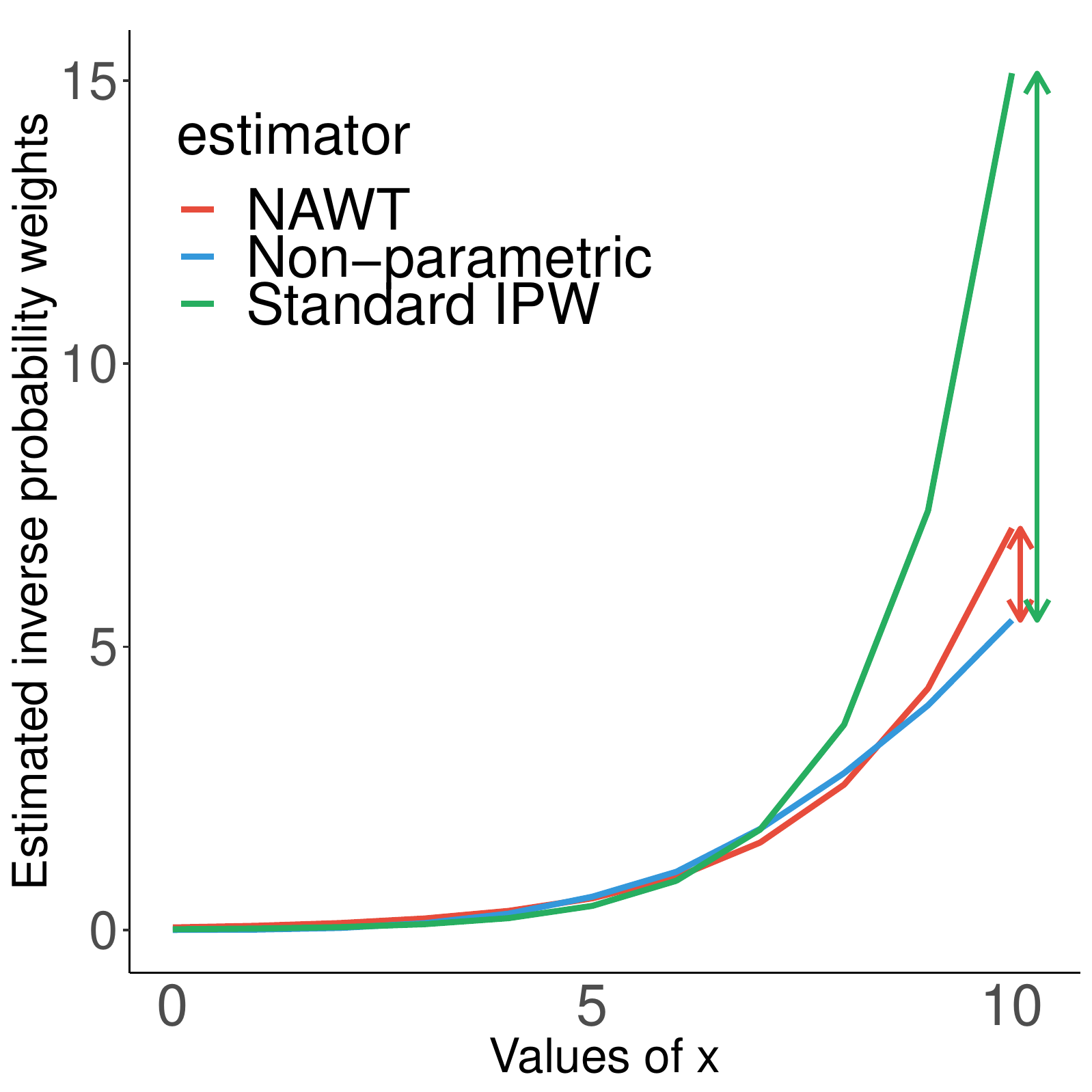}
\caption{The left panel of the figure shows estimated propensity scores and the right panel shows resulting estimated inverse probability weights for the ATT estimation, where the red, blue, green curves represent estimates using the navigated weighting (NAWT), non-parametric, and standard parametric propensity score estimation, respectively.} \label{fig_comparison}
\end{figure}

Figure~\ref{fig_comparison} demonstrates how the NAWT approximates non-parametrically estimated propensity scores and resulting inverse probability weights and shows differences from the standard parametric estimation in the ATT estimation. I consider a scenario where there are $200,000$ units with only one discrete covariate $x_i \in \{ 0, 1, \ldots, 10 \}$ and the probability of being treated is: $\Pr(t_i = 1) = 1 / (1 + \exp(6.5 - 3.5 \log(0.5 + x_i)))$. Since there is only one discrete covariate, the propensity score can be non-parametrically estimated by computing the proportion of treated units for each value of $x$, which is highly difficult with multidimensional and continuous $\mathbf{x}$. The standard IPW is based on the (misspecified) logistic regression for propensity score estimation. The NAWT also utilizes the (misspecified) logistic model but it prioritizes units with large estimated propensity scores by weighting the score with $\omega(\pi_\beta(\mathbf{x}_i)) = \pi_\beta(\mathbf{x}_i)^2$.

The left panel of the figure shows estimated propensity scores and the right panel shows resulting estimated inverse probability weights for the ATT estimation, where the red, blue, green curves represent estimates using the NAWT, non-parametric, and standard parametric propensity score estimation, respectively. In the left panel, for units with large propensity scores, the standard parametric propensity score estimates deviate from the non-parametric estimates, whereas the NAWT estimates are quite close to them. These differences are amplified in estimated inverse probability weights as shown in the right panel. For units with large inverse probability weights, the resulting inverse probability weights with the standard parametric propensity score estimation are far from the non-parametric estimates, whereas those with the NAWT approximate the non-parametric estimates quite closely.

\subsection{Related methods}\label{sec_related}

The key idea of the NAWT is to improve the estimation by tweaking the propensity score estimation depending on the pre-specified quantity of interest while being totally agnostic about outcome values nor models. This is different from the methods recently proposed in the framework of empirical likelihoods that incorporate outcome models for efficiency and other desirable properties \parencites{Graham2012, Tan2010}. 

This also differs from another robust and efficient propensity score estimation method, the CBPS, which estimates propensity scores so that covariates are balanced between treatment groups \parencites{Imai2014, Zhao2019}. The CBPS does not exploit outcome values in estimating propensity scores but its robustness and efficiency depend on how close the linear combination of balanced covariates approximates the true outcome model \parencites{Fan2016}. 

However, the general form of the NAWT includes the just-identified CBPS with the logistic model as the special case where $\omega(\pi_\beta(\mathbf{x}_i)) = 1 / (\pi_\beta(\mathbf{x}_i) (1 - \pi_\beta(\mathbf{x}_i)))$ for the ATE estimation and $\omega(\pi_\beta(\mathbf{x}_i)) = 1 / (1 - \pi_\beta(\mathbf{x}_i))$ for the ATT estimation:
\begin{align}
s_\mathrm{CBPS, ATE}(\beta, t_i, \mathbf{x}_i) &= \left( \frac{t_i}{\pi_\beta(\mathbf{x}_i)} - \frac{1 - t_i}{1 -  \pi_\beta(\mathbf{x}_i)} \right) \mathbf{x}_i \\
s_\mathrm{CBPS, ATT}(\beta, t_i, \mathbf{x}_i) &= \left( t_i - \frac{(1 - t_i) \pi_\beta(\mathbf{x}_i)}{1 -  \pi_\beta(\mathbf{x}_i)} \right) \mathbf{x}_i , 
\end{align}
considering $\pi'_\beta(\mathbf{x}_i) = \pi_\beta(\mathbf{x}_i) (1 - \pi_\beta(\mathbf{x}_i)) \mathbf{x}_i$ for the logistic model. This demonstrates that the CBPS weights the score function in the same spirits as the NAWT so that it puts more importance on units which should have large estimated inverse probability weights. Although the CBPS has not been justified from this perspective, considering it as a special case of the NAWT help understand why the CBPS gains efficiency and robustness even when the true outcome model is not a linear combination of covariates balanced via the CBPS and improve it further to gain potential efficiency and robustness in that case.

Another doubly robust method, the targeted maximum likelihood estimation (TMLE), gains desirable properties by focusing on estimation of parameters of interest (target parameters) like the NAWT \parencites{Schuler2017, van2010}. The important difference between the NAWT and the TMLE is that the NAWT utilizes propensity scores to improve the \textit{propensity score} estimation, whereas the TMLE uses them to improve the \textit{outcome model} estimation. Thus, the TMLE is also seen as a method relying on the outcome model though it is a doubly robust estimation. In contrast, the NAWT gains efficiency and robustness without assuming any outcome models nor exploiting outcome values in the propensity score estimation, which is useful even with little knowledge about the outcome model and in line with the original spirit of propensity scores \parencites{Rubin2007}. 

The NAWT is also different from the weight trimming technique for avoiding large inverse probability weights, which results in changing the quantity of interest \parencites{Crump2009, Yang2018}. Instead of defining the question best answered given a sample \parencites{Li2018}, the NAWT improves the estimation for the pre-specified quantity of interest.

\section{Extension: the NAWT with covariate balance conditions} \label{sec_extension}

The robustness of the NAWT to the model misspecification can be further improved by incorporating the covariate balancing conditions. Details are shown in Appendix~\ref{sec_nawtcbps}. In addition to the weighted score condition \eqref{eq_nawtatt}, we can utilize the following covariate balance conditions.
\begin{align}
\sum c_{\mathrm{ATT}}(\beta, t_i, \mathbf{x}_i) &= 0 \\
c_{\mathrm{ATT}}(\beta, t_i, \mathbf{x}_i) 
&\equiv \left( t_i - \frac{(1 - t_i) \pi_\beta(\mathbf{x}_i)}{1 -  \pi_\beta(\mathbf{x}_i)} \right) \tilde{\mathbf{x}}_i ,
\end{align}
where $\tilde{\mathbf{x}}_i$ are some functions of covariates to balance, which can include $\mathbf{x}_i$ or $\mathbf{x}_i^2$ etc. Using the score and covariate balance conditions results in more conditions than the number of parameters and the parameters are estimated via the over-identified GMM estimation:
\begin{equation}
\hat{\beta}_\mathrm{GMM} \equiv \argmin_{\beta \in B} \bar{g}(\beta, \mathbf{t}, \mathbf{X})^\tr \mathbf{A} \bar{g}(\beta, \mathbf{t}, \mathbf{X}) ,
\end{equation}
where
\begin{align}
\bar{g}(\beta, \mathbf{t}, \mathbf{X}) &\equiv \frac{1}{n} \sum g(\beta, t_i, \mathbf{x}_i) \\
g(\beta, t_i, \mathbf{x}_i) &= \left( \begin{array}{ll}
s_{\mathrm{ATT}}(\beta, t_i, \mathbf{x_i}) \\
c_{\mathrm{ATT}}(\beta, t_i, \mathbf{x_i}) 
\end{array}
\right) ,
\end{align}
for the ATT estimation, and $\mathbf{A}$ is some positive definite symmetric weighting matrix. Note that when we utilize only the covariate balance conditions, it becomes the same method as the just-identified CBPS, where $g(\beta, t_i, \mathbf{x}_i) = c_{\mathrm{ATT}}(\beta, t_i, \mathbf{x}_i)$ and $\mathbf{A} = \mathbf{I}$. As the properties of the GMM, $\hat{\beta}_\mathrm{GMM}$ is $\sqrt{n}$-consistent and converges to a normal distribution. 

After estimating propensity scores, we can estimate the quantity of interest by using the IPW estimators, such as the Horvitz–Thompson estimator, weighted difference-in-means estimator, doubly robust IPW estimator, and weighted least squares estimator, and its variance can be estimated by bootstrapping.

The advantage of using both the weighted score conditions and covariate balance conditions is that it may perform better when the propensity score model is misspecified thanks to its over-identified GMM property, but the disadvantage is that it may not perform well with a small sample.

\section{Simulation studies} \label{sec_simulation}

In this section, I apply the proposed method, the NAWT, to simulation data, to demonstrate how much it improves the performance of the standard IPW estimation and to compare it with the performance of the IPW using the CBPS to estimate propensity scores. The IPW method has large variances and it is also vulnerable to propensity score model misspecification, and the CBPS is proposed to solve these problems. Therefore, I conduct a simulation in which: (a) correct propensity score model, and two types of propensity score model misspecification (b) and (c) to test the validity of the NAWT for each of the ATT and ATE estimation.

Specifically, I use the following data-generating process. There are $1,000$ units and each unit $i$ has four covariates $\mathbf{x}_i = (x_{i1}, x_{i2}, x_{i3}, x_{i4})$, each of which is independently and identically distributed according to the standard normal distribution. Some units are assigned treatment $t_i = 1$ and the others are not $t_i = 0$. The true outcome model is $y_i \sim \mathcal{N}(\mu, 1)$ and $\mu = 210 + \tau t_i + 27.4 x_{i1} + 13.7 x_{i2} + 13.7 x_{i3} + 13.7 x_{i4}$, where the quantity of interest $\tau = 10$.

The true treatment assignment model is $t_i \sim \textrm{Bernoulli}(\pi_{i})$ and $\pi_{i} = \logistic(x_{i1} -0.5 x_{i2} + 0.25 x_{i3} + 0.1 x_{i4})$ for scenarios (a) and (b) and $\pi_{i} = \logistic(-x_{i1} + 0.5 x_{i2} - 0.25 x_{i3} - 0.1 x_{i4})$ for scenario (c), where $\logistic(\cdot) = 1 / (1 + \exp(-(\cdot)))$, resulting $\Pr(t_i = 1) = 0.5$. Finally, in scenarios (b) and (c), only the non-linear transforms of the covariates $\mathbf{x}_i^\star = (x_{i1}^\star, x_{i2}^\star, x_{i3}^\star, x_{i4}^\star) = (\exp(x_{i1} / 2), x_{i2} / (1 + \exp(x_{i1})) + 10, (x_{i1} x_{i3} / 25 + 0.6)^3, (x_{i1} + x_{i4} + 20)^2)$ can be observed, which results in exactly the same misspecification in scenario (b) as that used in the existing studies \parencites{Imai2014, Kang2007}. In scenarios (b) and (c), the propensity score model is misspecified because the true propensity score model is not a logistic function with $\mathbf{x}_i^*$ but one with $\mathbf{x}_i$ as linear predictors. Hence, the estimates are expected to be biased, but the NAWT and IPW with the CBPS are expected to mitigate this bias. For the ATE estimation, the NAWT utilizes the separate propensity score estimation for the potential outcomes with and without treatment, which performs better than the combined estimation (See Section~\ref{subsec_ate} for details and Appendix \ref{sec_combined} for the results of the combined estimation).

I use the weighted difference-in-means estimator for the quantities of interest and conduct $2,000$ Monte Carlo simulations and calculate the bias and root-mean-squared error (RMSE) for each propensity score estimator (the NAWT with $\omega(\pi_\beta(\mathbf{x}_i)) = \pi_\beta(\mathbf{x}_i)^2$, the standard IPW, and the IPW with the just-identified CBPS) in each scenario ((a), (b), and (c)) for each quantity of interest (the ATT and ATE).

\begin{figure}[p]
\centering
\includegraphics[width=0.325\textwidth]{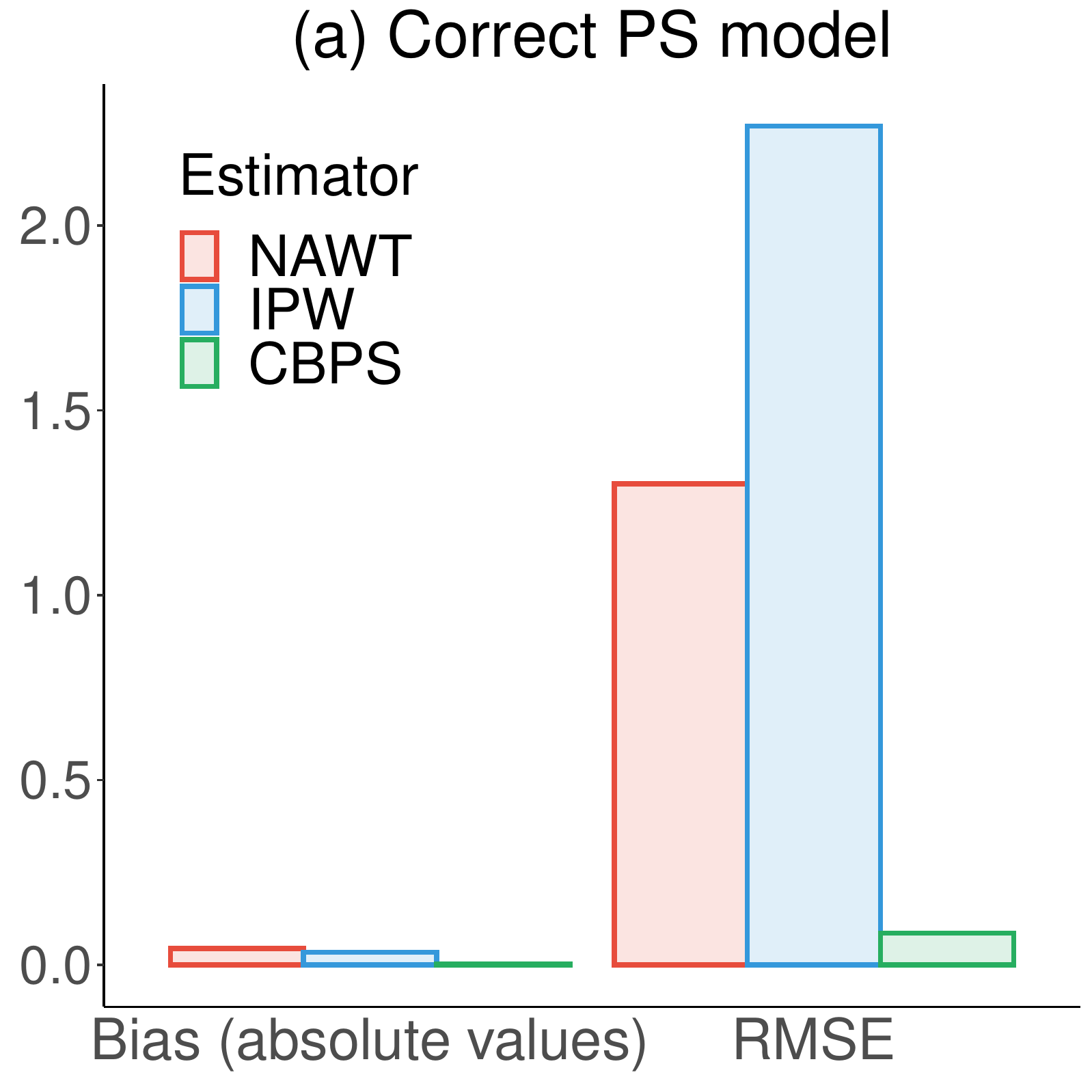}
\includegraphics[width=0.325\textwidth]{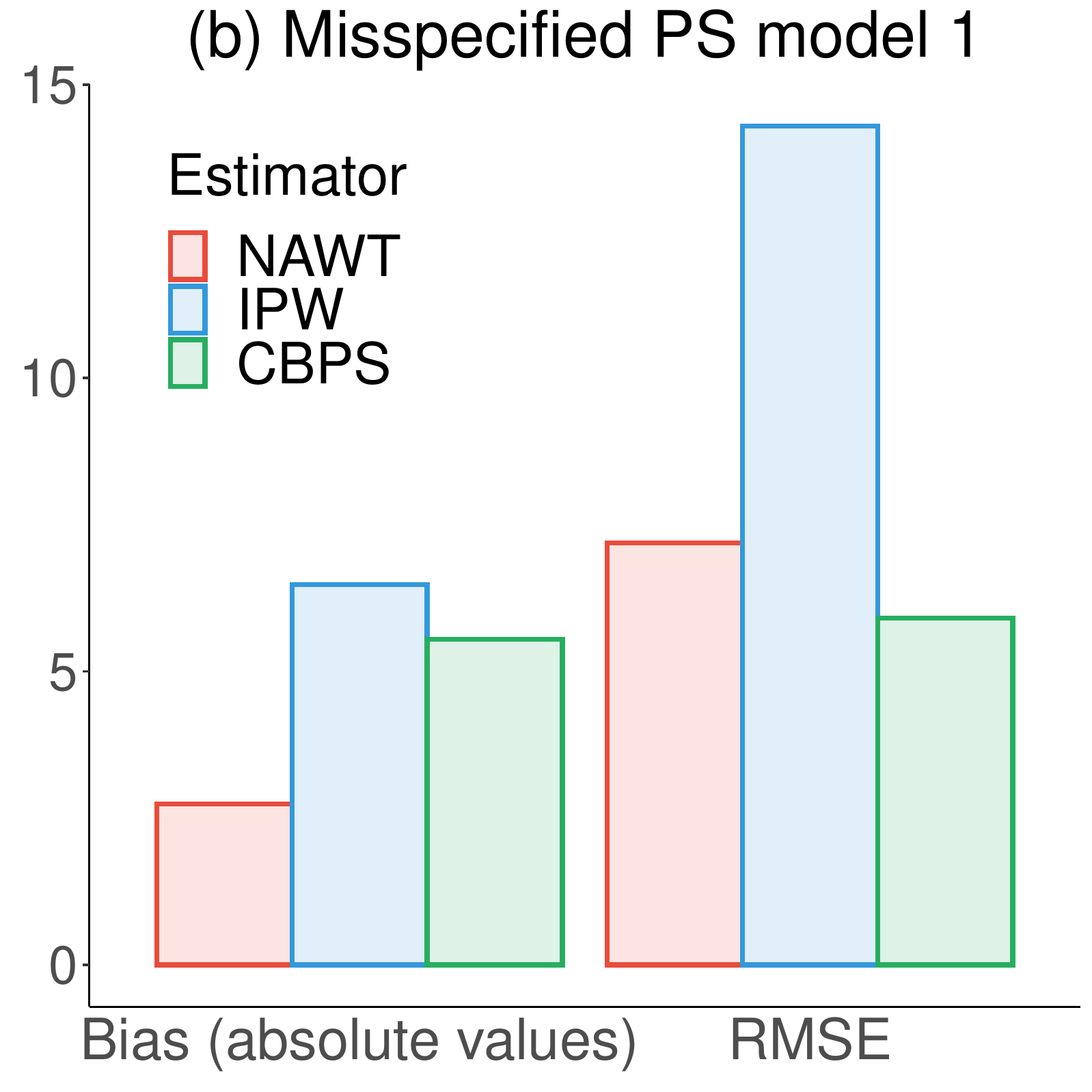}
\includegraphics[width=0.325\textwidth]{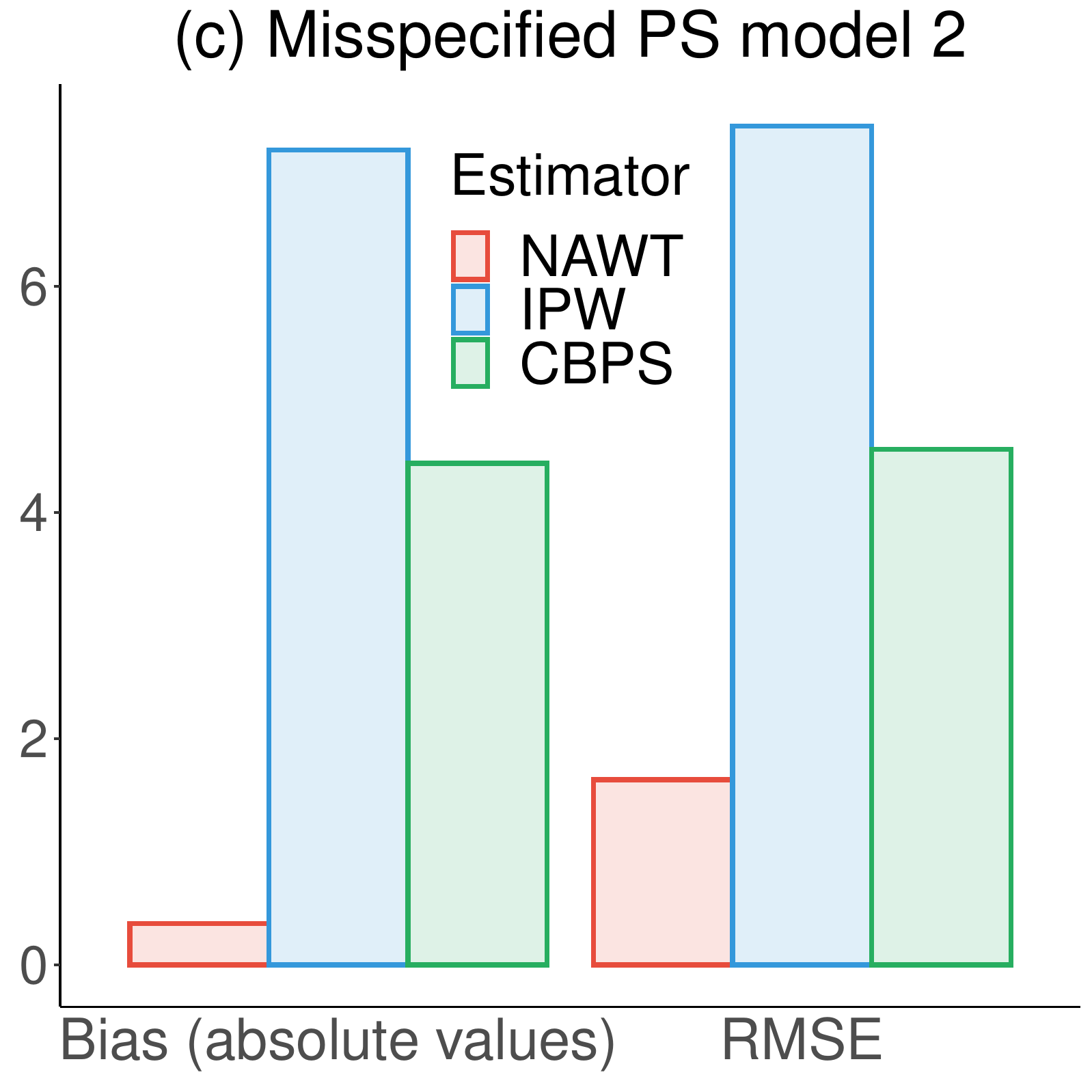}
\caption{The bias in absolute values and the RMSE for the ATT estimation using the NAWT, standard IPW, and IPW with the CBPS under the following scenarios: (a) correct propensity score model and two types of propensity score model misspecification (b) and (c). The NAWT outperforms the standard IPW in terms of the RMSE in all the scenarios, and it depends on the situation whether the NAWT works better than the IPW with CBPS in terms of both the bias and RMSE.} \label{fig_biasrmseatt}

\vspace{35pt}

\includegraphics[width=0.325\textwidth]{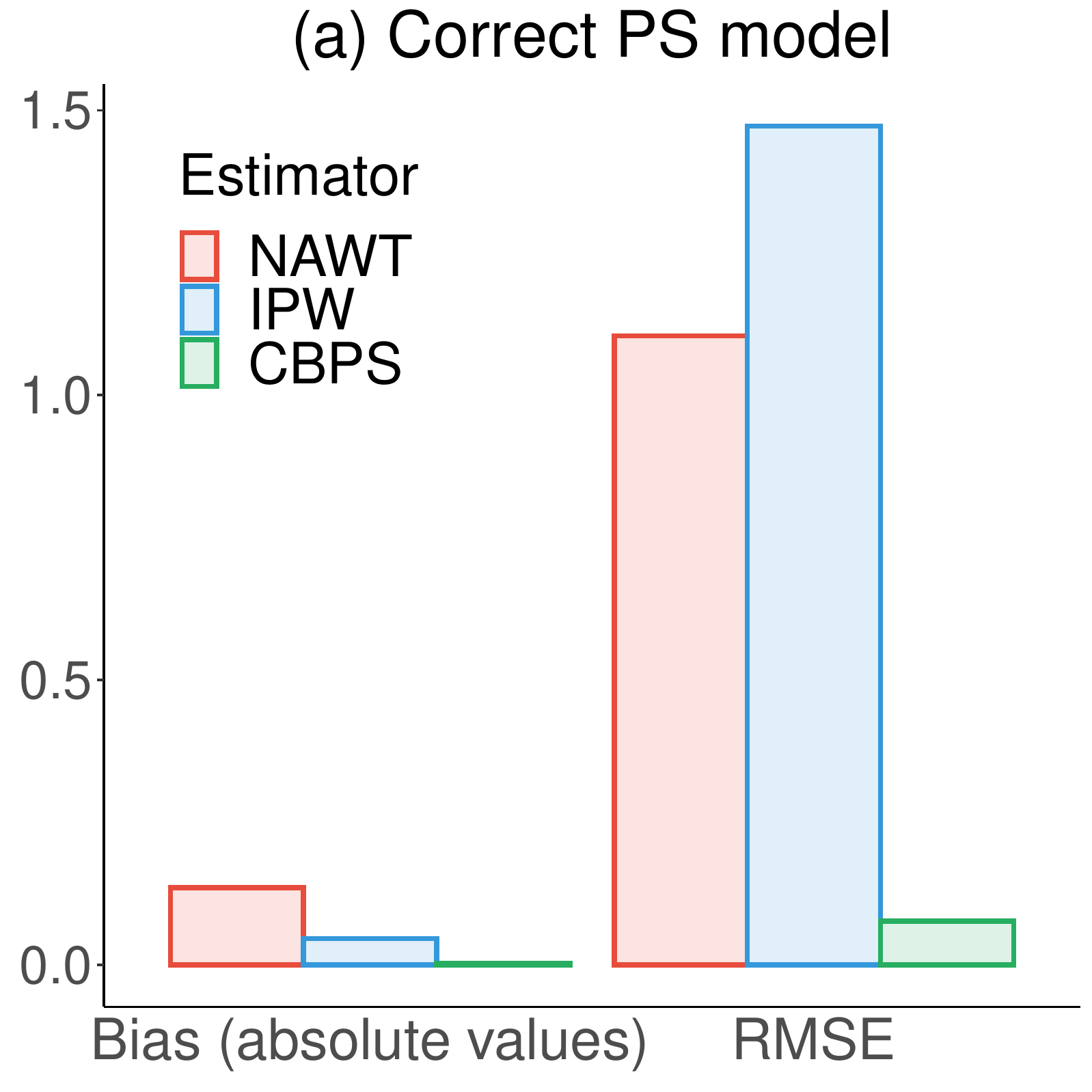}
\includegraphics[width=0.325\textwidth]{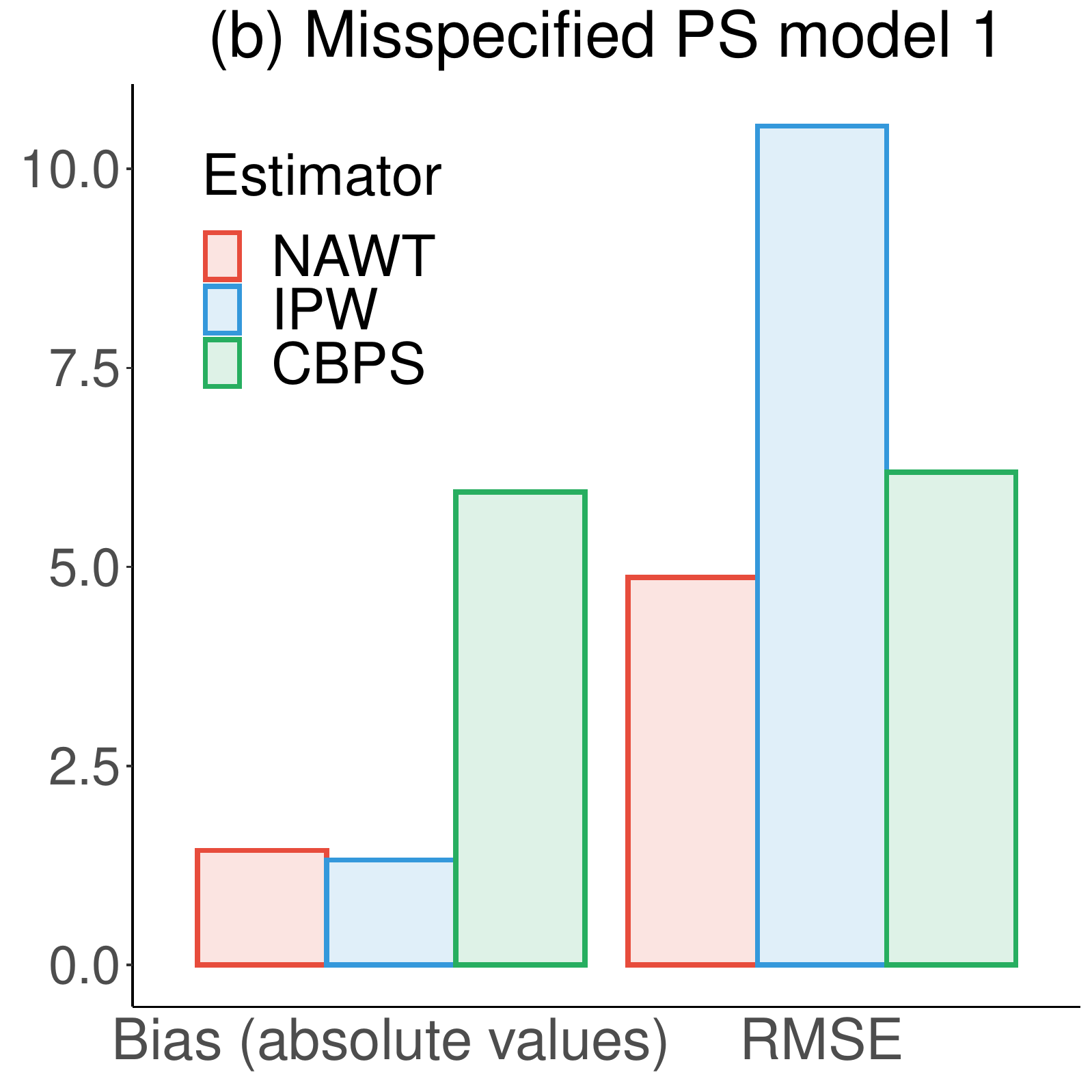}
\includegraphics[width=0.325\textwidth]{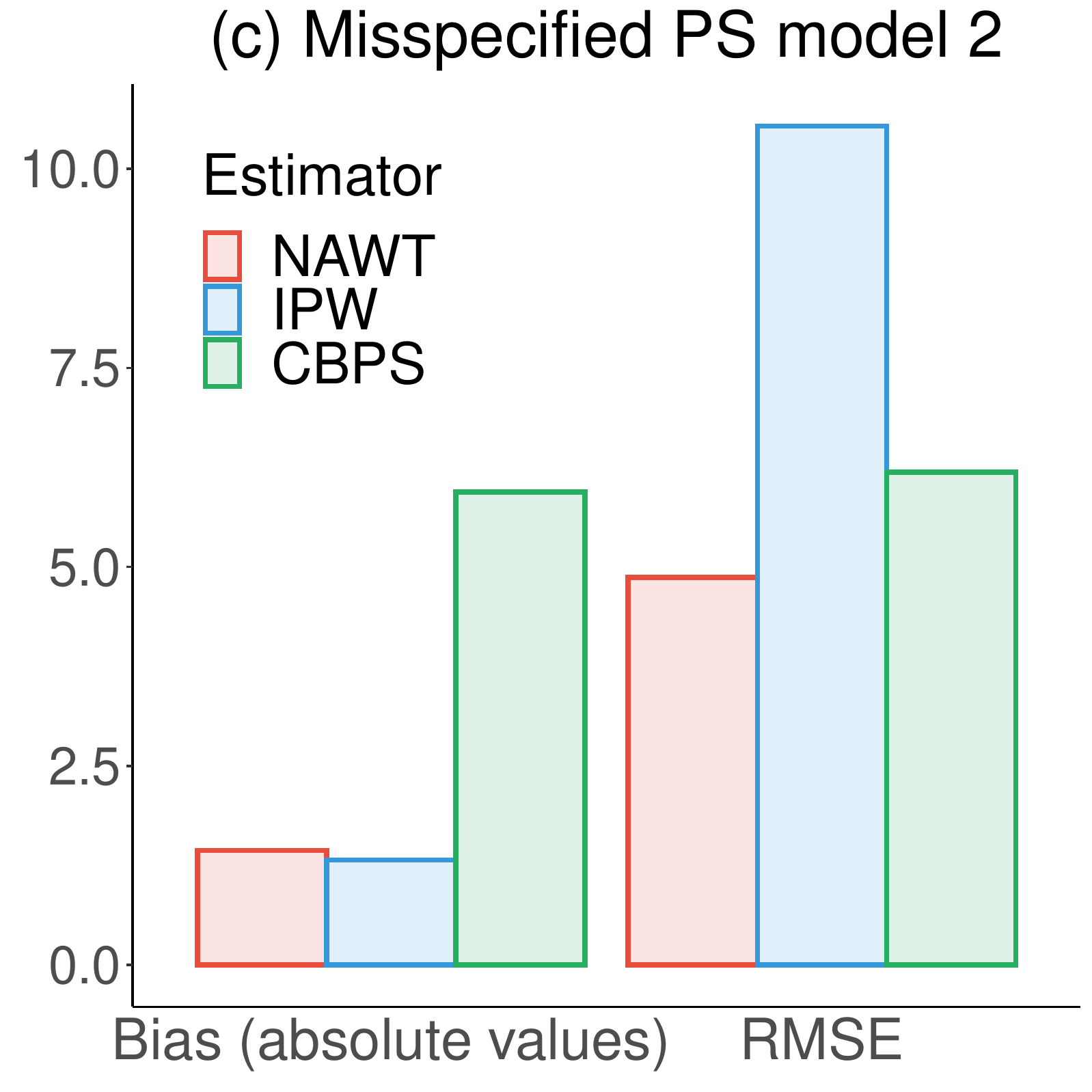}
\caption{The bias in absolute values and the RMSE for the ATE estimation using the NAWT (the separate estimation), standard IPW, and IPW with the CBPS under the following scenarios: (a) correct propensity score model and two types of propensity score model misspecification (b) and (c). The NAWT outperforms the standard IPW in terms of the RMSE in all the scenarios, and it depends on the situation whether the NAWT works better than the IPW with CBPS in terms of both the bias and RMSE.} \label{fig_biasrmseate}
\end{figure}

The summary of the results is shown in Figures~\ref{fig_biasrmseatt} and \ref{fig_biasrmseate}, Figures~\ref{fig_estatt} and \ref{fig_estate} in Appendix~\ref{sec_dist} show the distribution of the estimates, and Table~\ref{tb_simulation} in Appendix~\ref{sec_simtable} presents the details. The results demonstrate that the NAWT has negligible biases when the propensity score model is correctly specified and dramatically improves estimation compared with the standard IPW. The NAWT has smaller RMSEs than the standard IPW when (a) the propensity score model is correctly specified and smaller biases than the standard IPW when (b)(c) the propensity score model is misspecified. In terms of the RMSE, it depends on the situation whether or not the NAWT works better than the IPW with the CBPS because the performance of the CBPS depends on how close the linear combination of the balanced covariates can approximate the true outcome model. Since the true outcome model is a linear combination of the covariates in scenario (a), the IPW with the CBPS works better than the NAWT, but when the propensity score model is misspecified and the outcome model is not the linear combination of the covariates (scenarios (b) and (c)), the NAWT works better than the IPW with the CBPS in scenario (b) but not in scenario (c). Surprisingly, when the propensity score model is misspecified in scenarios (b) and (c), the NAWT overwhelms the IPW with the CBPS in terms of the bias. Considering that the CBPS is proposed and used to mitigate the bias due to propensity score model misspecification, these results impressively confirm the robustness of the NAWT. See Appendix~\ref{sec_simefffinite} for more discussion when the NAWT mitigates the bias.

For the ATE estimation, the separate propensity score estimation for the potential outcomes with and without treatment slightly outperforms the combined estimation in terms of the RMSE when the propensity score model is correctly specified. Moreover, the separate estimation has much smaller RMSEs than the combined estimation when the propensity score model is misspecified, demonstrating that the separate estimation is preferable to the combined estimation.

In summary, the NAWT has negligible biases when the propensity score model is correctly specified and is much more robust and efficient than the standard IPW. Besides, it is also more robust than the IPW with the CBPS which is proposed to mitigate the bias due to the propensity score model misspecification.

\section{Empirical example} \label{sec_empirical}

This section examines how much the NAWT improves the standard IPW with empirical data. I analyze the data from \textcite{Lalonde1986} which is extensively utilized to test the validity of causal inference methods. The goal of the original study was to evaluate a job training program in a randomized controlled trial (`the National Supported Work Demonstration Program'). \textcite{Lalonde1986} replaced the experimental control group with other untreated observations, survey data from the Panel Study for Income Dynamics (PSID) and the Current Population Survey, and found it difficult to recover experimental benchmark with various estimators.

Various studies investigated whether causal inference methods can replicate the experimental benchmark \parencites{Dehejia1999, Diamond2013, Hainmueller2012, Imai2014, Smith2005}. Whereas most of them examined matching methods, this section investigates the validity of the NAWT compared with the standard IPW and CBPS in one of the most difficult scenarios where survey data from the PSID (2490 observations) is used as the control observations with original \textcite{Lalonde1986}'s experimental sample (297 treated and 425 untreated observations).

The pretreatment covariates include age, years of education, race (black, Hispanic, or white), marital status, high school degree, earnings in 1974, earnings in 1975, and employment status in 1975, and the outcome of interest is earnings in 1978. I estimate the propensity score, which is the conditional probability of being in the experimental sample given the pretreatment covariates, with the NAWT with $\omega(\pi_\beta(\mathbf{x}_i)) = \pi_\beta(\mathbf{x}_i)^2$, the standard logistic regression, and the CBPS. To check the sensitivity to the propensity score model specification, I utilize three different models. The first one (Linear) includes all the pretreatment covariates, the second one (Quadratic) includes all the pretreatment covariates and squared terms of age and education, and the third one (Simple) includes only continuous variables (age, years of education, earnings in 1974, and earnings in 1975). Since earnings in 1974 are missing in some observations, I deal with them in two ways: using only complete data or setting missing values to zero \parencites{Dehejia1999, Imai2014}. The parameter of interest is the ATT, but it is estimated as the average difference between the experimental treatment observations and the weighted non-experimental PSID observations with the weighted difference-in-means estimator. The experimental benchmark is \$866 with a standard error of \$488.

\begin{table}[t]
\renewcommand{\tabcolsep}{3.5pt}
\centering
\caption{Comparison between the ATT estimates with the NAWT, the standard IPW, and the CBPS with LaLonde (1986) data set}\label{tb_lalonde}
\begin{tabular}{ccccccccccccccc}
\hline
 && \multicolumn{5}{c}{Full sample} && && \multicolumn{5}{c}{Complete-data sample} \\
\cline{3-7} \cline{11-15}
 && NAWT && Standard IPW && CBPS && && NAWT && Standard IPW && CBPS \\
\hline
\\
\multirow{2}{*}{Linear} && $1351.4$ && $162.3$ && $398.3$ && && $1094.4$ && $173.1$ && $359.7$ \\
 && $(861.4)$ && $(811.5)$ && $(637.3)$ && && $(846.5)$ && $(825.5)$ && $(649.5)$ \\
\\
\multirow{2}{*}{Quadratic} && $962.2$ && $-151.6$ && $321.4$ && && $768.6$ && $-98.0$ && $288.6$ \\
 && $(950.5)$ && $(830.6)$ && $(647.0)$ && && $(903.9)$ && $(850.5)$ && $(661.7)$ \\
\\
\multirow{2}{*}{Simple} && $1421.0$ && $-2038.8$ && $-684.0$ && && $1243.3$ && $-2231.4$ && $-725.2$ \\
 && $(1115.2)$ && $(697.7)$ && $(718.7)$ && && $(1150.1)$ && $(723.3)$ && $(757.4)$ \\
\\
\hline
\multicolumn{15}{p{.99\textwidth}}{\small Note: The experimental benchmark is \$866 with a standard error of \$488. I compare three methods to estimate propensity scores for the IPW: the NAWT, the standard logistic regression, and the CBPS. Each method is estimated with combinations of three model specifications (the Linear, Quadratic and Simple models) and two versions of the sample (the full sample with missing values in earnings in 1974 imputed as zero and the complete-data sample). The standard errors are in parentheses. Across all specifications and samples, the NAWT improves the performance of the standard IPW and outperforms the CBPS.}
\end{tabular}
\end{table}

Table~\ref{tb_lalonde} presents the results, where each number represents the ATT estimate and the bootstrap standard errors are shown in parentheses. Across all specifications and samples, the NAWT improves the performance of the standard IPW and outperforms the CBPS. Even in the third model where the propensity score model includes only a few covariates and should be severely misspecified, the bias of the NAWT is small (\$357 or \$535) whereas the biases of the standard IPW and the CBPS are quite large (\$1570--3118). Another clear pattern is that the standard errors of the NAWT estimates are larger than those of the standard IPW and CBPS estimates when the propensity score model is more severely misspecified in the third model. The large differences in the standard errors may be a useful signal of the propensity score model misspefcification.

\section{Conclusions}\label{subsec_adaptive}

The IPW is broadly utilized to address missing data problems including causal inference because it can eliminate the dependence of missingness on observed covariates without relying on correct specification of outcome models under the conditional ignorability assumption. However, existing research has pointed out that the IPW may have an excessively large variance due to extreme estimated weights and be highly vulnerable to the misspecification of the propensity score model. To address these problems, this study proposed the \underline{na}vigated \underline{w}eigh\underline{t}ing (NAWT), which improves efficiency and robustness by utilizing estinating equations suitable for a specific pre-specified parameter of interest (e.g., the ATT and ATE). The NAWT includes the standard IPW and the CBPS as special cases. Large-sample properties of the NAWT were investigated and its finite sample improvement of the performance of estimation compared with the standard IPW and the CBPS was demonstrated through simulation studies and an empirical example.

The key idea of the NAWT is that tailoring the propensity score estimation for the pre-specified parameter of interest by prioritizing important units determined by the parameter of interest itself. It uses a parametric model but approximates the non-parametric propensity score estimates for these important units because the IPW with the non-parametric propensity score estimation is asymptotically efficient and has no bias due to model misspecification. This enables the NAWT to enjoy the best of both worlds: it performs well with a finite sample and it is efficient and robust to model misspecification. Importantly, the NAWT improves the standard IPW especially when the IPW suffers from large variances due to excessively large inverse probability weights for units whose outcomes are also deviated from the mean.

Finally, I show some future directions for the improvement and the application of the NAWT. As I demonstrated that the NAWT can incorporate covariate balance conditions, it is natural to extend it to incorporate kernel balance conditions, which makes the NAWT more flexible \parencites{Hazlett2016, Wong2017, Zhao2019}. Since the NAWT is an extension of the standard IPW, it may be combined with attractive methods for the standard IPW, such as estimating propensity scores with regularization via the ridge or LASSO and inverse probability weight trimming. Although this study concentrates on the logistic model for the propensity score estimation, the idea of the NAWT may apply to recently proposed machine learning techniques for the propensity score estimation, such as the decision tree, random forest, and generalized random forest, where we may improve the algorithm via weighting a purity measure, such as the Gini impurity, by a function of propensity scores.

Although the NAWT is more robust and efficient than the standard IPW in a broad range of the data generating process, we may gain additional efficiency by specifying the weighting function to be optimized for the covariate distribution in the sample. The simplest way other than the adaptive method examined in this study is to select a weighting function which minimizes the directed Kullbuck entropy divergence between estimated inverse probability weights $w_i$ and the base weights $b_i$, which is defined by $h(w_i) = w_i \log(w_i / b_i)$, where $b_i$ is usually uniform weights \parencites{Hainmueller2012}. This entropy divergence decreases as the estimated weights approach to the uniform weights and becomes exactly 0 when $w_i = b_i$ for all $i$, which implies that the NAWT with the weighting function minimizing this divergence retains information in the sample and thus improves efficiency. Unlike the original entropy balancing method by \textcite{Hainmueller2012}, the NAWT with this adaptive specification of the weighting function explicitly models propensity scores and optimizes the estimation by searching wide varieties of weighting functions without any outcome models.

I am currently exploring these potential directions and other application of the NAWT.

\printbibliography

\clearpage
\newpage

\appendix

\section*{Appendix}

\section{Numerical illustration on the pseudo-log-likelihood function of the NAWT}
\label{sec_llnawt}

\begin{figure}[t]
\centering
\includegraphics[width=0.325\textwidth]{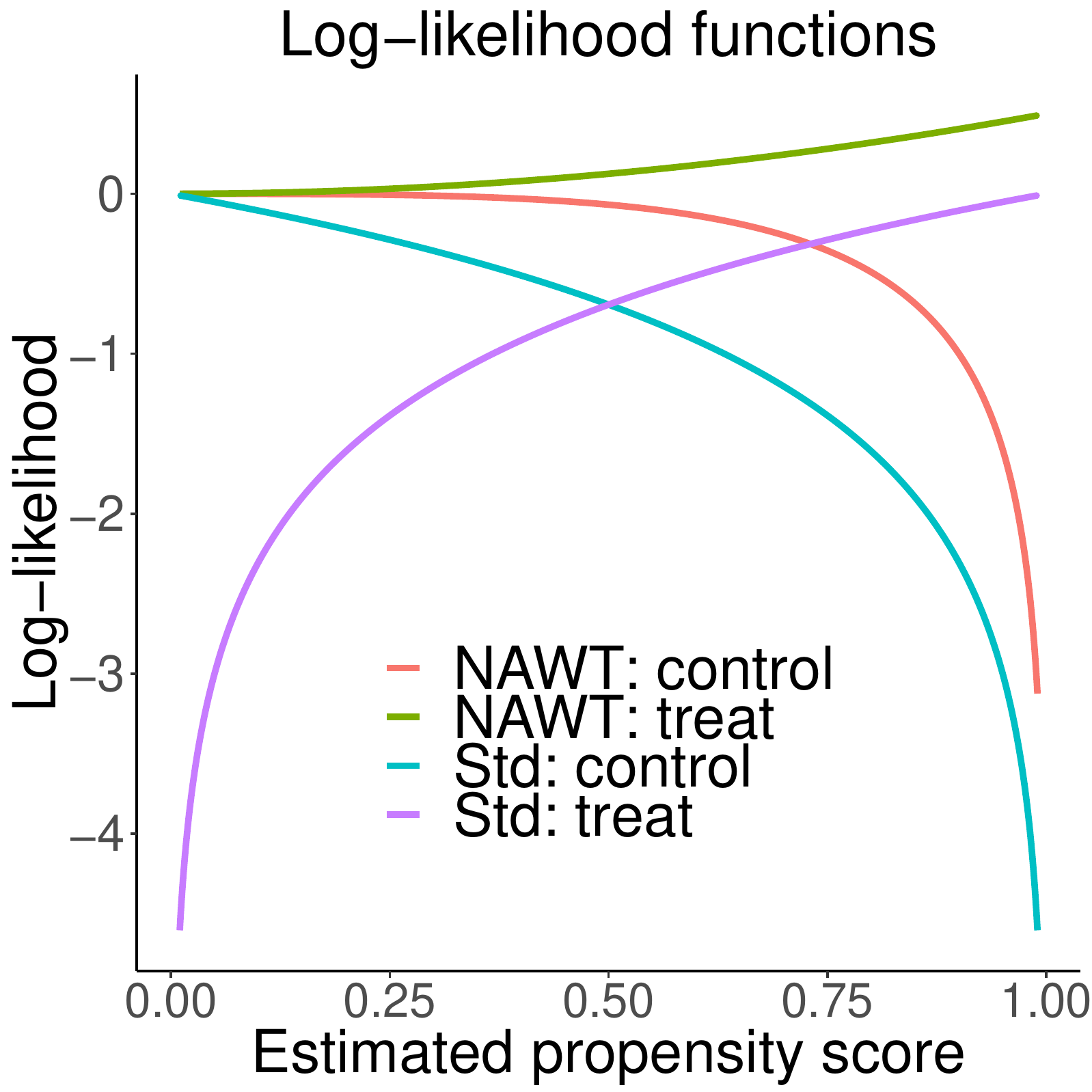}
\includegraphics[width=0.325\textwidth]{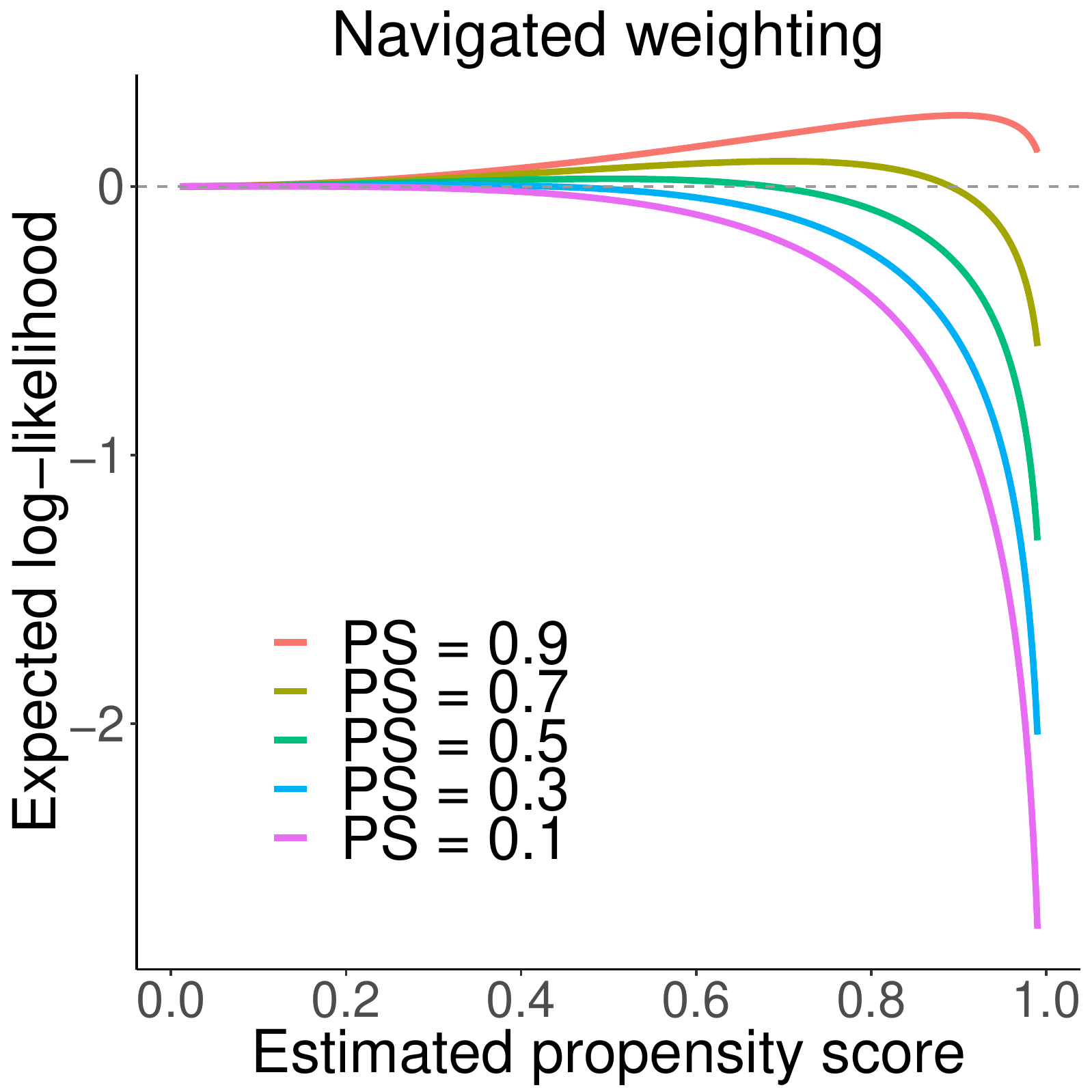}
\includegraphics[width=0.325\textwidth]{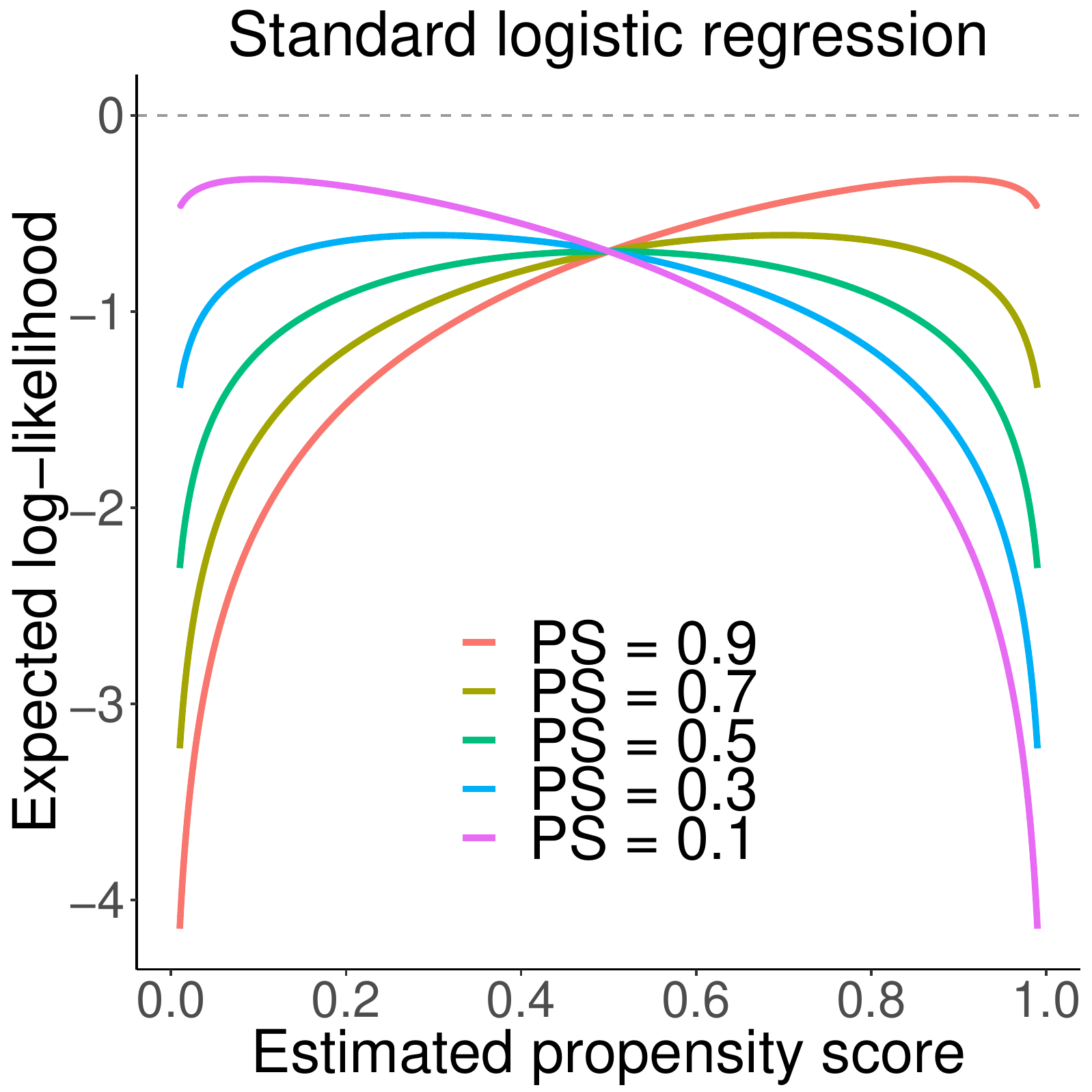}
\caption{The left panel of the figure shows the (pseudo-) log-likelihoods of the navigated weighting (NAWT) for the ATT estimation with $\omega(\pi_\beta(\mathbf{x}_i)) = \pi_\beta(\mathbf{x}_i)^2$ and standard logistic regression, and the center and right panels show the expected (pseudo-) log-likelihoods of the NAWT with $\omega(\pi_\beta(\mathbf{x}_i)) = \pi_\beta(\mathbf{x}_i)^2$ and standard logistic regression, respectively. In the left panel, the green and purple curves represent the (pseudo-) log-likelihood for the treatment units and the red and blue curves represent the (pseudo-) log-likelihood for the control units whose estimated propensity scores are shown along the x-axis from $0.01$ to $0.99$. In the center and right panels, the expected (pseudo-) log-likelihood functions for the units whose true propensity scores are $0.9$, $0.7$, $0.5$, $0.3$, and $0.1$ are represented by the red, yellow, green, blue, and purple curves, respectively, along with estimated propensity scores on the x-axis.} \label{fig_ell}
\end{figure}

To convey the intuition behind the NAWT, Figure~\ref{fig_ell} presents the (pseudo-) log-likelihoods of the NAWT for the ATT estimation with $\omega(\pi_\beta(\mathbf{x}_i)) = \pi_\beta(\mathbf{x}_i)^2$ and standard logistic regression in the left panel, and the expected (pseudo-) log-likelihoods of the NAWT with $\omega(\pi_\beta(\mathbf{x}_i)) = \pi_\beta(\mathbf{x}_i)^2$ and standard logistic regression, respectively in the center and right panels. In the left panel, the green and purple curves represent the (pseudo-) log-likelihood for the treatment units and the red and blue curves represent the log-likelihood for the control units whose estimated propensity scores are shown along the x-axis from $0.01$ to $0.99$. The (pseudo-) log-likelihoods for treatment units are
\begin{align}
l_{\mathrm{ATT}, 1}(\pi_\beta(\mathbf{x}_i)) &= \frac{\pi_\beta(\mathbf{x}_i)^2}{2} \\
l_{\mathrm{MLE}, 1}(\pi_\beta(\mathbf{x}_i)) &= \log(\pi_\beta(\mathbf{x_i})) ,
\end{align}
the (pseudo-) log-likelihoods for control units are
\begin{align}
l_{\mathrm{ATT}, 0}(\pi_\beta(\mathbf{x}_i)) &= -\frac{\pi_\beta(\mathbf{x}_i)^{1 + 2} {_2F_1}(1, 1 + 2, 2 + 2, \pi_\beta(\mathbf{x}_i))}{1 + 2} \\
l_{\mathrm{MLE}, 0}(\pi_\beta(\mathbf{x}_i)) &= \log(1 - \pi_\beta(\mathbf{x_i})) ,
\end{align}
and the expected (pseudo-) log-likelihoods are
\begin{align}
l_{\mathrm{ATT}, e}(\pi_\beta(\mathbf{x}_i)) &= \pi_\beta(\mathbf{x}_i) \ l_{\mathrm{ATT}, 1}(\pi_\beta(\mathbf{x}_i)) + (1 - \pi_\beta(\mathbf{x}_i)) \ l_{\mathrm{AO}, 0}(\pi_\beta(\mathbf{x}_i)) \\
l_{\mathrm{MLE}, e}(\pi_\beta(\mathbf{x}_i)) &= \pi_\beta(\mathbf{x}_i) \ l_{\mathrm{MLE}, 1}(\pi_\beta(\mathbf{x}_i)) + (1 - \pi_\beta(\mathbf{x}_i)) \ l_{\mathrm{MLE},0}(\pi_\beta(\mathbf{x}_i)) .
\end{align}

In the left panel, the green curve which represents the pseudo-log-likelihood of the NAWT for treatment units does not increase much even when estimated propensity scores increase, which demonstrates that the NAWT for the ATT estimation is influenced by the treatment units quite a little. In contrast, the pseudo-log-likelihood for control units, the red curve, steeply decreases as estimated propensity scores approach 1, which indicates that the estimation of the NAWT is dominated by the control units. On the other hand, the standard MLE symmetrically places importance on treatment and control units as shown in its symmetric curves in the left and right panels.

In the center and right panels, the expected (pseudo-) log-likelihood functions for the units whose true propensity scores are $0.9$, $0.7$, $0.5$, $0.3$, and $0.1$ are represented by the red, yellow, green, blue, and purple curves, respectively, along with estimated propensity scores on the x-axis. the center panel shows that the NAWT places more weights on units with \textit{large estimated propensity scores} and its estimation is also dominated by units with \textit{small true propensity scores} as they heavily drop where estimated propensity scores are large, which may sound counter-intuitive. However, these results are reasonable because the probability of $t_i = 0$ for units with small true propensity scores are much higher than units with large true propensity scores and thus it heavily decreases the pseudo-log-likelihood if their propensity scores are estimated as large. These results imply that the estimation of the NAWT for the ATT estimation is anchored by units with small true propensity scores so that they do not have large inverse probability weights, which leads to robust and efficient estimation.

\section{The NAWT for the missing outcome problem}\label{sec_missing}
Here, I consider the missing outcome problem where each unit has an outcome $y_i \in \mathbb{R}$ but we cannot observe outcomes of some of them and a missingness indicator $m_i \in \{0, 1\}$ is introduced to denote the missing units which takes $m_i = 1$ for units with missing outcomes and $m_i = 0$ for those with non-missing outcomes. Note that we can observe covairates of all the units, including the missing units. The quantity of interest here is the average outcome (AO): 
\begin{equation}
\mu \equiv \mathbb{E}[y_i].
\end{equation}
To identify this quantity, I make the following two assumptions. The first one is the conditional ignorability of missing assumption, or the missing-at-random assumption, which says that the missing is ignorable conditional on the observed covariates. This implies that the missing and non-missing units have the same expected outcome conditional on the covariates.
\begin{assum}[Conditional ignorability of missingness]\label{as_conditionalignorabilitym}
\begin{equation} \label{eq_conditionalignorabilitym}
m_i \ \indep \ y_i \mid \mathbf{x}_i = \mathbf{x} .
\end{equation}
\end{assum}

I introduce the propensity score for missingness, which is the probability of missingness given covariates $\pi(\mathbf{x}) \equiv \Pr(m_i = 1 \mid \mathbf{x}_i = \mathbf{x})$. The second assumption is the positivity assumption that the probability of non-missing is bounded away from 0.
\begin{assum}[Positivity of the non-missing probability]\label{as_positivitym}
\begin{equation}
1 - \pi(\mathbf{x}) > 0.
\end{equation}
\end{assum}
The IPW produces the pseudo-population that would have been observed if there had been no missingness by re-weighting non-missing units with the inverse probability of non-missing conditional on the covairates $w(\mathbf{x}) \equiv 1 / (1 - \pi(\mathbf{x}_i))$.

This case (missing outcome problem) can be seen as the special case of the causal inference case. Specifically, we can think of the missing outcome problem as the special case of the ATE estimation where the potential outcomes with treatment are 0 for all units. Moreover, the estimation of the average outcome (AO) can be seen as the estimation of the average outcome for both $m_i = 1$ and $m_i = 0$, which can be estimated by estimating average outcome for $m_i = 1$ using non-missing units $m_i = 0$ with inverse probability weights $\pi_\beta(\mathbf{x}_i) / (1 - \pi_\beta(\mathbf{x}_i))$ and estimating average outcome for $m_i = 0$ using non-missing units $m_i = 0$ with (inverse probability) weights $1$; resulting $1 + \hat{\pi}_\beta(\mathbf{x}_i) / (1 - \hat{\pi}_\beta(\mathbf{x}_i)) = 1 / (1 - \hat{\pi}_\beta(\mathbf{x}_i))$ as weights for the non-missing units. This implies that we should use the same weighted score function for the propensity score estimation for the AO estimation as the ATT estimation.

We can estimate the quantity of interest via the M-estimation using the weighted score condition and the conditions for the AO estimation. For example, the joint conditions for estimating the AO with the difference-in-means type estimator are:
\begin{align}
\sum_{i = 1}^n s_{\mathrm{AO}}(\beta, m_i, \mathbf{x}_i) &= 0 \\
\sum_{i = 1}^n q_{\mathrm{AO}}(\mu, \beta, m_i, \mathbf{x}_i) &= 0 ,
\end{align}
where 
\begin{align}
s_{\mathrm{AO}}(\beta, m_i, \mathbf{x}_i) 
&\equiv \left( \frac{m_i}{\pi_\beta(\mathbf{x}_i)} - \frac{1 - m_i}{1 -  \pi_\beta(\mathbf{x}_i)} \right) \omega(\pi_\beta(\mathbf{x}_i)) \pi'_\beta(\mathbf{x}_i) \\
q_{\mathrm{AO}}(\mu, \beta, m_i, \mathbf{x}_i) 
&\equiv \frac{(1 - m_i) (y_i - \mu)}{1 - \pi_\beta(\mathbf{x}_i)} .
\end{align}

\section{Asymptotic distribution of the NAWT estimates (ATT)}\label{sec_asymptoticdist}

I consider asymptotic variance-covariance matrix of the weighted difference-in-means estimator for the ATT estimation here:
\begin{equation}
\sqrt{n} \left( \begin{array}{l}
\hat{\beta}_\mathrm{ATT} - \beta \\
\hat{\tau}_\mathrm{ATT} - \tau_\mathrm{ATT}
\end{array}
\right) \xrightarrow[d]{} \mathcal{N}(0, \mathbf{H}^{-1} \mathbf{\Sigma} \mathbf{H}^{-\tr}) ,
\end{equation}
where hessian $\mathbf{H}$ is
\begin{align}
\mathbf{H} &\equiv \left( \begin{array}{ll}
\mathbf{H}_{\beta\beta} & \mathbf{0} \\
\mathbf{H}_{\tau\beta} & \mathbf{H}_{\tau\tau}
\end{array}
\right) ,
\end{align}
where $\mathbf{H}_{\tau\beta}$ is
\begin{align}
\mathbf{H}_{\tau\beta} 
&\equiv \mathbb{E} \left[ \frac{\partial q_{\mathrm{ATT}}(\tau, \beta, \mathbf{t}, \mathbf{X})}{\partial \beta} \right] \\
\begin{split}
&= -n \mathbb{E} \left[ \frac{1}{\sum \frac{(1 - t_i) \pi_\beta(\mathbf{x}_i)}{1 - \pi_\beta(\mathbf{x}_i)}} \frac{(1 - t_i) \pi_\beta(\mathbf{x}_i) (y_i - \mu_0)}{1 - \pi_\beta(\mathbf{x}_i)} \mathbf{x}_i^\tr \right] \\
&= -\frac{n}{n_1} \mathbb{E} \left[ \pi_\beta(\mathbf{x}_i) (y_i - \mu_0) \mathbf{x}_i^\tr \right] ,
\end{split}
\end{align}
$\mathbf{H}_{\tau\tau}$ is
\begin{align}
\mathbf{H}_{\tau\tau} 
&\equiv \mathbb{E} \left[ \frac{\partial q_{\mathrm{ATT}}(\tau, \beta, \mathbf{t}, \mathbf{X})}{\partial \tau} \right] \\
\begin{split}
&= -1 ,
\end{split}
\end{align}
and $\mathbf{\Sigma}$ is
\begin{align}
\mathbf{\Sigma} &\equiv \left( \begin{array}{ll}
\mathbf{\Sigma}_{\beta\beta} & \mathbf{\Sigma}_{\beta\tau} \\
\mathbf{\Sigma}_{\tau\beta} & \mathbf{\Sigma}_{\tau\tau}
\end{array}
\right) \\
&= \mathbb{E} \left[ \left( s_{\mathrm{ATT}}(\beta, t_i, \mathbf{x}_i), \ q_{\mathrm{ATT}}(\tau, \beta, t_i, \mathbf{x}_i) \right) \left( s_{\mathrm{ATT}}(\beta, t_i, \mathbf{x}_i), \ q_{\mathrm{ATT}}(\tau, \beta, t_i, \mathbf{x}_i) \right)^\tr \right] .
\end{align}

\section{Asymptotic distribution of the NAWT estimates (ATE)}\label{sec_asymptoticdistate}

I consider asymptotic variance-covariance matrix of the weighted difference-in-means estimator for the ATE estimation here:
\begin{equation}
\sqrt{n} \left( \begin{array}{l}
\hat{\beta}_\mathrm{ATE}^0 - \beta \\
\hat{\beta}_\mathrm{ATE}^1 - \beta \\
\hat{\mu}_0 - \mu_0 \\
\hat{\mu}_1 - \mu_1 \\
\hat{\tau}_\mathrm{ATE} - \tau_\mathrm{ATT}
\end{array}
\right) \xrightarrow[d]{} \mathcal{N}(0, \mathbf{H}^{-1} \mathbf{\Sigma} \mathbf{H}^{-\tr}) ,
\end{equation}
where hessian $\mathbf{H}$ is
\begin{align}
\mathbf{H} &\equiv \left( \begin{array}{lllll}
\mathbf{H}_{\beta^0 \beta^0} & \mathbf{0} & \mathbf{0} & \mathbf{0} & \mathbf{0} \\
\mathbf{H}_{\mu_0 \beta^0} & \mathbf{H}_{\mu_0 \mu_0} & \mathbf{0} &  \mathbf{0} & 0 \\
\mathbf{0} & \mathbf{0} & \mathbf{H}_{\beta^1 \beta^1} & \mathbf{0} & \mathbf{0} \\
\mathbf{0} & \mathbf{0} & \mathbf{H}_{\mu_1 \beta^1} & \mathbf{H}_{\mu_1 \mu_1} & 0 \\
\mathbf{0} & \mathbf{H}_{\tau \mu_0} & \mathbf{0} & \mathbf{H}_{\tau \mu_1} & \mathbf{H}_{\tau \tau}
\end{array}
\right) ,
\end{align}
where 
\begin{align}
\mathbf{H}_{\mu_0 \beta^0} 
&= -\mathbb{E} \left[ \pi_\beta^0(\mathbf{x}_i) (y_i(0) - \mu_0) \mathbf{x}_i^\tr \right] \\
\mathbf{H}_{\mu_1 \beta^1} 
&= -\mathbb{E} \left[ \pi_\beta^1(\mathbf{x}_i) (y_i(1) - \mu_1) \mathbf{x}_i^\tr \right] \\
\mathbf{H}_{\mu_0 \mu_0} = \mathbf{H}_{\mu_1 \mu_1} = \mathbf{H}_{\tau \mu_0} = \mathbf{H}_{\tau \tau} &= -1 \\
\mathbf{H}_{\tau \mu_1} &= 1 .
\end{align}

\newpage
\section{Large-sample performance of the NAWT}\label{sec_simeff}

Equation~\eqref{eq_relimpact} suggests that units with outcomes largely deviated from the mean, as well as units with large estimated propensity scores, are important on the ATT estimation. To demonstrate how these two factors affect the performance of the NAWT with a large sample, I conduct a thorough simulation in a broad range of settings.

Specifically, the following settings are examined. The true outcome model with and without treatment is $y_i(0) = b_{0,1} x_i + b_{0,2} x_i^2 + b_{0,3} x_i^3$ and $y_i(1) = b_{1,1} x_i + b_{1,2} x_i^2 + b_{1,3} x_i^3$, where $x_i$ is independently and identically distributed according to the standard normal distribution truncated at $-4$ and 4, and the binary treatment is assigned with probability $\pi_i = 1 / (1 + \exp(-x_i))$. There are 495 scenarios for each of the ATT and ATE estimation, where the scenarios consist of all the combination of the coefficients $b_{0,1} \in \{ -5, -4, \ldots, 4, 5 \}$, $b_{0,2} \in \{ -1, -0.5, 0, 0.5, 1 \}$, $b_{0,3} \in \{ -1, 0, 1 \}$, and $(b_{1,1}, b_{1,2}, b_{1,3}) \in \{ (0, 0, 1), (0, 1, 0), (-1, 0, 1) \}$. To approximate large-sample performance, the number of units in the simulation is 10 million. The performances of the NAWT with weighting functions $\omega(\pi_\beta(\mathbf{x}_i)) = \pi_\beta(\mathbf{x}_i)^\alpha$ with $\alpha = 4$ and $\alpha = 2$ are compared with the standard IPW ($\alpha = 0$) for the ATT estimation, and the performances of the NAWT with weighting functions $\omega(\pi_\beta(\mathbf{x}_i)) = \pi_\beta(\mathbf{x}_i)^\alpha$ and $\omega(\pi_\beta(\mathbf{x}_i)) = (1 - \pi_\beta(\mathbf{x}_i))^\alpha$ with $\alpha = 4$ and $\alpha = 2$ are compared with the standard IPW for the ATE estimation.

\begin{figure}[t]
\centering
\includegraphics[width=0.325\textwidth]{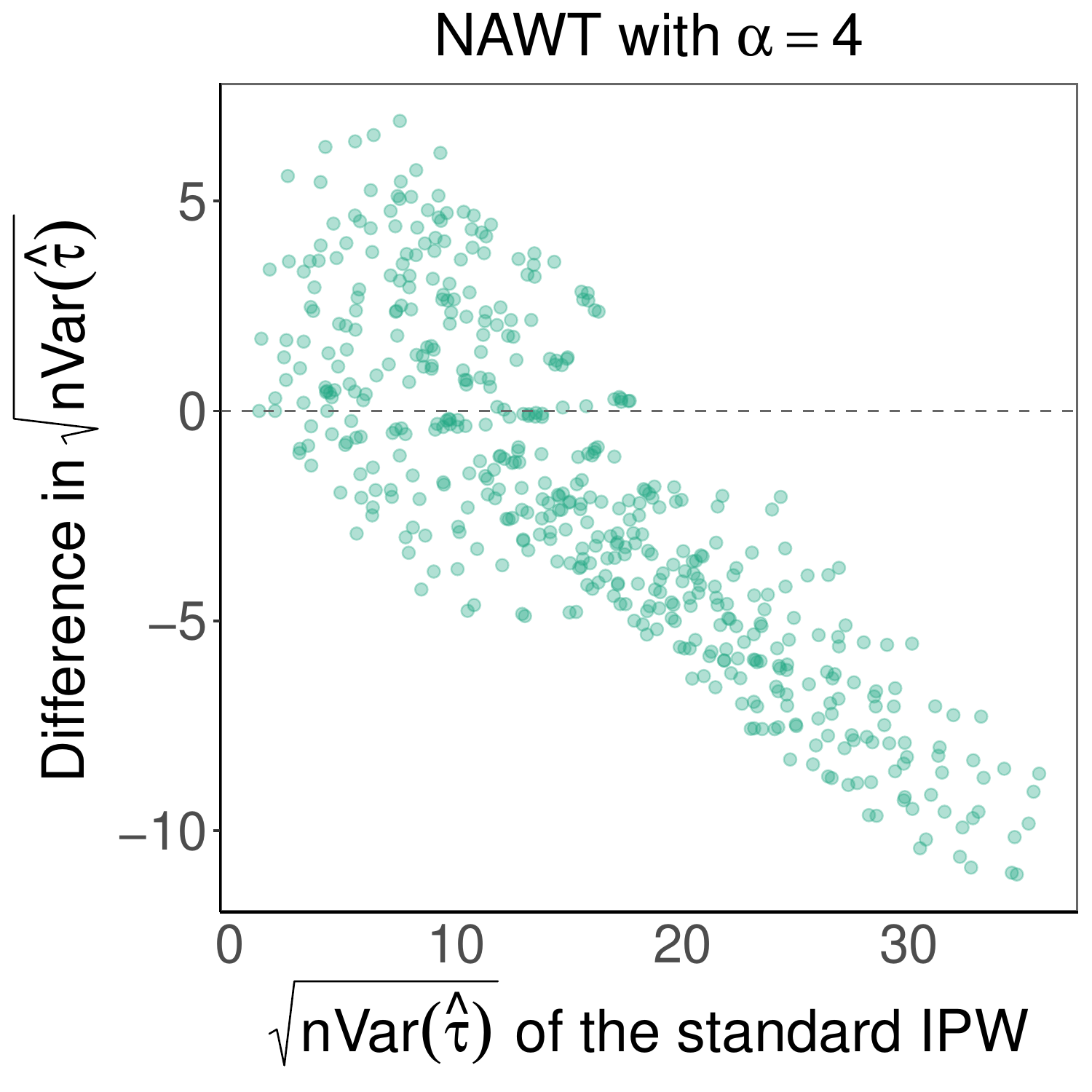}
\hspace{20pt}
\includegraphics[width=0.325\textwidth]{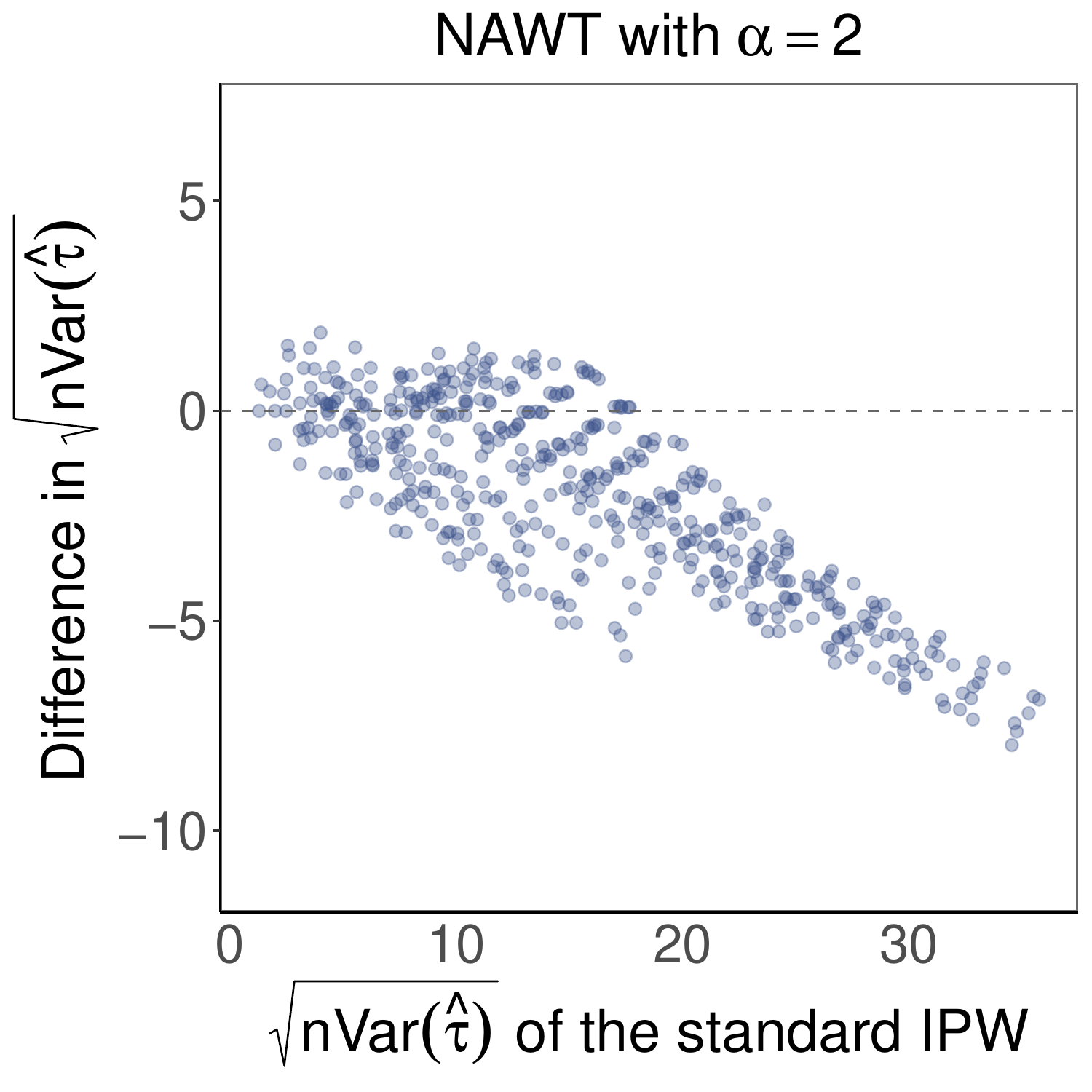}
\vspace{-8pt}
\caption{The difference in the large-sample standard errors between the NAWT and IPW for the ATT estimation against those of the standard IPW.} \label{fig_attdif}
\end{figure}

\begin{figure}[hp]
\centering
\includegraphics[width=0.935\textwidth]{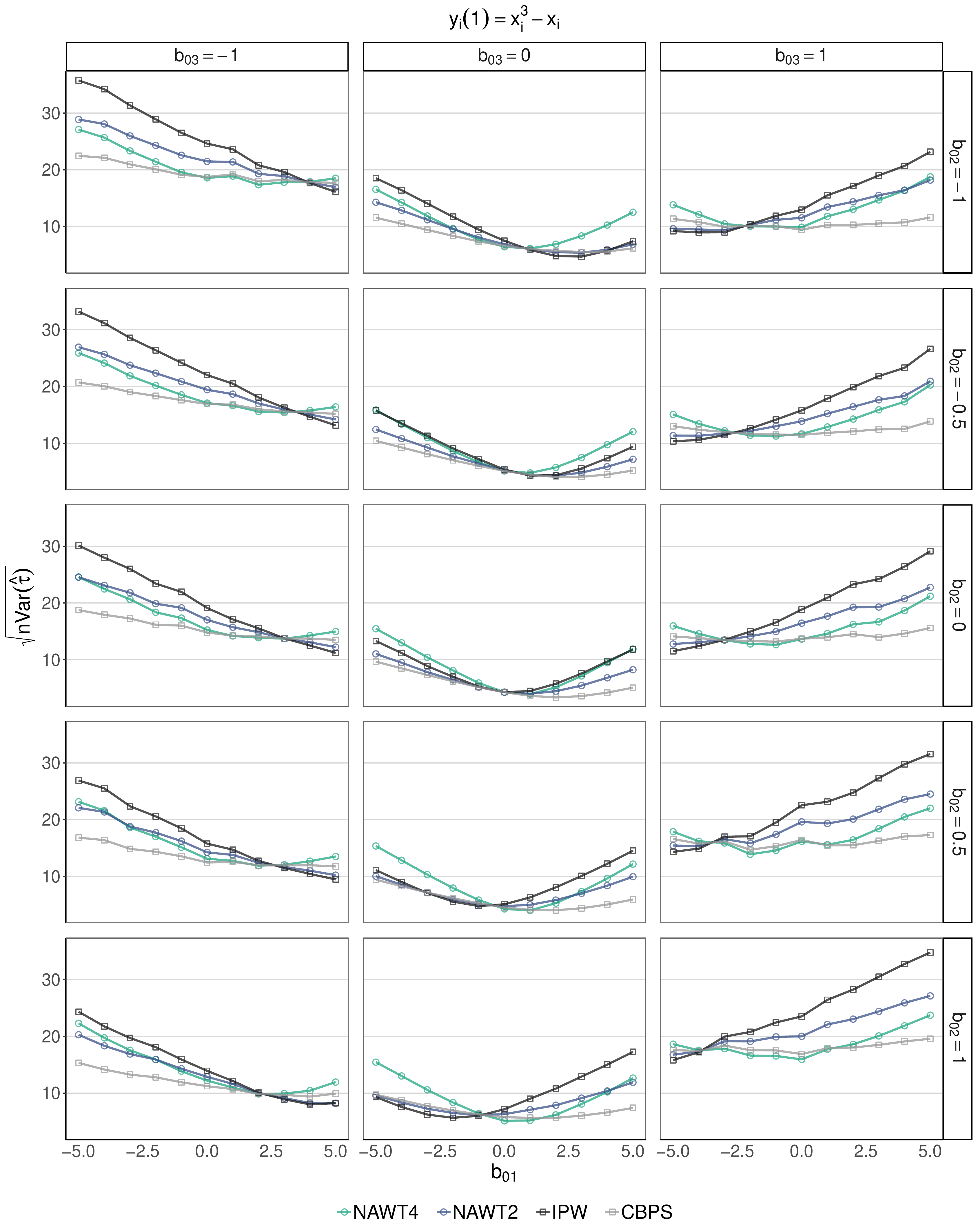}
\vspace{-12pt}
\caption{The estimates for the large-sample variance of the ATT using the NAWT with $\alpha = 4$ (NAWT4) and $\alpha = 2$ (NAWT2), standard IPW, and IPW with the CBPS under the various settings for the true outcome model for the controlled when the true outcome model for the treated is $y_i(1) = x_i^3 - x_i$.} \label{fig_varatt3}
\end{figure}

\begin{figure}[hp]
\centering
\includegraphics[width=0.935\textwidth]{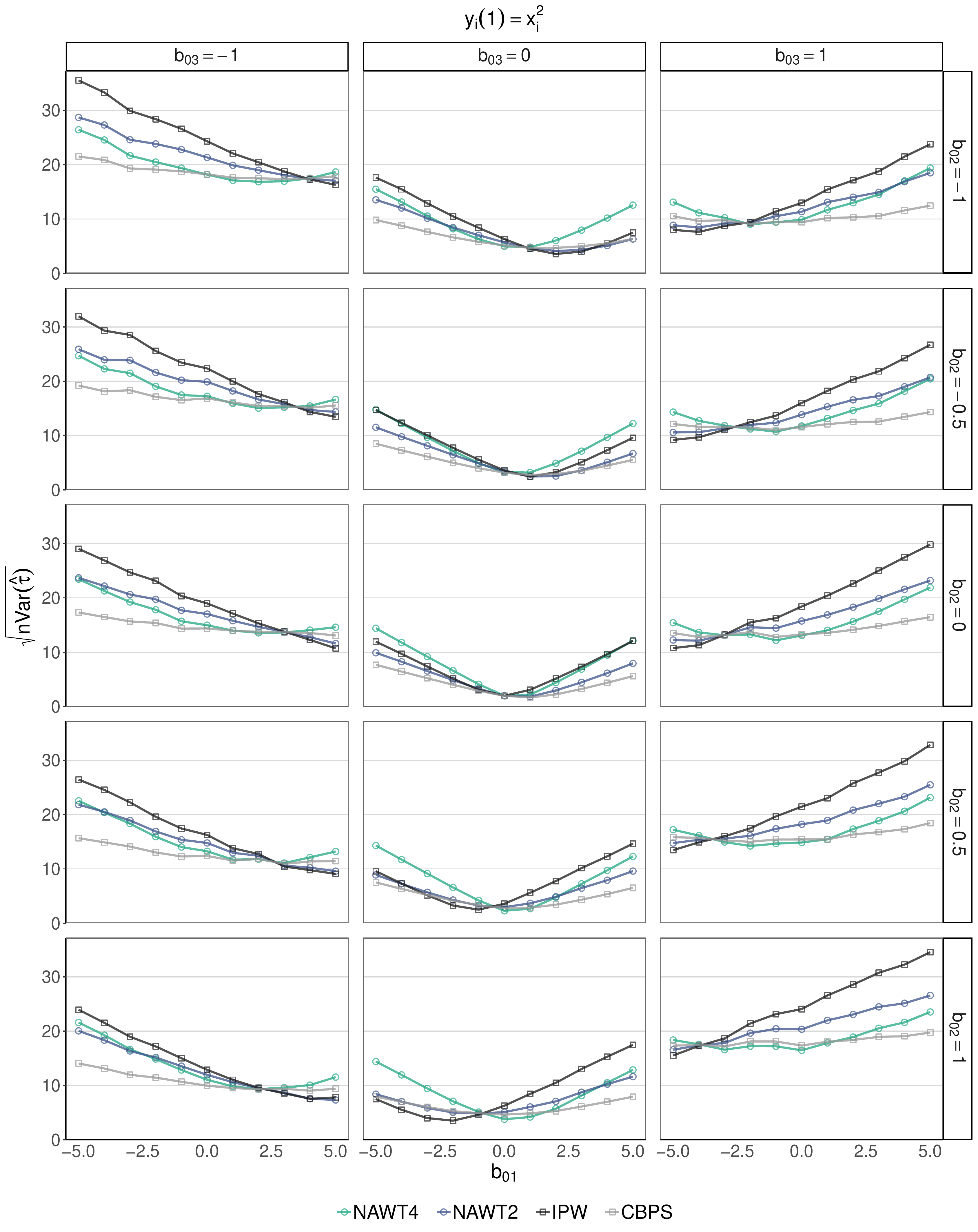}
\vspace{-12pt}
\caption{The estimates for the large-sample variance of the ATT using the NAWT with $\alpha = 4$ (NAWT4) and $\alpha = 2$ (NAWT2), standard IPW, and IPW with the CBPS under the various settings for the true outcome model for the controlled when the true outcome model for the treated is $y_i(1) = x_i^2$.} \label{fig_varatt2}
\end{figure}

\begin{figure}[hp]
\centering
\includegraphics[width=0.935\textwidth]{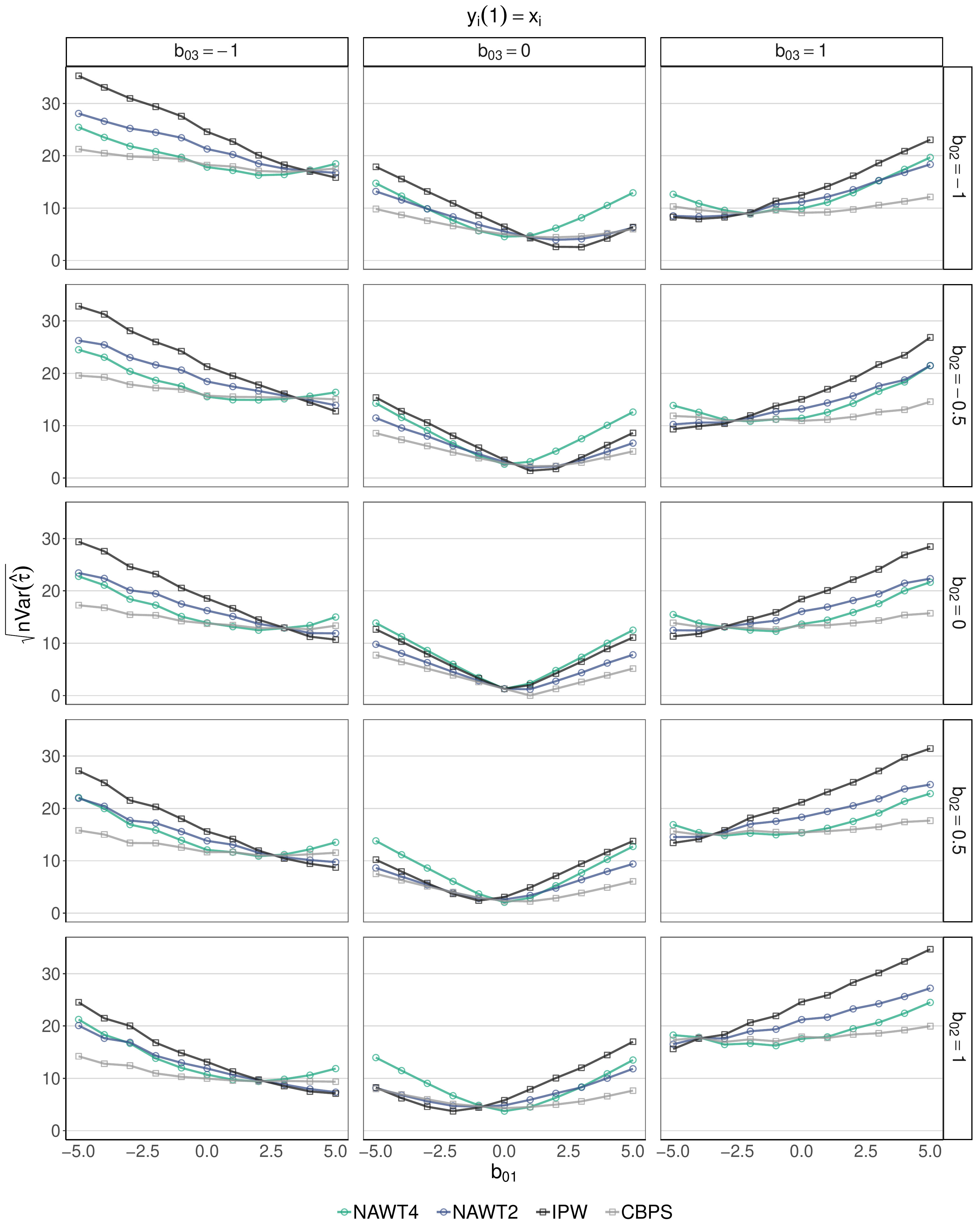}
\vspace{-12pt}
\caption{The estimates for the large-sample variance of the ATT using the NAWT with $\alpha = 4$ (NAWT4) and $\alpha = 2$ (NAWT2), standard IPW, and IPW with the CBPS under the various settings for the true outcome model for the controlled when the true outcome model for the treated is $y_i(1) = x_i$.} \label{fig_varatt1}
\end{figure}

The summary of the results for the ATT estimation are shown in Figure~\ref{fig_attdif} and the details are shown in Figure~\ref{fig_varatt3}--\ref{fig_varatt1}. Figure~\ref{fig_attdif} shows the difference in the large-sample standard errors (square root of the variance multiplied by the number of observations) between the NAWT and IPW for the ATT estimation in the y-axis against those of the standard IPW in the x-axis, where negative values in y-axis indicate smaller standard errors with the NAWT than the standard IPW. Both the NAWT with $\alpha = 4$ and $\alpha = 2$ greatly improves the standard IPW when the IPW performs poorly when units with large propensity scores have average outcomes without treatment largely deviated from the mean. This result is not dependent on the true outcome model for the treated (Figure~\ref{fig_varatt3}--\ref{fig_varatt1}). When the NAWT improves the IPW, the NAWT with $\alpha = 4$ is likely to outperform the NAWT with $\alpha = 2$ but the NAWT with $\alpha = 4$ sometimes works poorly when the IPW performs well where units with small propensity scores have average outcomes without treatment largely deviated from the mean (e.g. when $b_{01} = -5$ and $b_{03} = 1$). These results imply that an adaptive method, where the weighting function for the NAWT is selected to minimize the sample variance, is expected to work well irrespective of the data generating process, which is examined in Appendix~\ref{sec_simefffinite}.

The summary of the results for the ATE estimation are shown in Figure~\ref{fig_attdif} and the details are shown in Figure~\ref{fig_varatt3}--\ref{fig_varatt1}. The results are similar to the ATT estimation but three points should be noted. First, the true outcome model for the treated affects the performance of the NAWT. Second, the NAWT performs much better than the standard IPW in the ATE estimation, especially with $\alpha = 4$. This result is largely due to the improvement in the estimation of the mean outcomes of the treated where units with large propensity scores have average outcomes with treatment largely deviated from the mean. Third, the NAWT performs at least as well as and sometimes better than the CBPS.

\begin{figure}[t]
\centering
\includegraphics[width=0.325\textwidth]{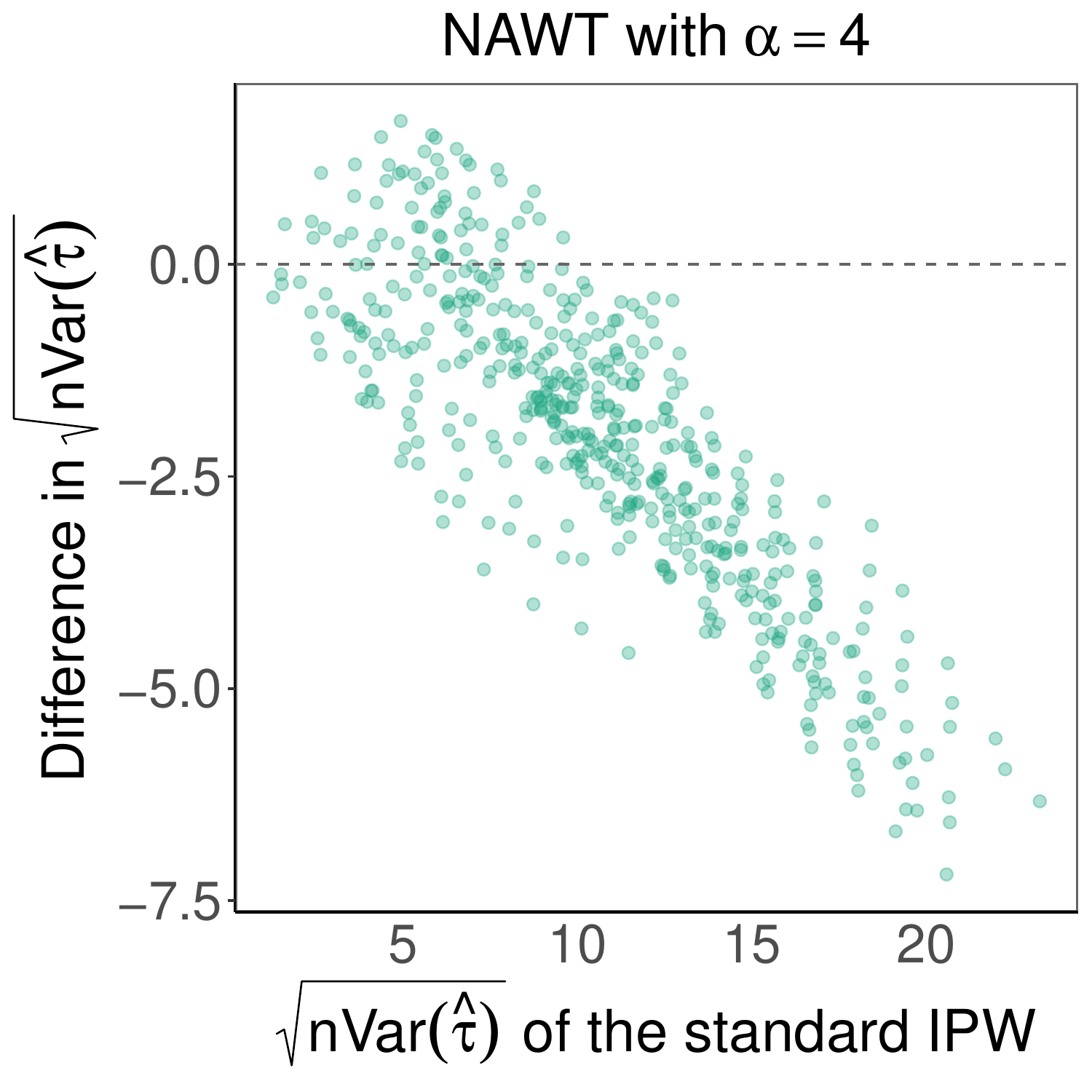}
\hspace{20pt}
\includegraphics[width=0.325\textwidth]{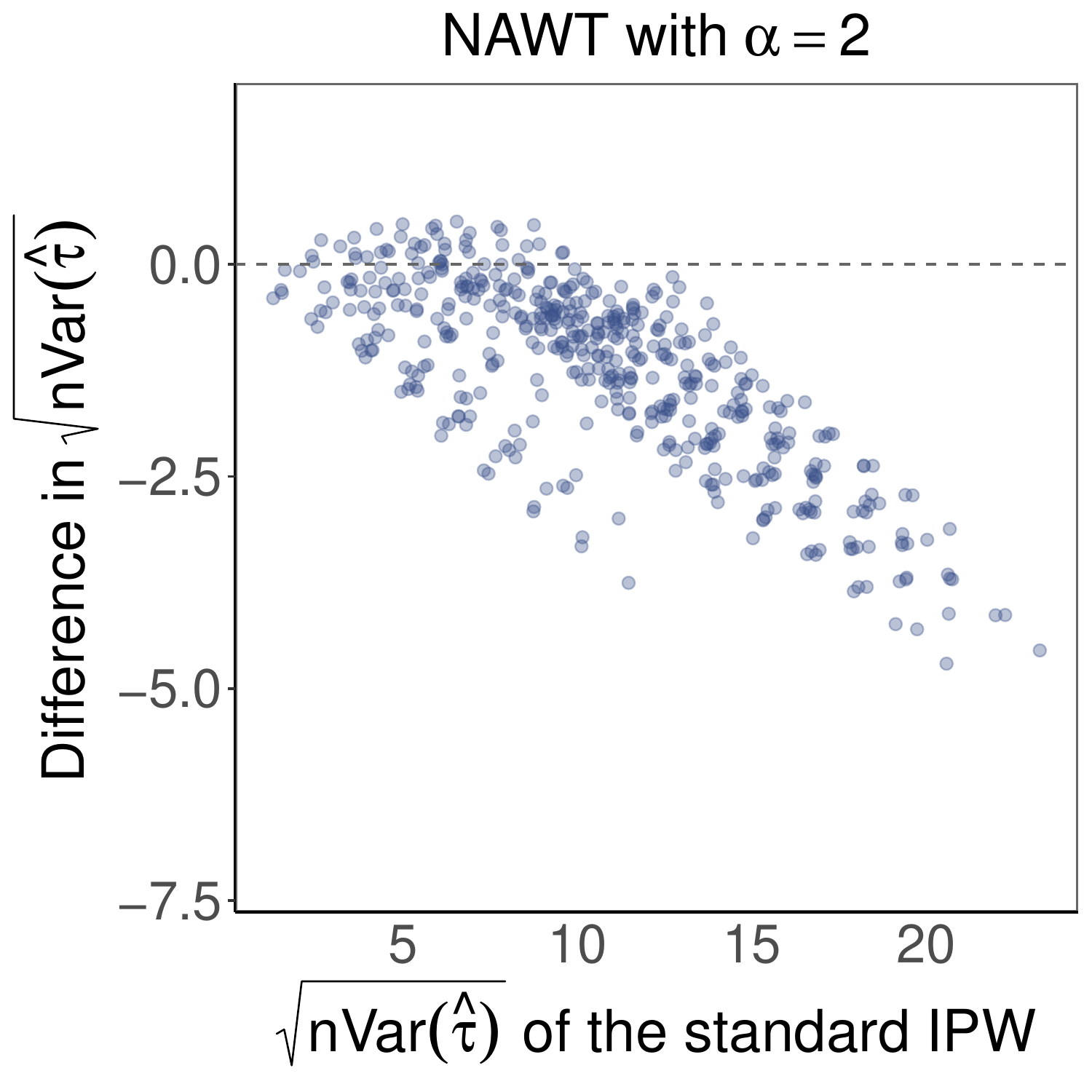}
\vspace{-8pt}
\caption{The difference in the large-sample standard errors between the NAWT and IPW for the ATE estimation against those of the standard IPW.} \label{fig_atedif}
\end{figure}

\begin{figure}[hp]
\centering
\includegraphics[width=0.935\textwidth]{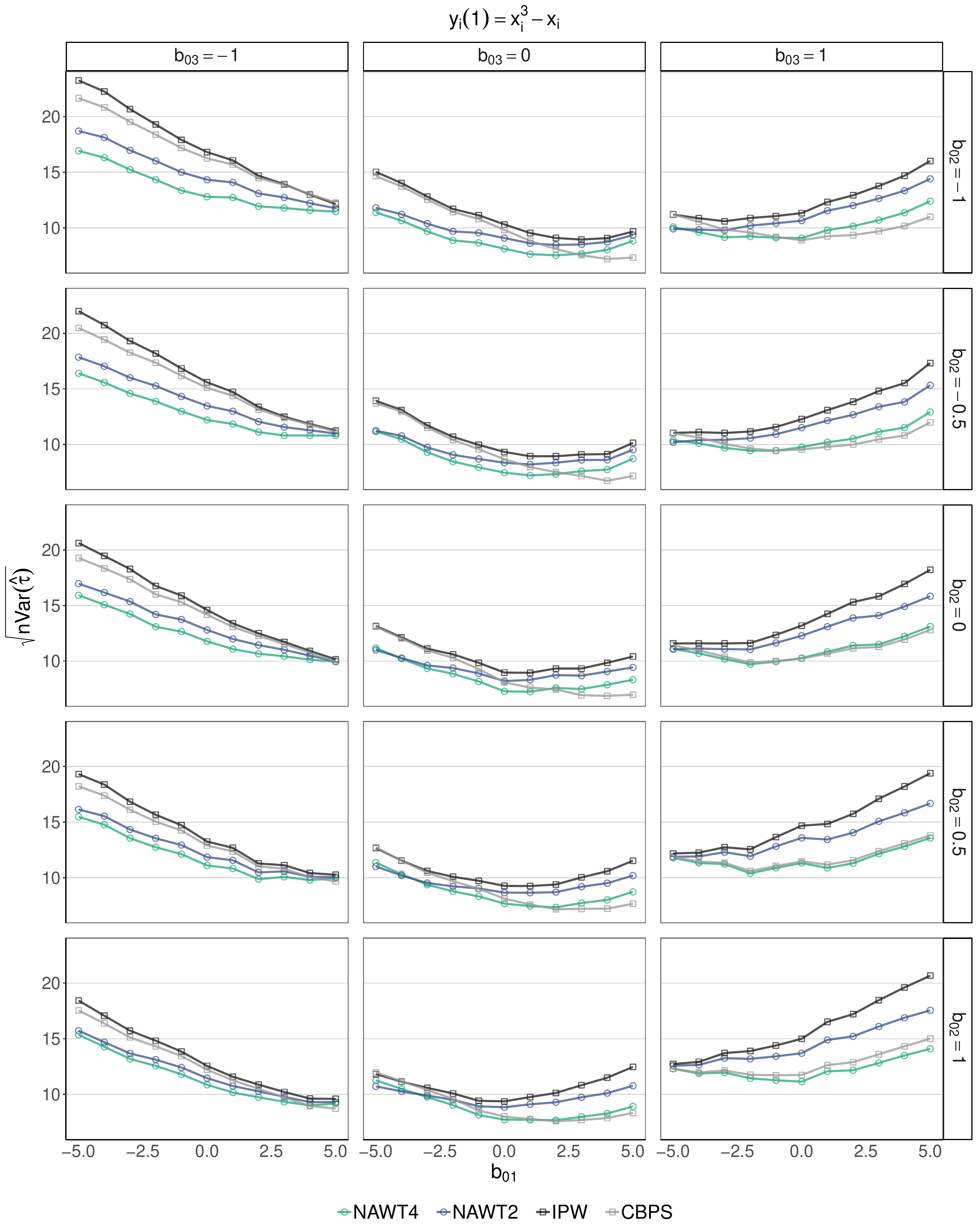}
\vspace{-12pt}
\caption{The estimates for the large-sample variance of the ATE using the NAWT with $\alpha = 4$ (NAWT4) and $\alpha = 2$ (NAWT2), standard IPW, and IPW with the CBPS under the various settings for the true outcome model for the controlled when the true outcome model for the treated is $y_i(1) = x_i^3 - x_i$.} \label{fig_varate3}
\end{figure}

\begin{figure}[hp]
\centering
\includegraphics[width=0.935\textwidth]{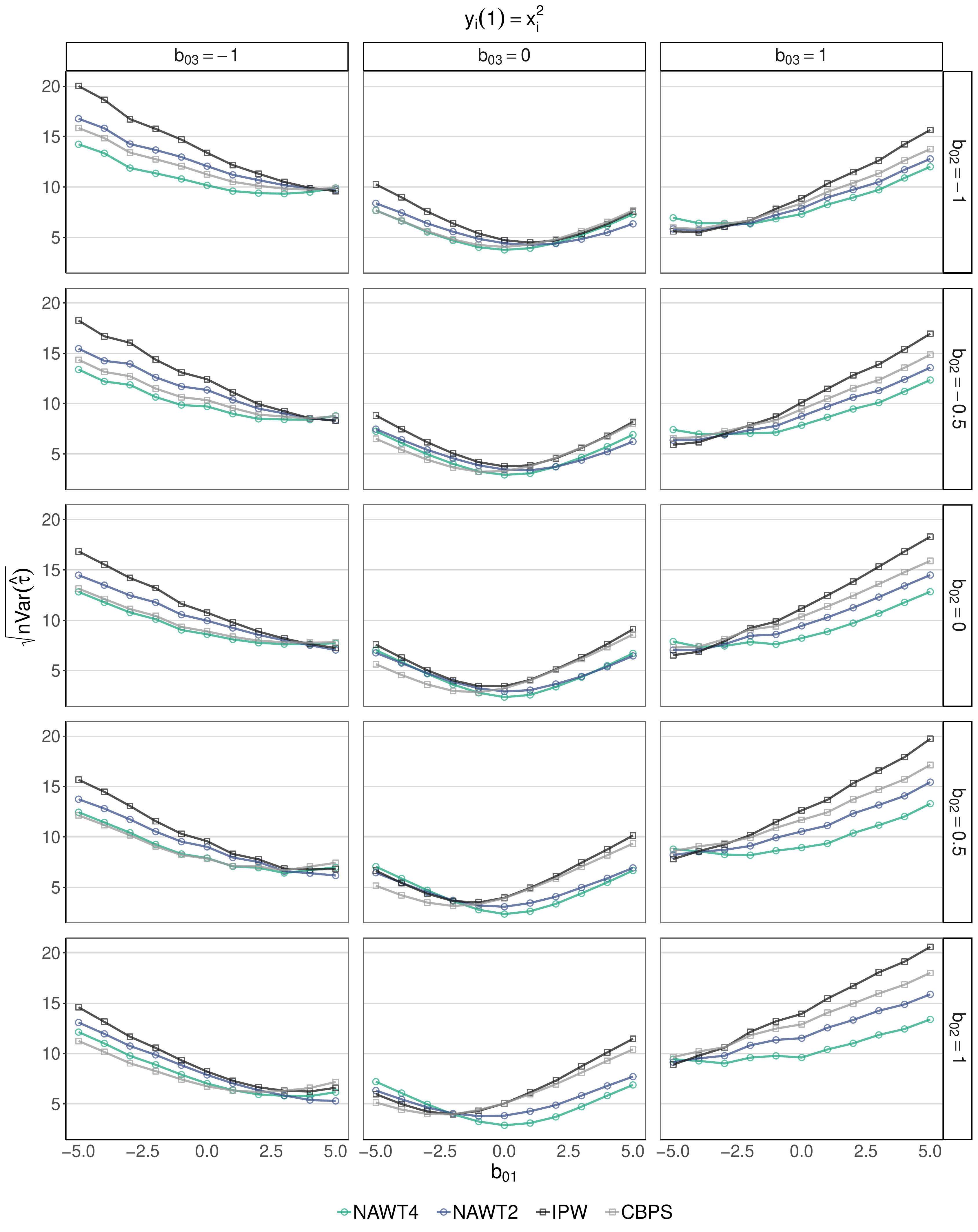}
\vspace{-12pt}
\caption{The estimates for the large-sample variance of the ATE using the NAWT with $\alpha = 4$ (NAWT4) and $\alpha = 2$ (NAWT2), standard IPW, and IPW with the CBPS under the various settings for the true outcome model for the controlled when the true outcome model for the treated is $y_i(1) = x_i^2$.} \label{fig_varate2}
\end{figure}

\begin{figure}[hp]
\centering
\includegraphics[width=0.935\textwidth]{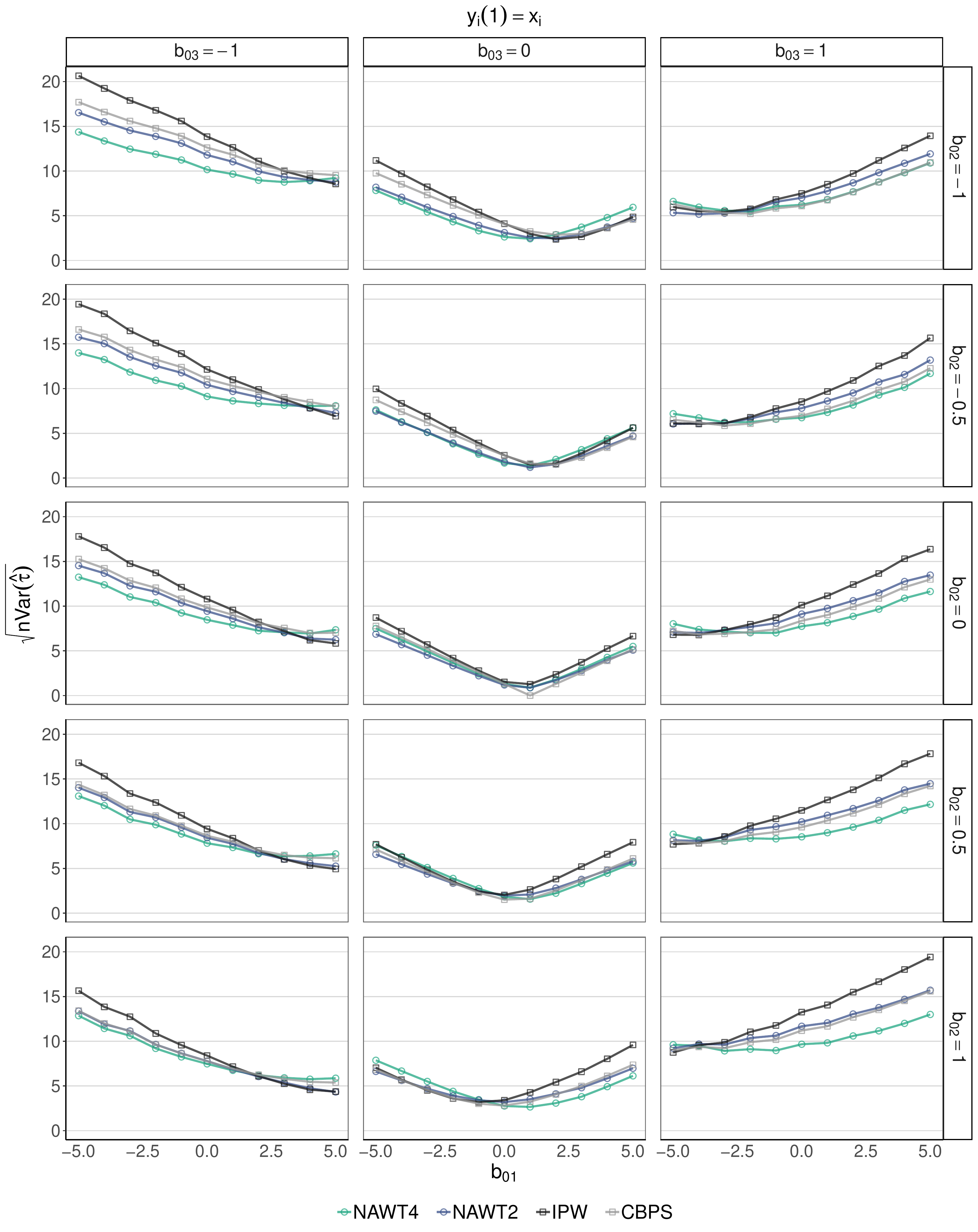}
\vspace{-12pt}
\caption{The estimates for the large-sample variance of the ATE using the NAWT with $\alpha = 4$ (NAWT4) and $\alpha = 2$ (NAWT2), standard IPW, and IPW with the CBPS under the various settings for the true outcome model for the controlled when the true outcome model for the treated is $y_i(1) = x_i$.} \label{fig_varate1}
\end{figure}

\clearpage
\section{Finite-sample performance of the NAWT}\label{sec_simefffinite}

To examine the finite-sample performance of the NAWT, this section compares the NAWT with the same weighting functions in Appendix~\ref{sec_simeff}; $\alpha = 4$ and $\alpha = 2$, and the standard IPW ($\alpha = 0$) for the ATT and ATE estimation. Besides, the performance of the adaptive NAWT, which chooses the weighting function based on the sample variance for the quantity of interest, is also examined. In this simulation study, the adaptive NAWT utilizes the weighting function which has the smallest sample variance among ones with $\alpha = \{0, 2, 4\}$.

The data generating process is the same as one used in Appendix~\ref{sec_simeff} except for the number of observations being 400. To investigate how the NAWT mitigates the bias due to propensity score model misspecification, I utilize two propensity score models for the estimation as well as the true one. The type~1 misspecification model uses $\exp(x_i / 3)$ instead of $x_i$, which results in the over-estimation of propensity scores for units with both ends of true propensity scores and the under-estimation for units with middle values of true propensity scores. The type~2 misspecification model uses $\sqrt(x_i + 4)$ instead of $x_i$, which results in the under-estimation of propensity scores for units with both ends of true propensity scores and the over-estimation for units with middle values of true propensity scores. I conduct $4,000$ Monte Carlo simulations and calculate the RMSE for each of the settings.

\begin{figure}[h]
\centering
\includegraphics[width=0.325\textwidth]{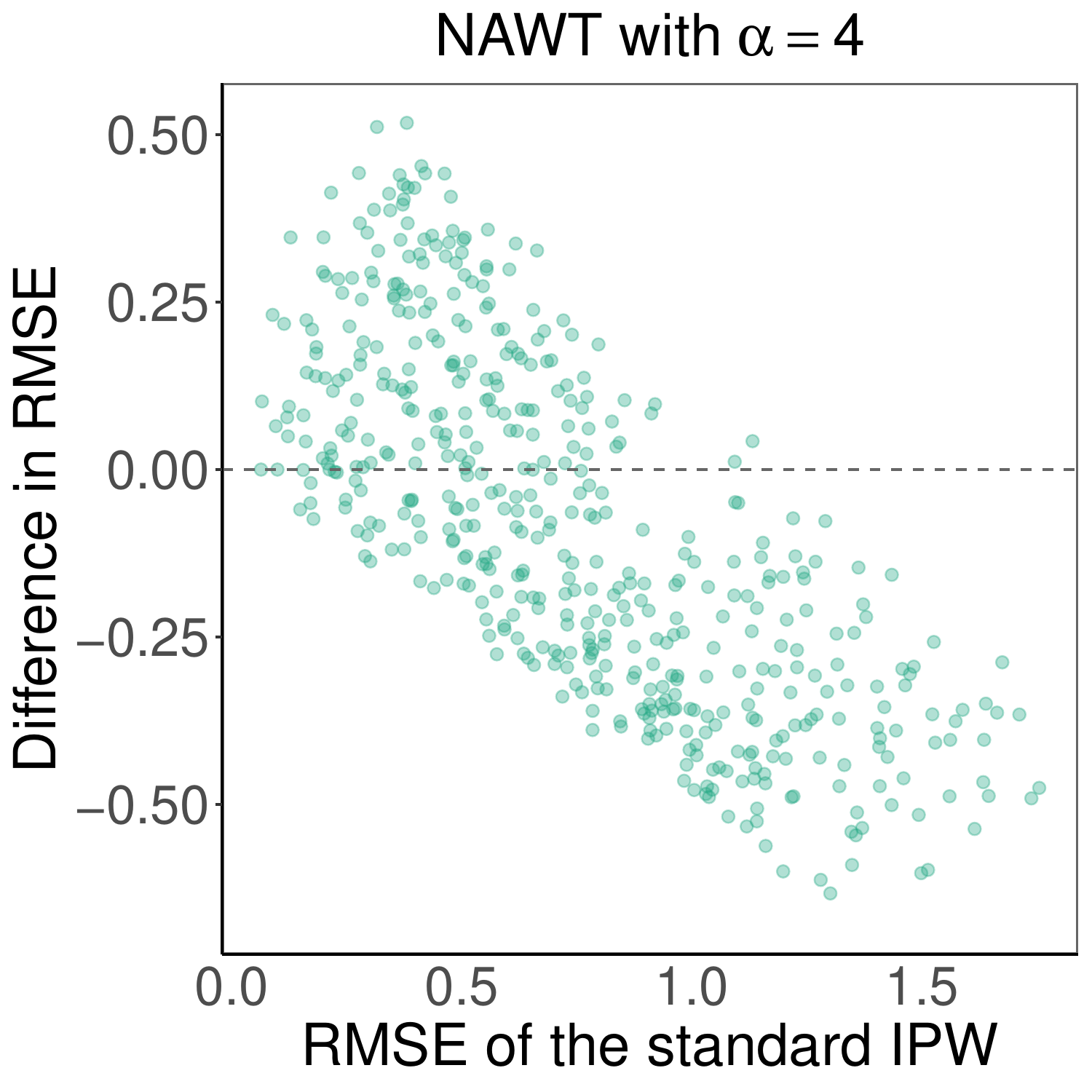}
\includegraphics[width=0.325\textwidth]{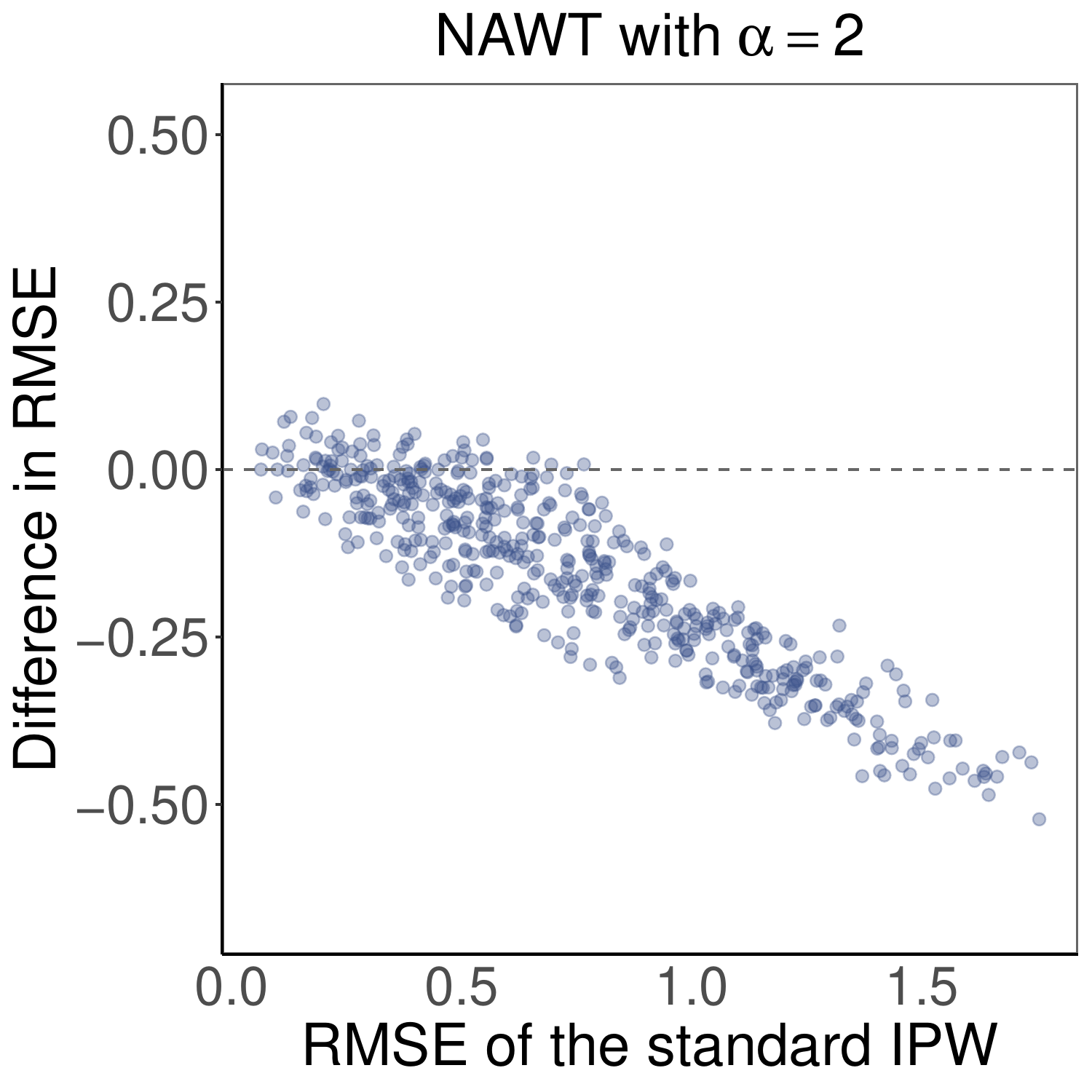}
\includegraphics[width=0.325\textwidth]{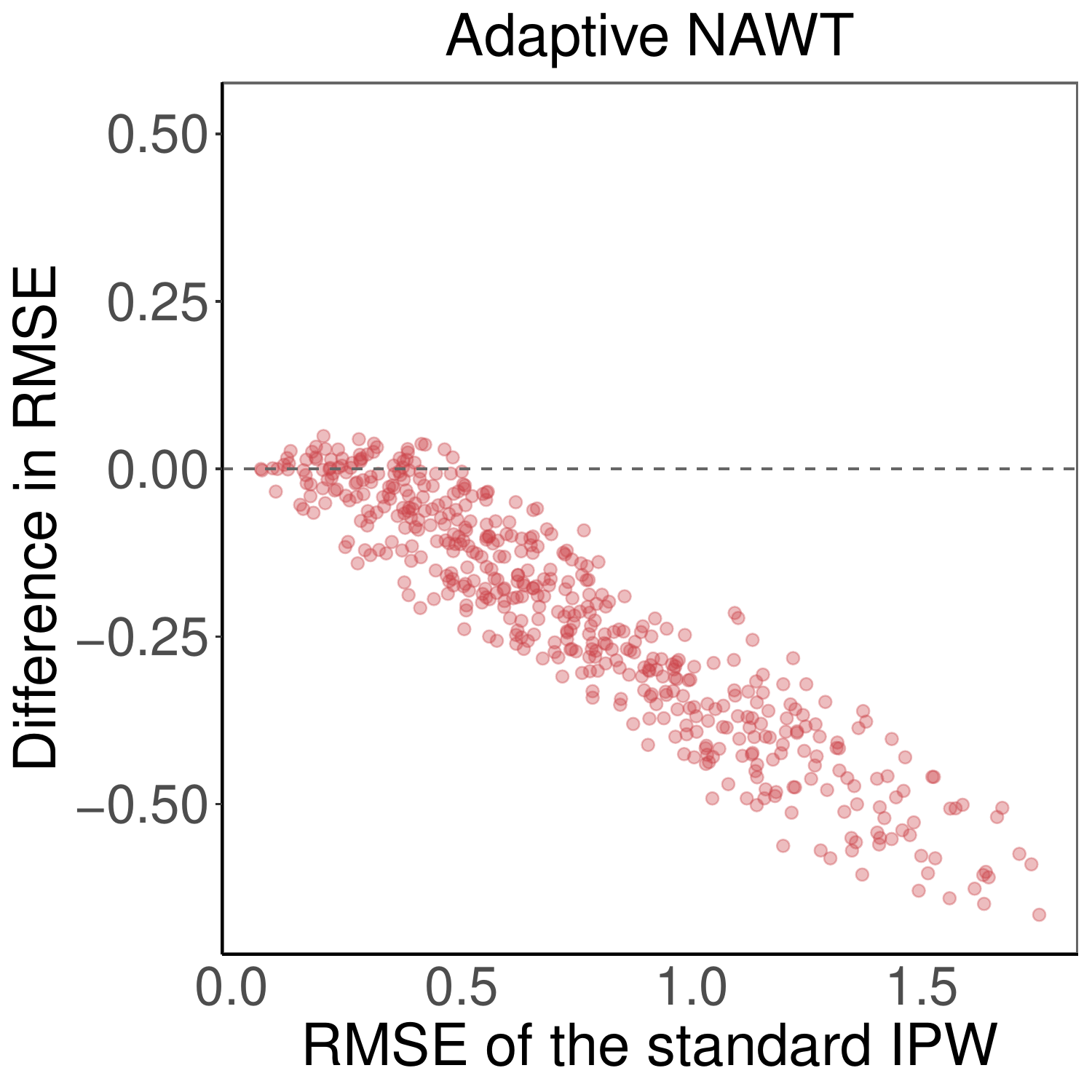}
\vspace{-8pt}
\caption{The difference in the RMSE between the NAWT and IPW for the ATT estimation against the RMSE of the standard IPW with the correct propensity score model.} \label{fig_attfindifcor}
\end{figure}

\begin{figure}[h]
\centering
\includegraphics[width=0.325\textwidth]{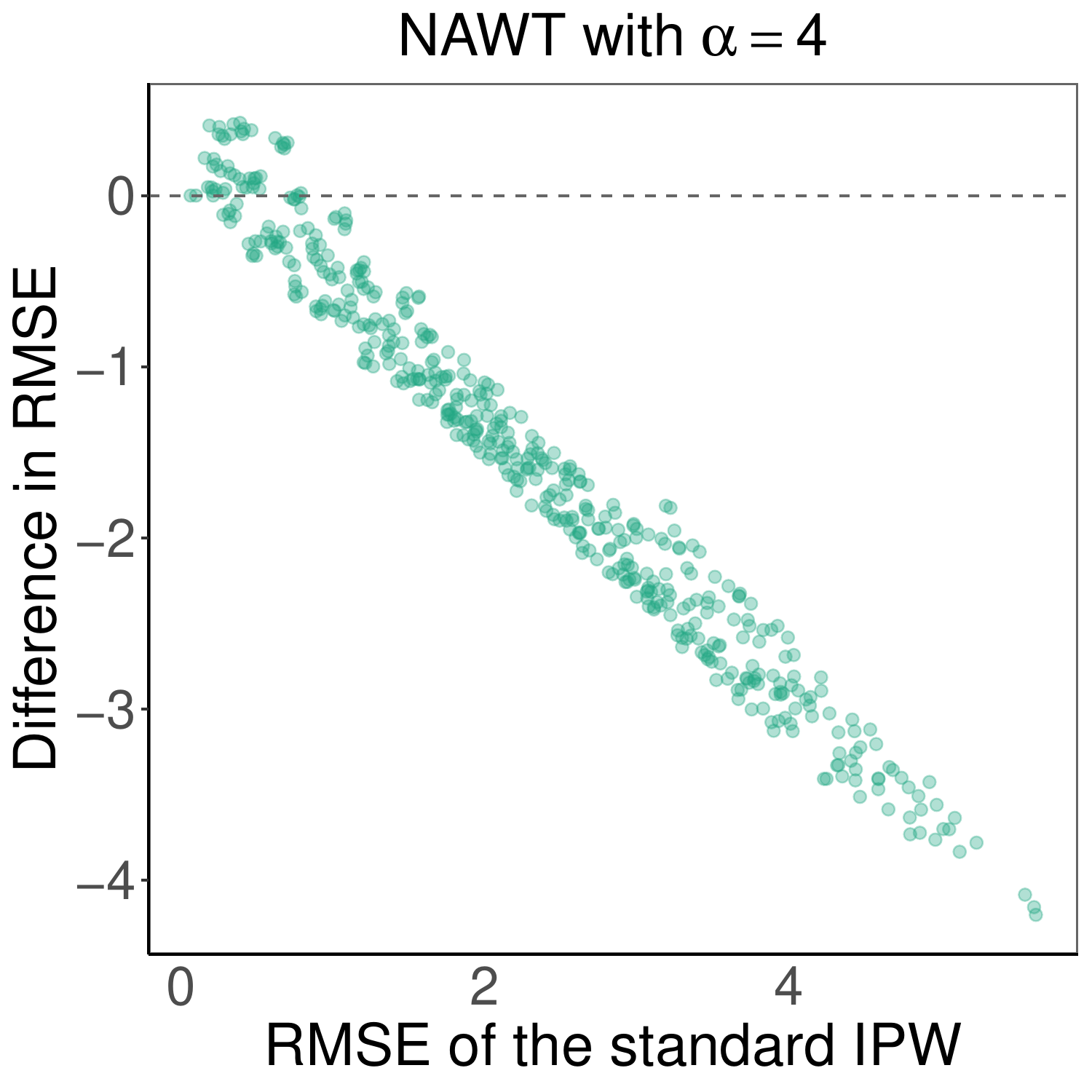}
\includegraphics[width=0.325\textwidth]{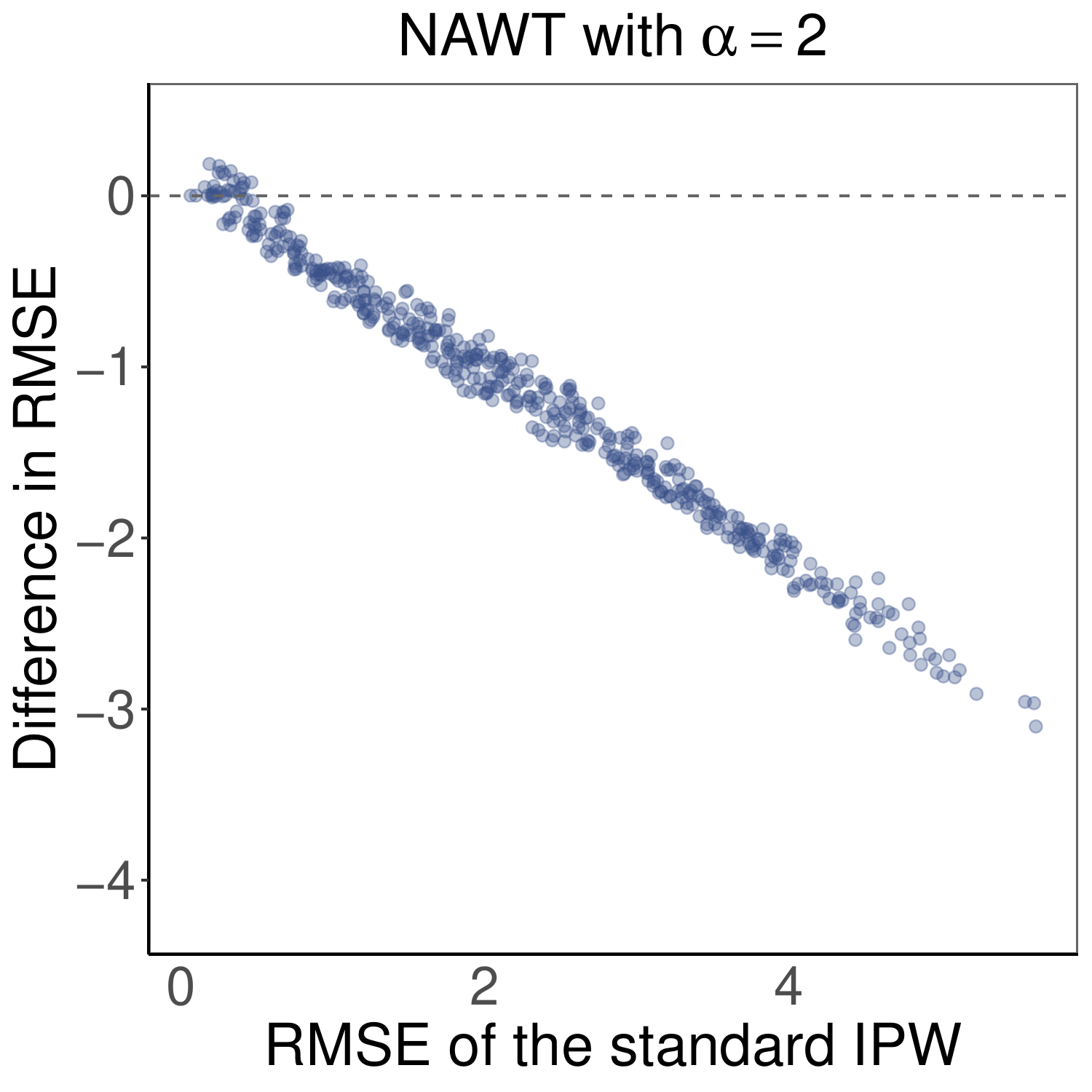}
\includegraphics[width=0.325\textwidth]{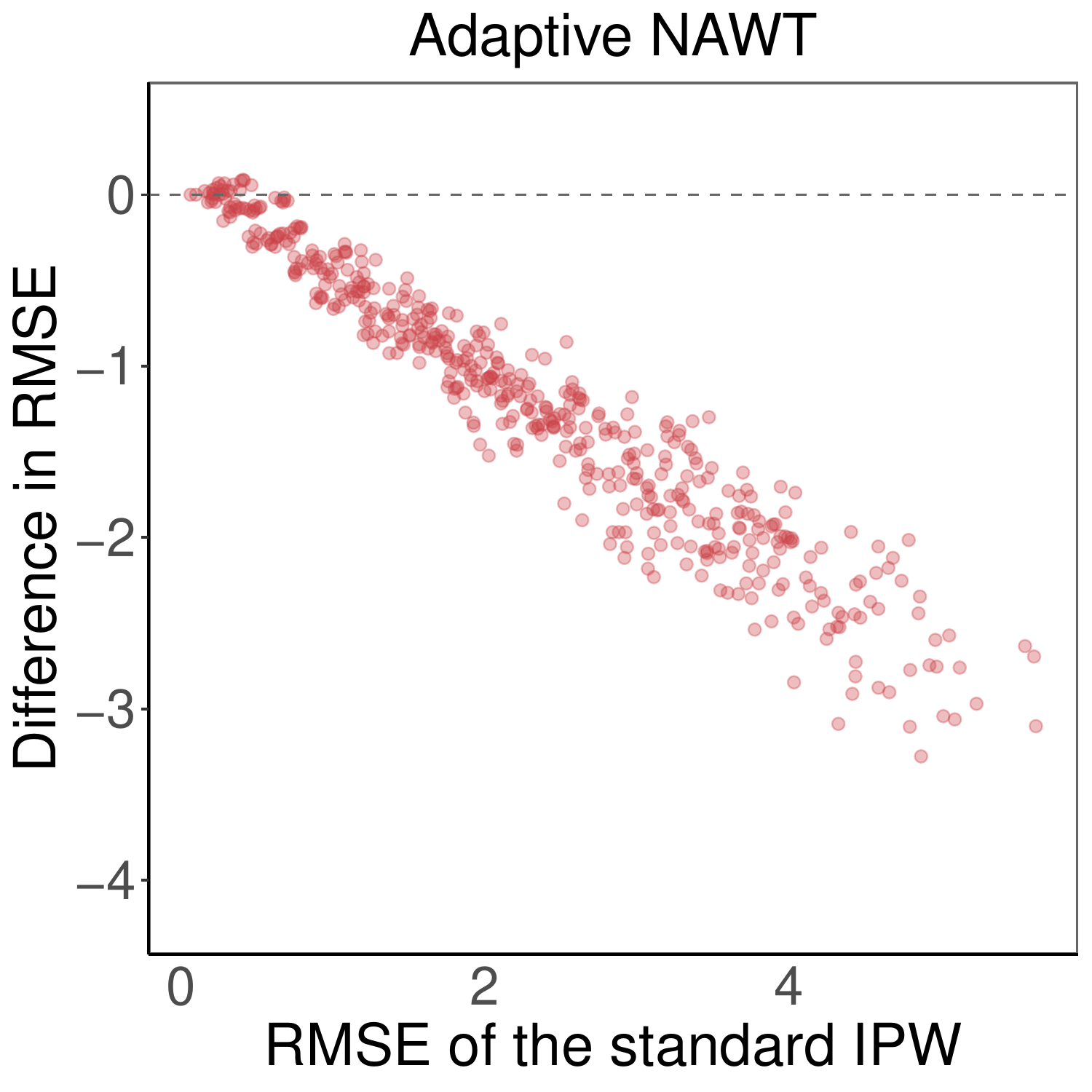}
\vspace{-8pt}
\caption{The difference in the RMSE between the NAWT and IPW for the ATT estimation against the RMSE of the standard IPW with the misspecified propensity score model (type~1).} \label{fig_attfindifmis1}
\end{figure}

\begin{figure}[h]
\centering
\includegraphics[width=0.325\textwidth]{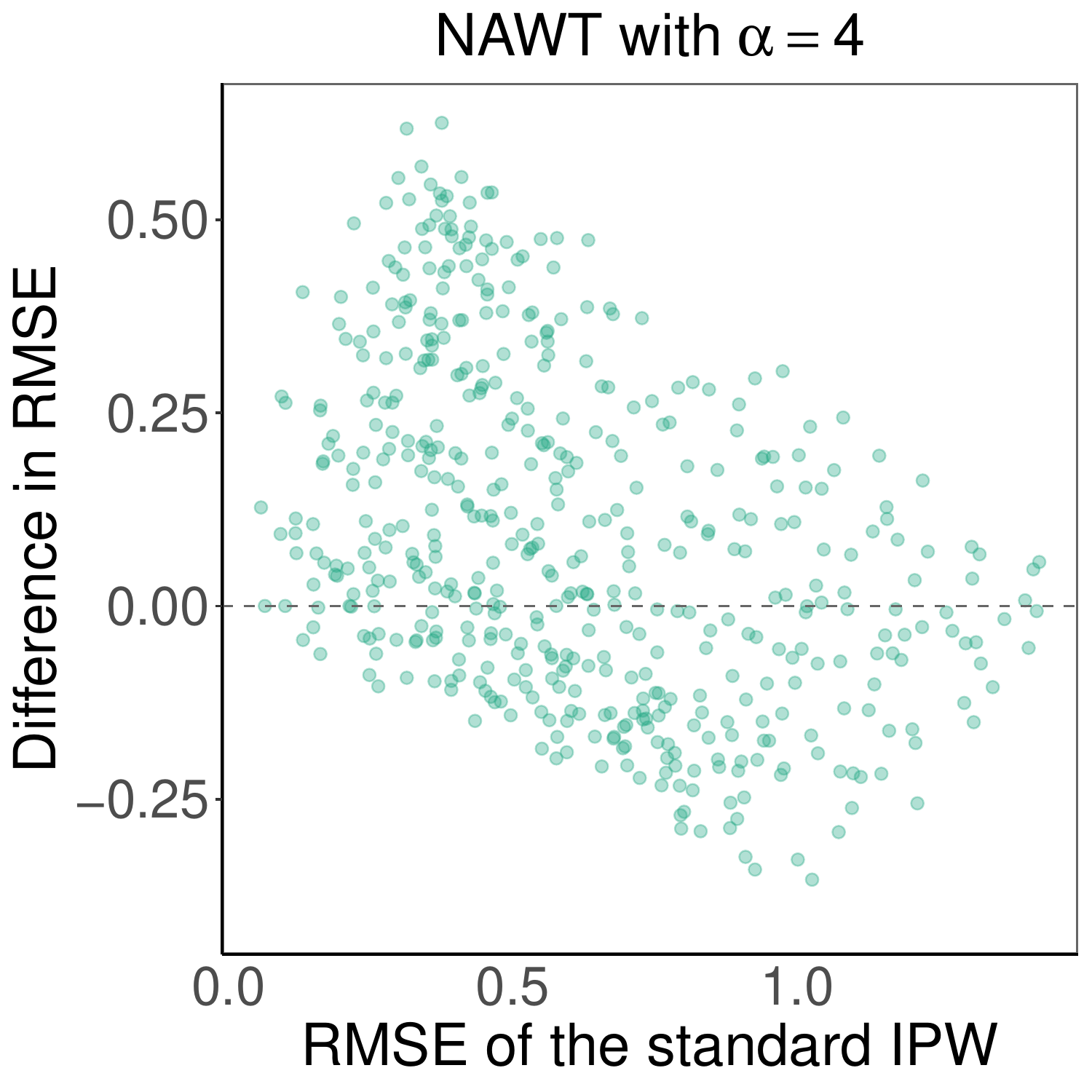}
\includegraphics[width=0.325\textwidth]{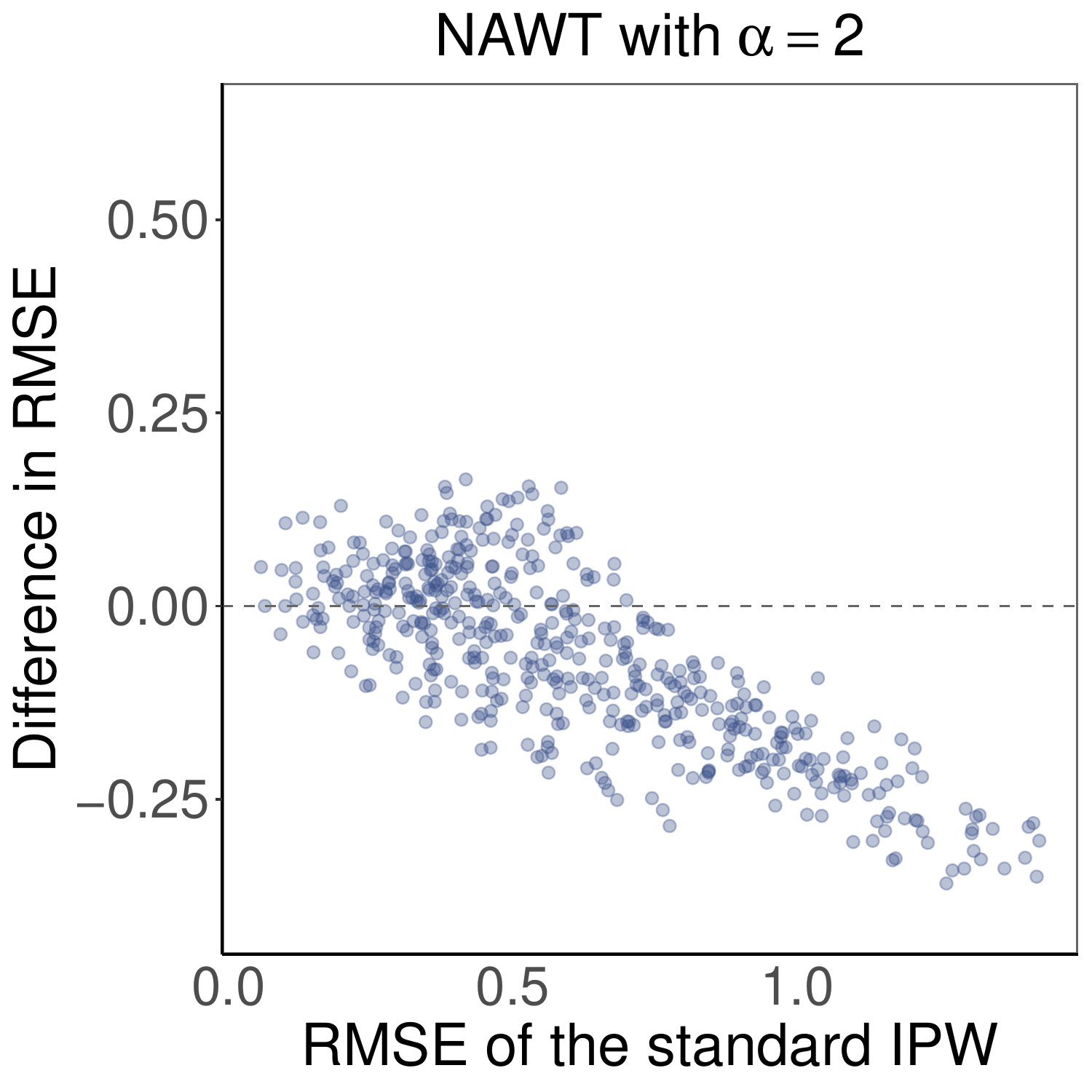}
\includegraphics[width=0.325\textwidth]{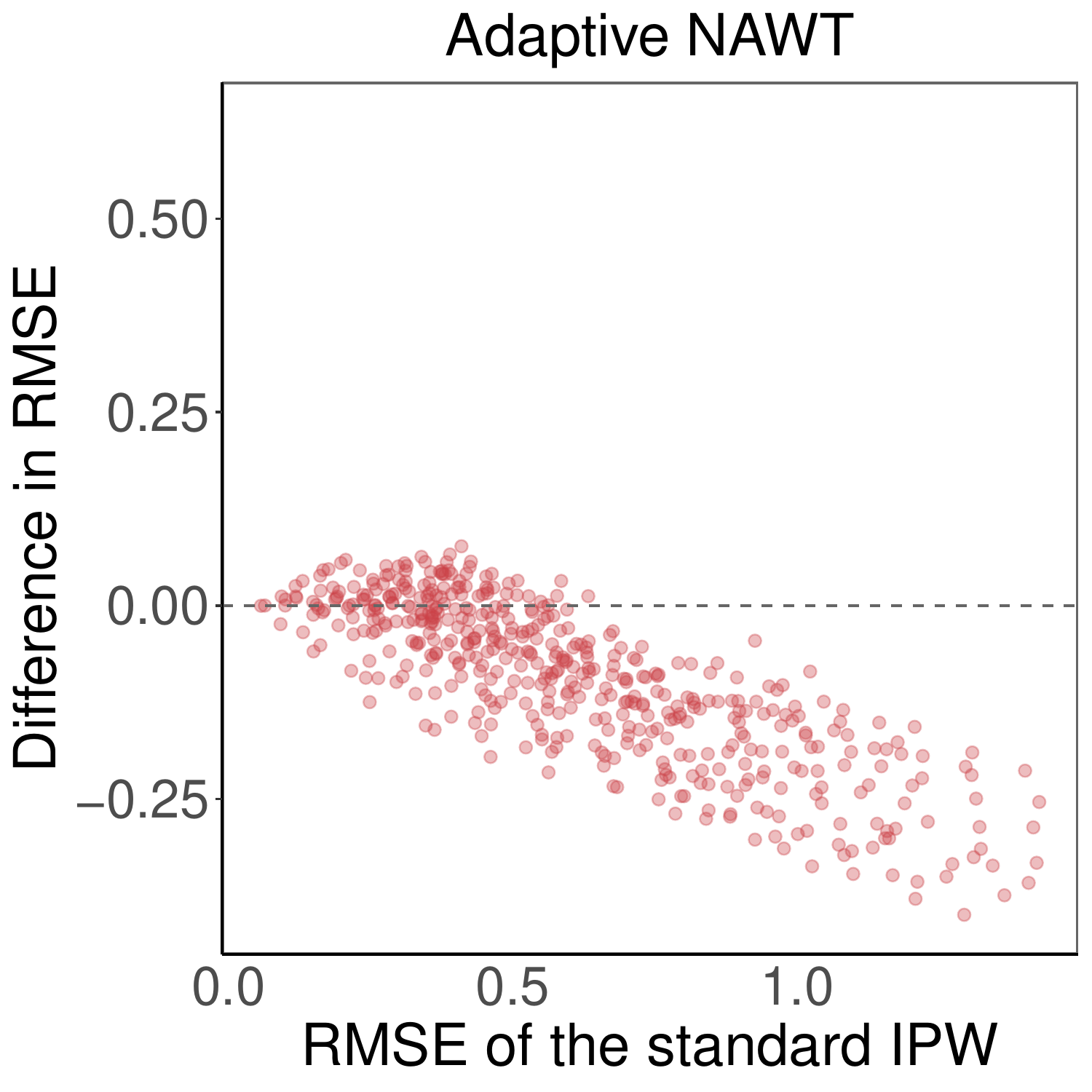}
\vspace{-8pt}
\caption{The difference in the RMSE between the NAWT and IPW for the ATT estimation against the RMSE of the standard IPW with the misspecified propensity score model (type~2).} \label{fig_attfindifmis2}
\end{figure}

\begin{figure}[hp]
\centering
\includegraphics[width=0.935\textwidth]{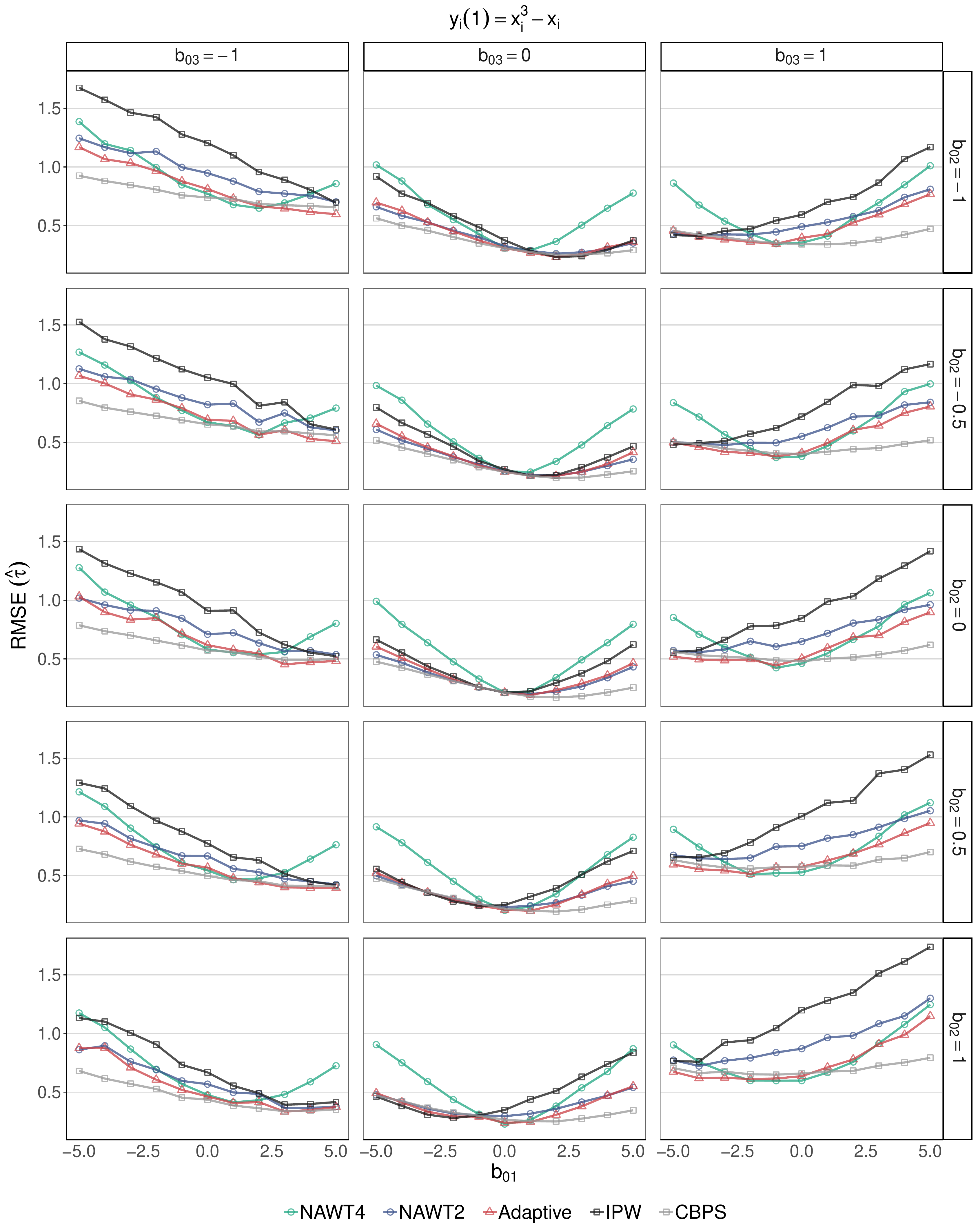}
\vspace{-12pt}
\caption{The finite-sample RMSE for the ATT estimation with the true propensity score model using the NAWT with $\alpha = 4$ (NAWT4) and $\alpha = 2$ (NAWT2), adaptive NAWT, standard IPW, and the CBPS under the various settings for the true outcome model for the controlled when the true outcome model for the treated is $y_i(1) = x_i^3 - x_i$.} \label{fig_attfincor3}
\end{figure}

\begin{figure}[hp]
\centering
\includegraphics[width=0.935\textwidth]{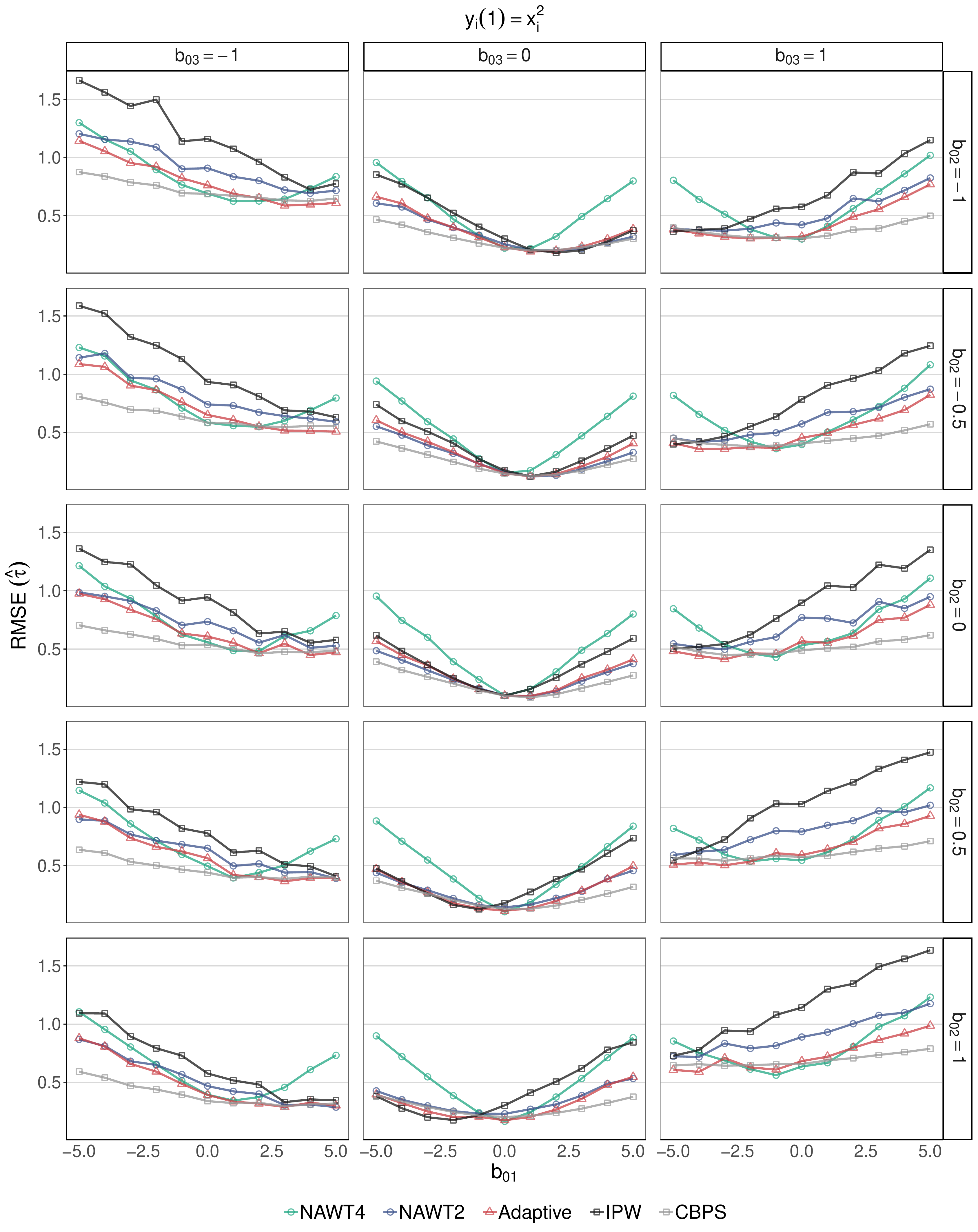}
\vspace{-12pt}
\caption{The finite-sample RMSE for the ATT estimation with the true propensity score model using the NAWT with $\alpha = 4$ (NAWT4) and $\alpha = 2$ (NAWT2), adaptive NAWT, standard IPW, and the CBPS under the various settings for the true outcome model for the controlled when the true outcome model for the treated is $y_i(1) = x_i^2$.} \label{fig_attfincor2}
\end{figure}

\begin{figure}[hp]
\centering
\includegraphics[width=0.935\textwidth]{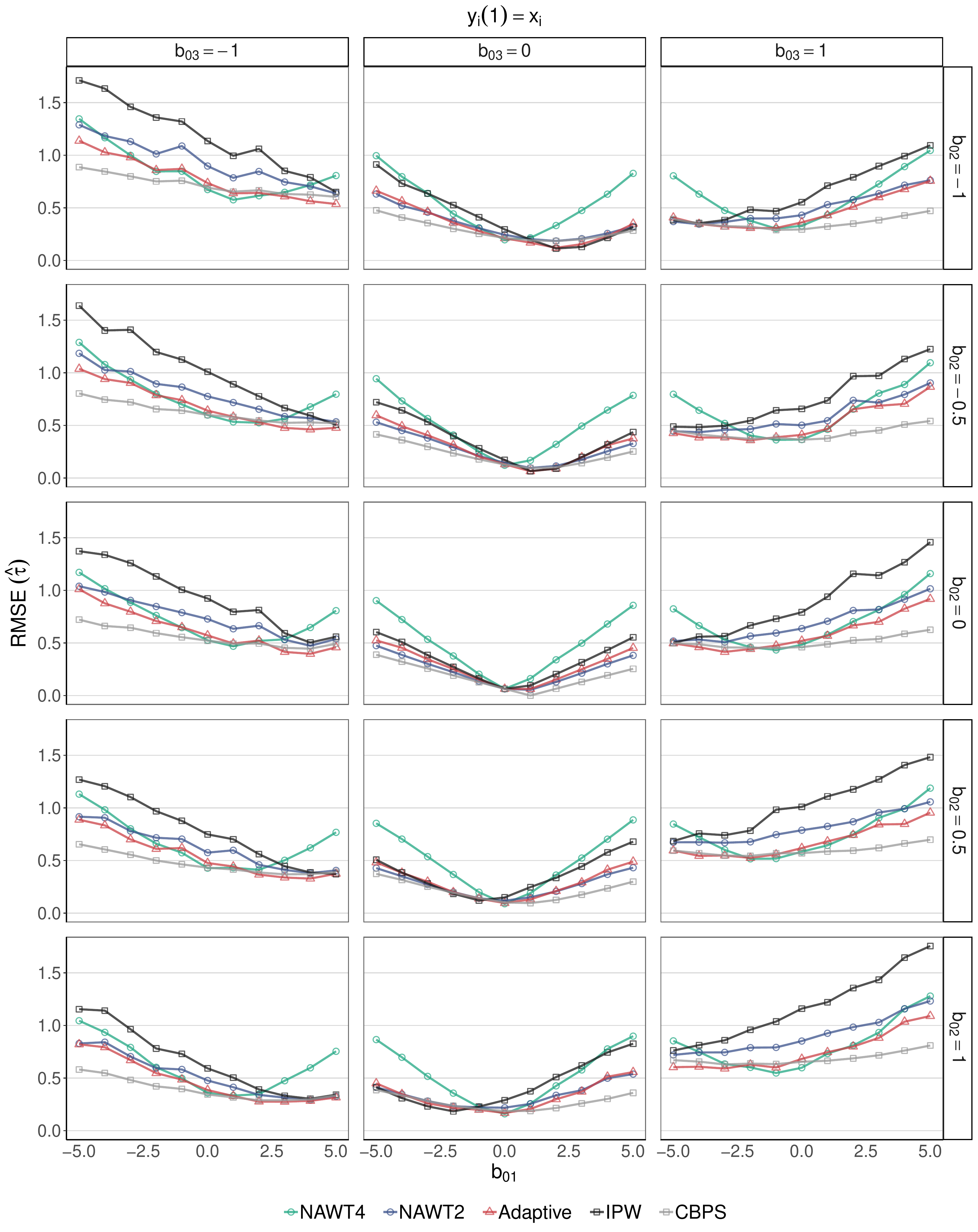}
\vspace{-12pt}
\caption{The finite-sample RMSE for the ATT estimation with the true propensity score model using the NAWT with $\alpha = 4$ (NAWT4) and $\alpha = 2$ (NAWT2), adaptive NAWT, standard IPW, and the CBPS under the various settings for the true outcome model for the controlled when the true outcome model for the treated is $y_i(1) = x_i$.} \label{fig_attfincor1}
\end{figure}

\begin{figure}[hp]
\centering
\includegraphics[width=0.935\textwidth]{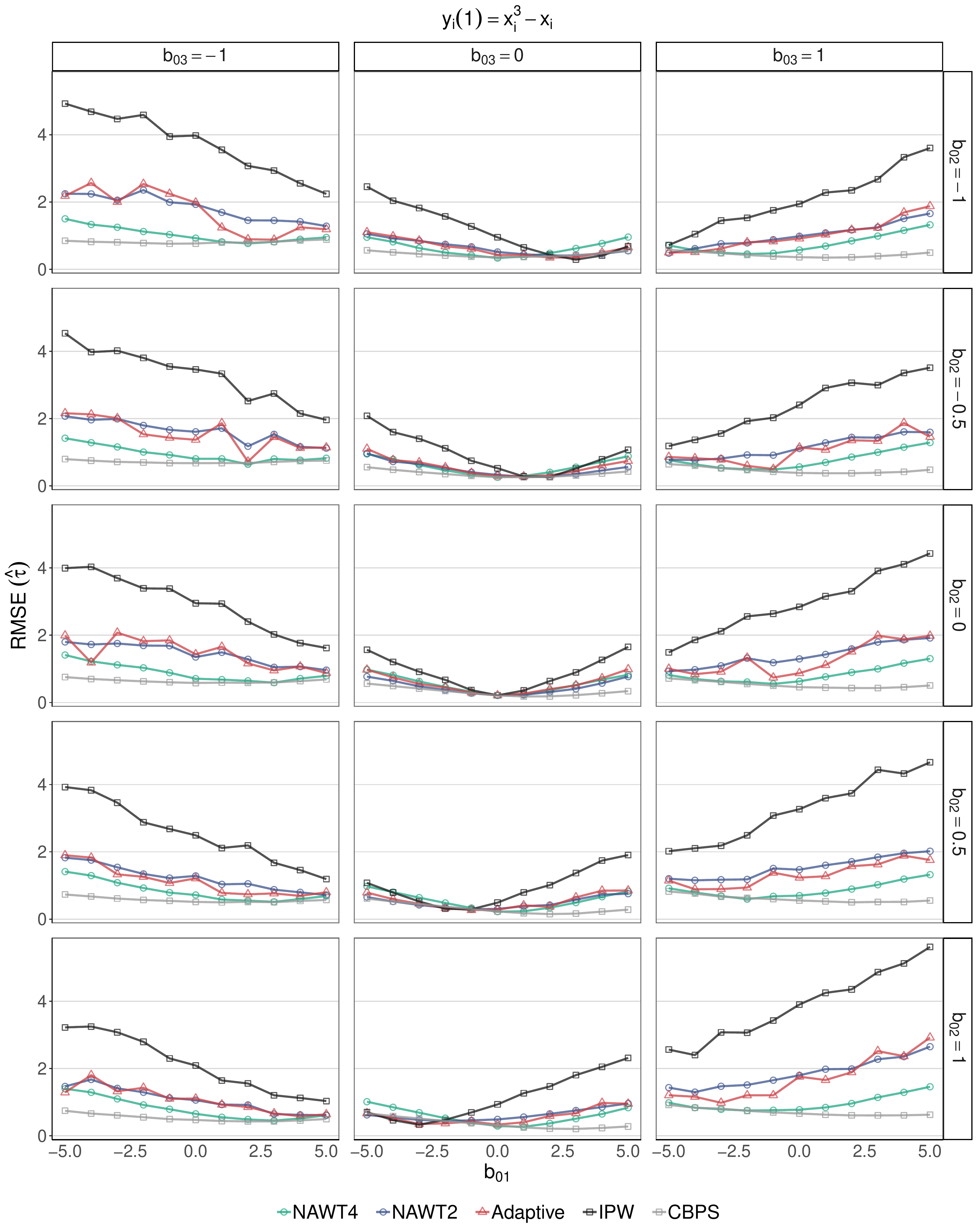}
\vspace{-12pt}
\caption{The finite-sample RMSE for the ATT estimation with the misspecified propensity score model (type~1) using the NAWT with $\alpha = 4$ (NAWT4) and $\alpha = 2$ (NAWT2), adaptive NAWT, standard IPW, and the CBPS under the various settings for the true outcome model for the controlled when the true outcome model for the treated is $y_i(1) = x_i^3 - x_i$.} \label{fig_attfinmis13}
\end{figure}

\begin{figure}[hp]
\centering
\includegraphics[width=0.935\textwidth]{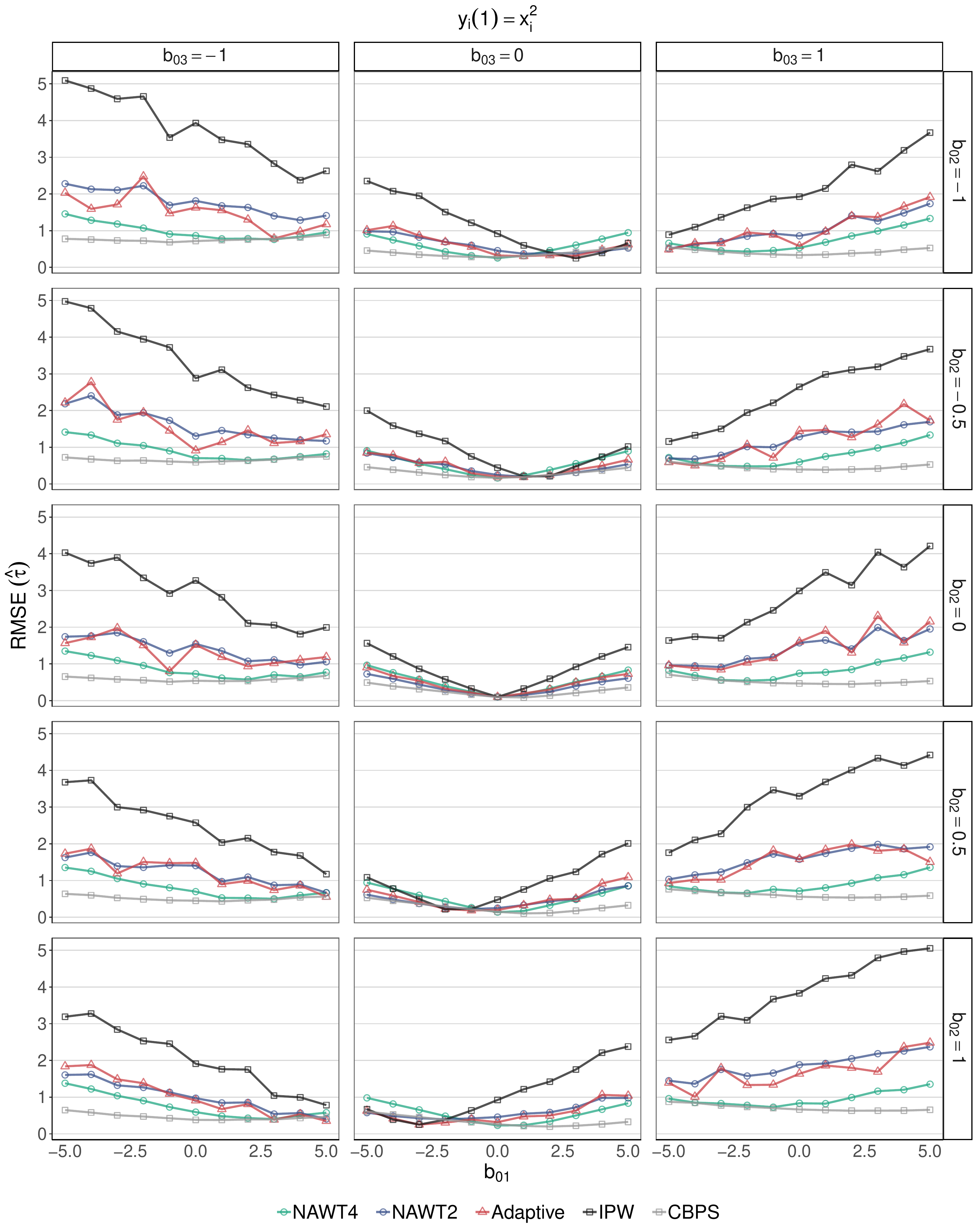}
\vspace{-12pt}
\caption{The finite-sample RMSE for the ATT estimation with the misspecified propensity score model (type~1) using the NAWT with $\alpha = 4$ (NAWT4) and $\alpha = 2$ (NAWT2), adaptive NAWT, standard IPW, and the CBPS under the various settings for the true outcome model for the controlled when the true outcome model for the treated is $y_i(1) = x_i^2$.} \label{fig_attfinmis12}
\end{figure}

\begin{figure}[hp]
\centering
\includegraphics[width=0.935\textwidth]{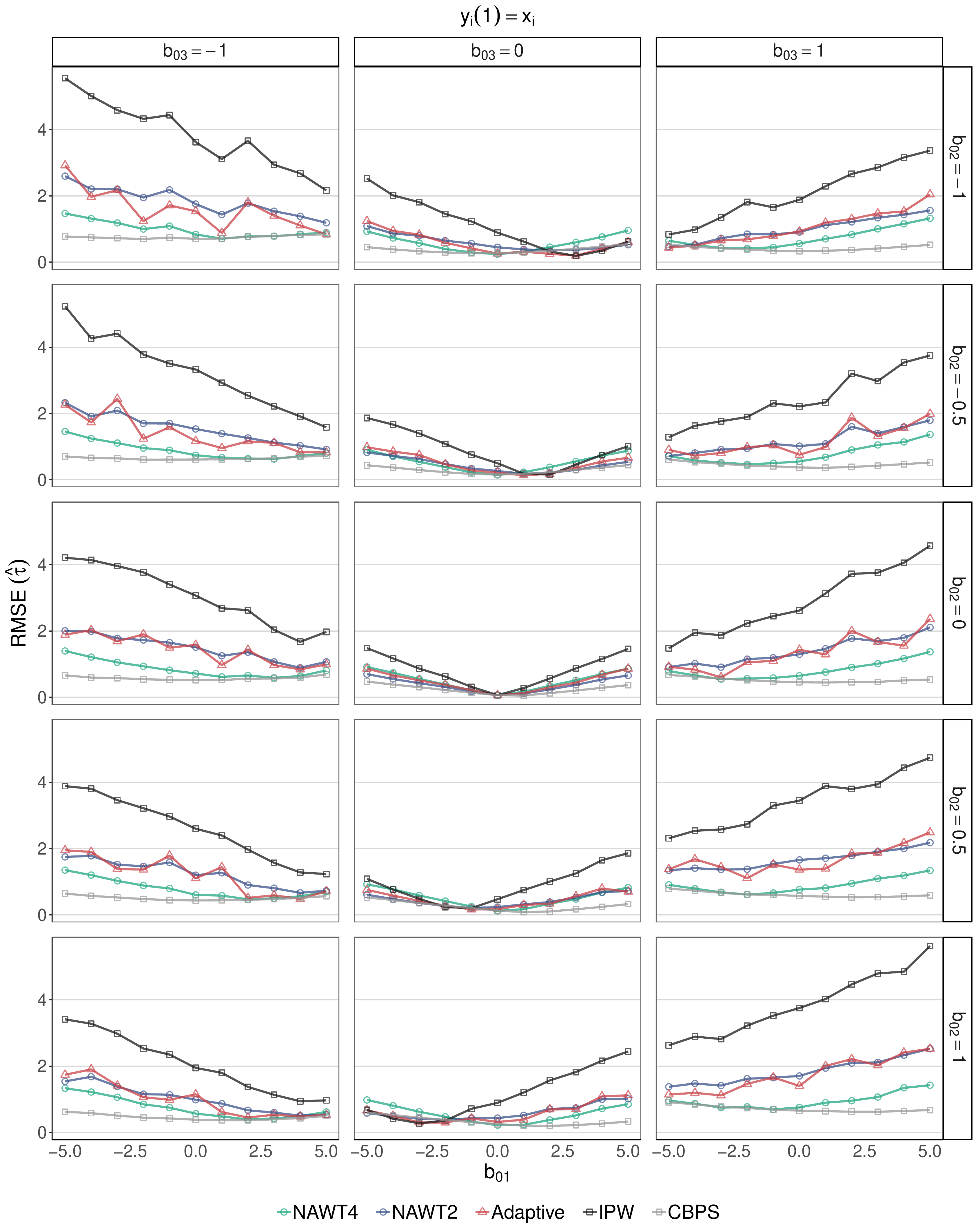}
\vspace{-12pt}
\caption{The finite-sample RMSE for the ATT estimation with the misspecified propensity score model (type~1) using the NAWT with $\alpha = 4$ (NAWT4) and $\alpha = 2$ (NAWT2), adaptive NAWT, standard IPW, and the CBPS under the various settings for the true outcome model for the controlled when the true outcome model for the treated is $y_i(1) = x_i$.} \label{fig_attfinmis11}
\end{figure}

\begin{figure}[hp]
\centering
\includegraphics[width=0.935\textwidth]{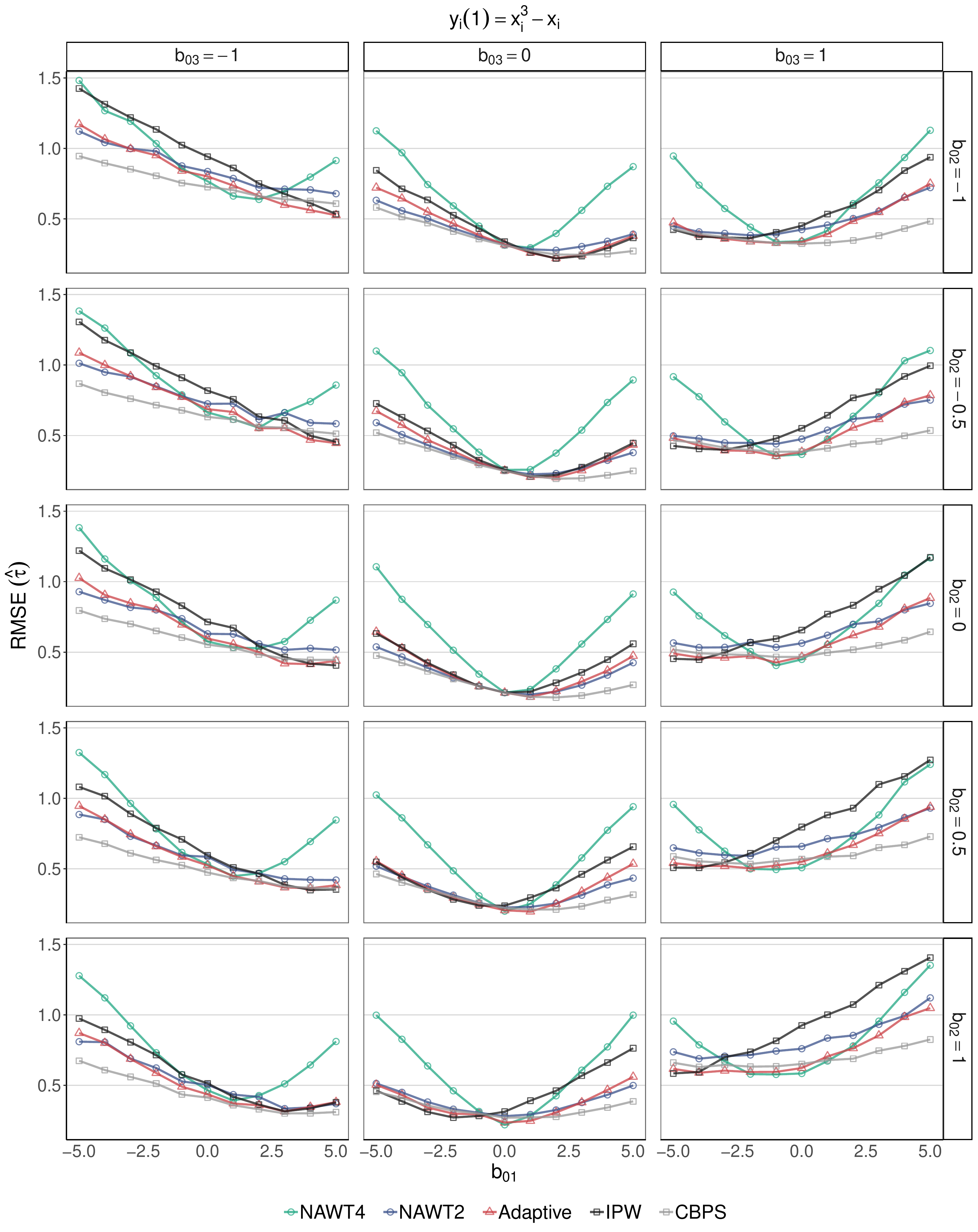}
\vspace{-12pt}
\caption{The finite-sample RMSE for the ATT estimation with the misspecified propensity score model (type~2) using the NAWT with $\alpha = 4$ (NAWT4) and $\alpha = 2$ (NAWT2), adaptive NAWT, standard IPW, and the CBPS under the various settings for the true outcome model for the controlled when the true outcome model for the treated is $y_i(1) = x_i^3 - x_i$.} \label{fig_attfinmis23}
\end{figure}

\begin{figure}[hp]
\centering
\includegraphics[width=0.935\textwidth]{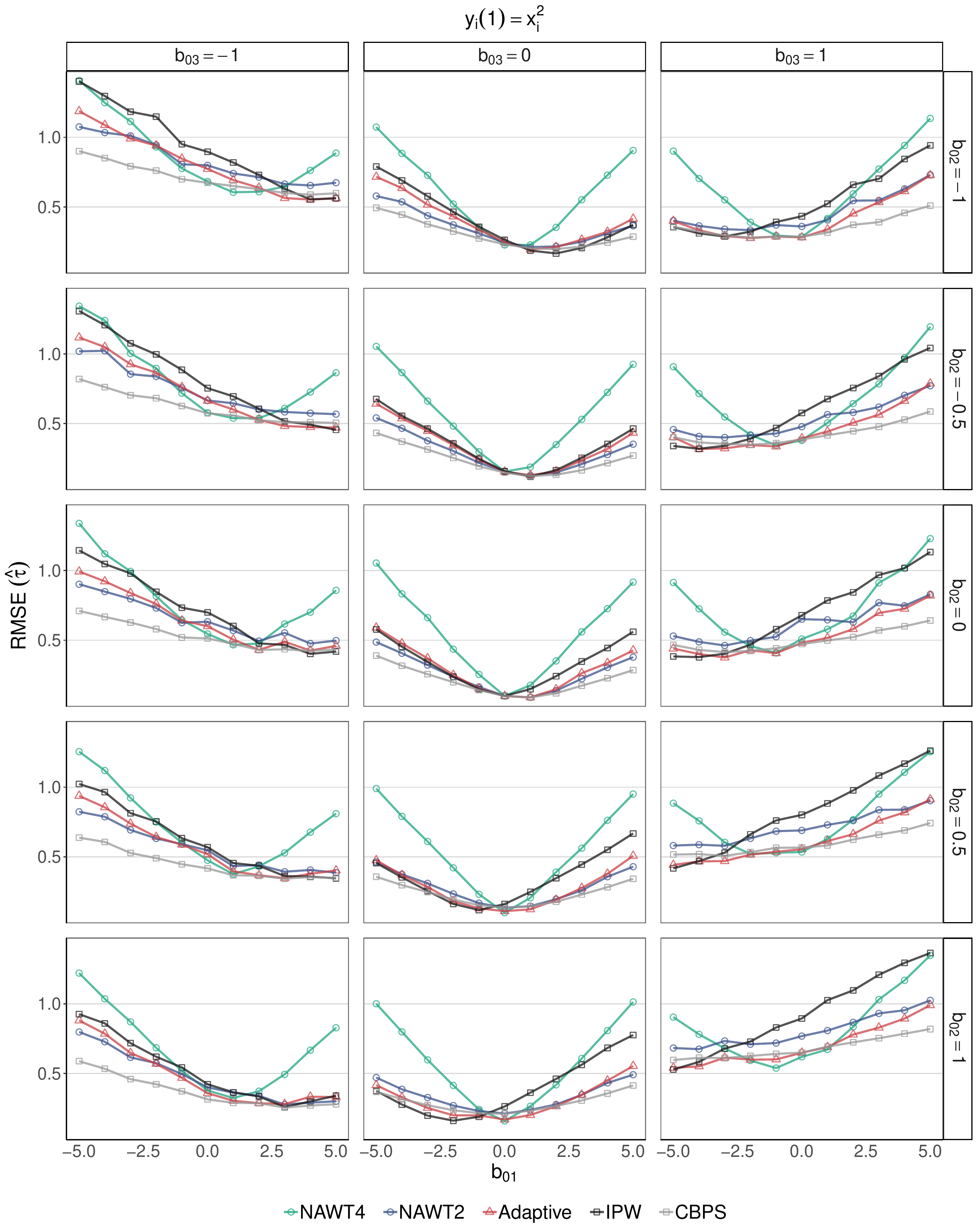}
\vspace{-12pt}
\caption{The finite-sample RMSE for the ATT estimation with the misspecified propensity score model (type~2) using the NAWT with $\alpha = 4$ (NAWT4) and $\alpha = 2$ (NAWT2), adaptive NAWT, standard IPW, and the CBPS under the various settings for the true outcome model for the controlled when the true outcome model for the treated is $y_i(1) = x_i^2$.} \label{fig_attfinmis22}
\end{figure}

\begin{figure}[hp]
\centering
\includegraphics[width=0.935\textwidth]{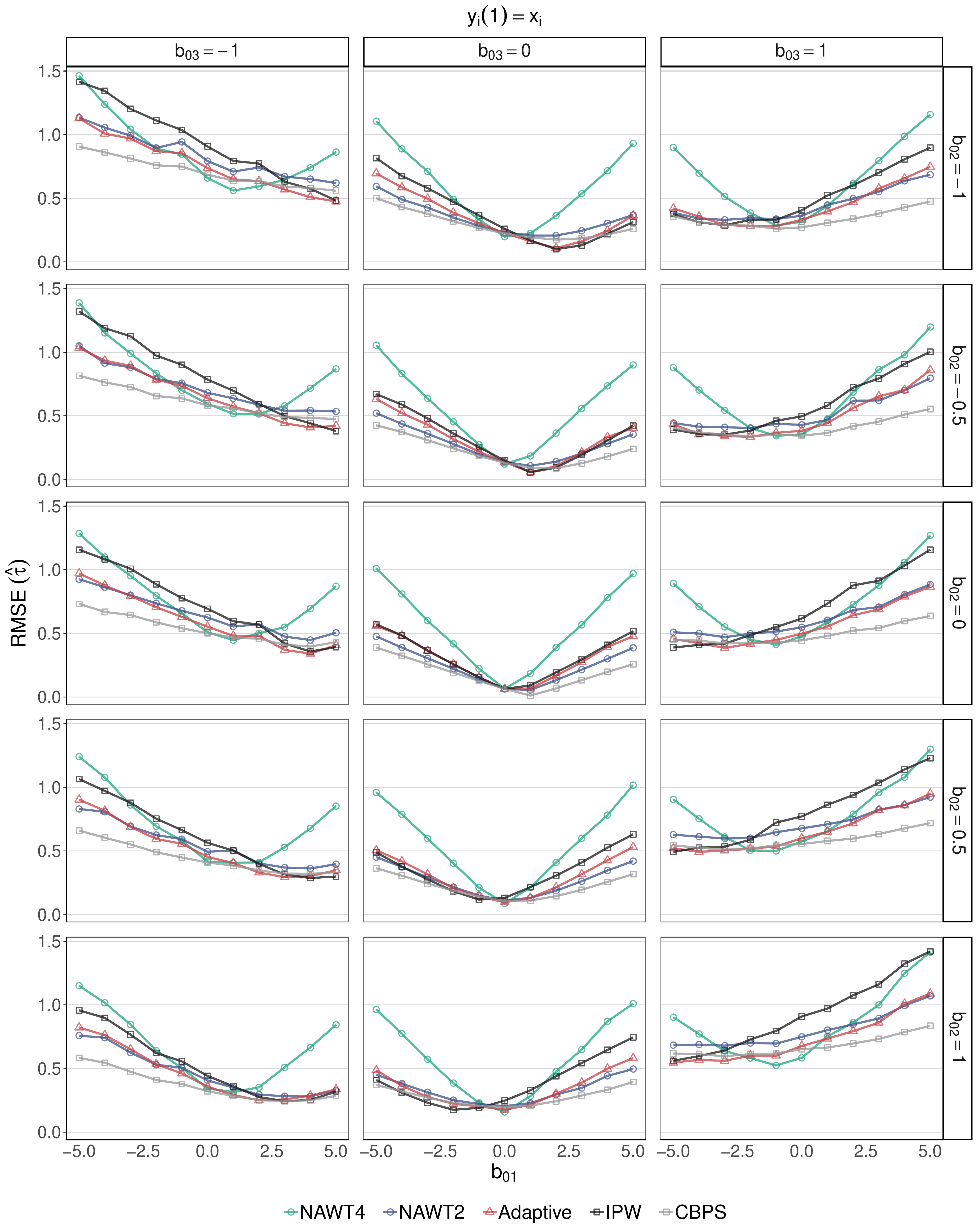}
\vspace{-12pt}
\caption{The finite-sample RMSE for the ATT estimation with the misspecified propensity score model (type~2) using the NAWT with $\alpha = 4$ (NAWT4) and $\alpha = 2$ (NAWT2), adaptive NAWT, standard IPW, and the CBPS under the various settings for the true outcome model for the controlled when the true outcome model for the treated is $y_i(1) = x_i$.} \label{fig_attfinmis21}
\end{figure}

The summary of the results for the ATT estimation are shown in Figure~\ref{fig_attfindifcor}--\ref{fig_attfindifmis2} and the details are shown in Figure~\ref{fig_attfincor3}--\ref{fig_attfinmis21}. Figure~\ref{fig_attfindifcor}--\ref{fig_attfindifmis2} shows the difference in the RMSE between the NAWT and IPW for the ATT estimation in the y-axis against those of the standard IPW in the x-axis, where negative values in y-axis indicate smaller RMSEs with the NAWT than the standard IPW. Like the large-sample cases, both the NAWT with $\alpha = 4$ and $\alpha = 2$ greatly improves the standard IPW when the IPW performs poorly when units with large propensity scores have average outcomes without treatment largely deviated from the mean irrespective of the true outcome model for the treated (Figure~\ref{fig_attfincor3}--\ref{fig_attfinmis21}). Again, the NAWT with $\alpha = 4$ is likely to outperform the NAWT with $\alpha = 2$ when the NAWT improves the IPW but the NAWT with $\alpha = 4$ sometimes works poorly when the IPW performs well where units with small propensity scores have average outcomes without treatment largely deviated from the mean. The adaptive NAWT adequately utilizes the best weighting function as expected, improves the standard IPW more than the NAWT with $\alpha = 4$ and $\alpha = 2$, and rarely works worse than the standard IPW. With the misspecified propensity score models, the NAWT reduces the bias and the adaptive NAWT works the best. Comparing the results among the different propensity score models, the NAWT improves the IPW more under more difficult scenarios with type~1 misspecification where inverse probability weights are over-estimated for units with large inverse probability weights. When the IPW works relatively well with the true or type~2 misspecified propensity score models, the NAWT with $\alpha = 4$ sometimes works worse than the IPW but the NAWT with $\alpha = 2$ and adaptive NAWT improve the IPW.

\begin{figure}[h]
\centering
\includegraphics[width=0.325\textwidth]{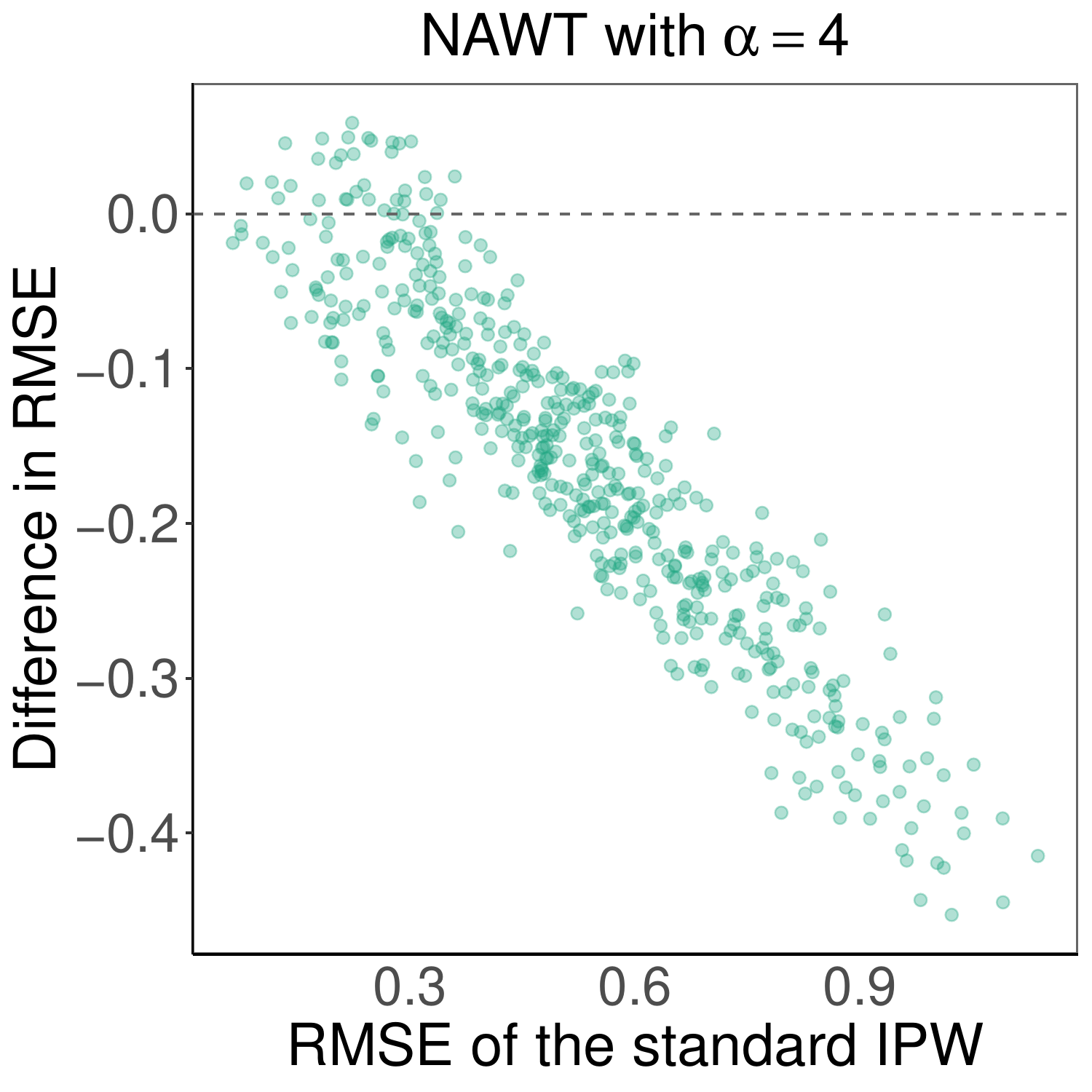}
\includegraphics[width=0.325\textwidth]{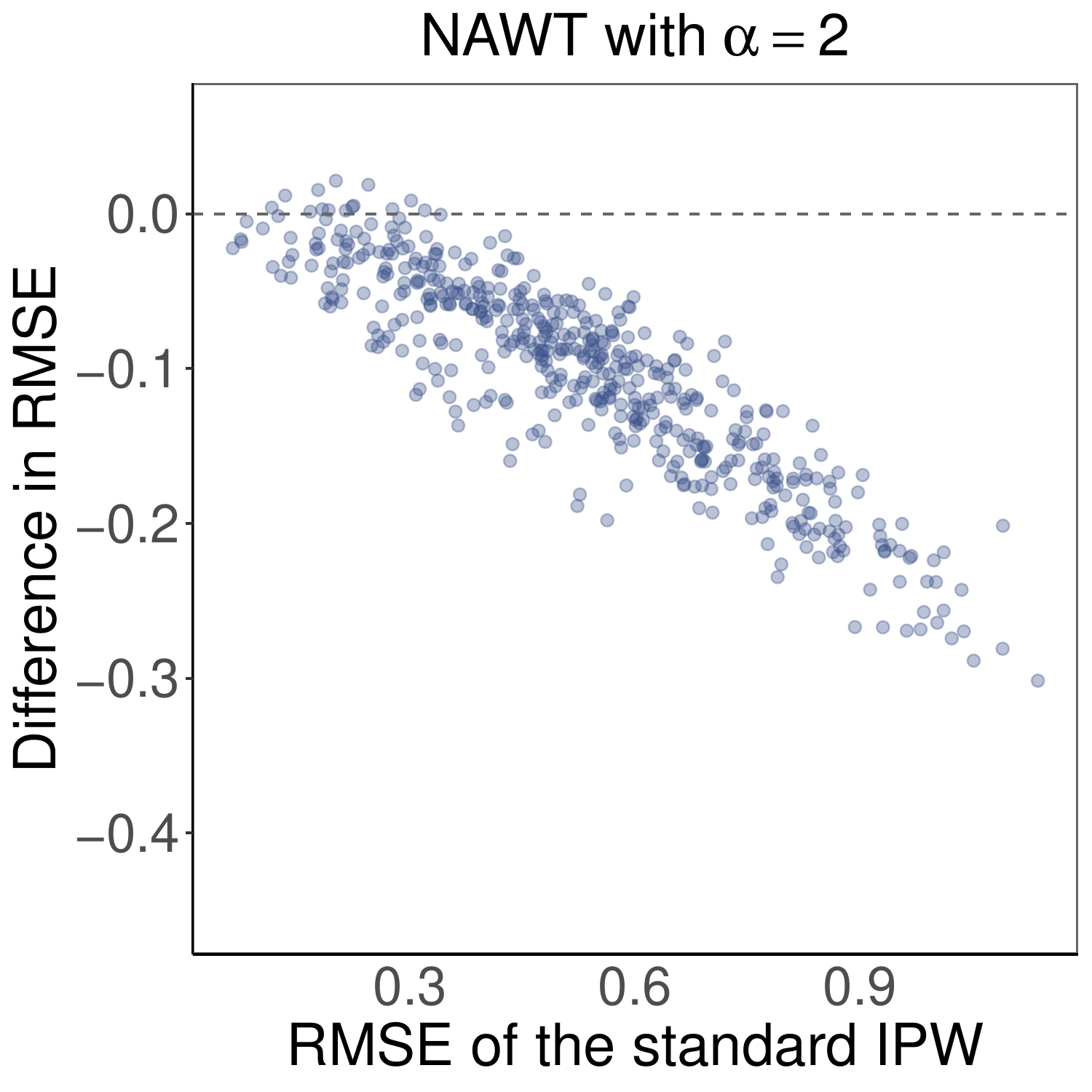}
\includegraphics[width=0.325\textwidth]{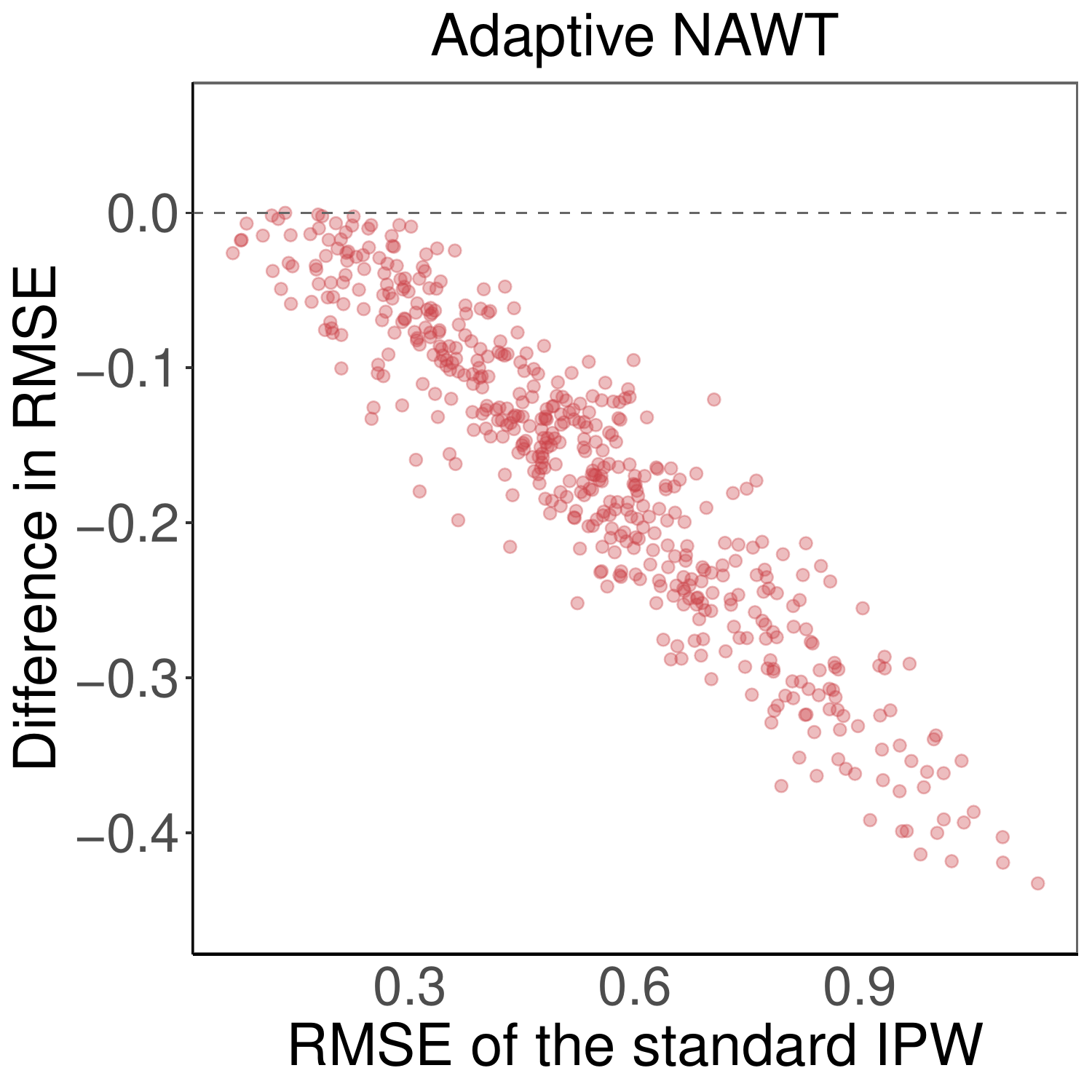}
\vspace{-8pt}
\caption{The difference in the RMSE between the NAWT and IPW for the ATE estimation against the RMSE of the standard IPW with the correct propensity score model.} \label{fig_atefindifcor}
\end{figure}

\begin{figure}[h]
\centering
\includegraphics[width=0.325\textwidth]{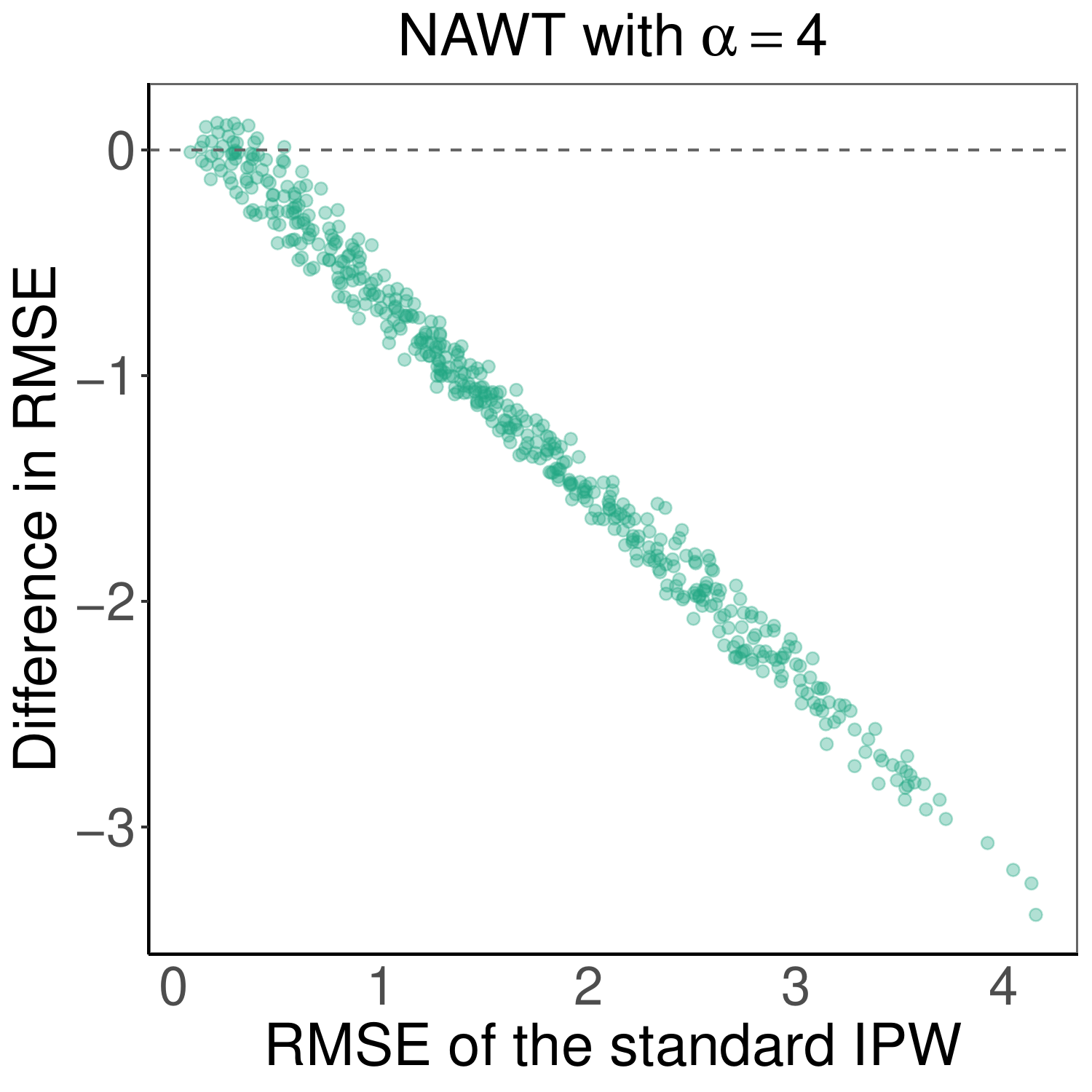}
\includegraphics[width=0.325\textwidth]{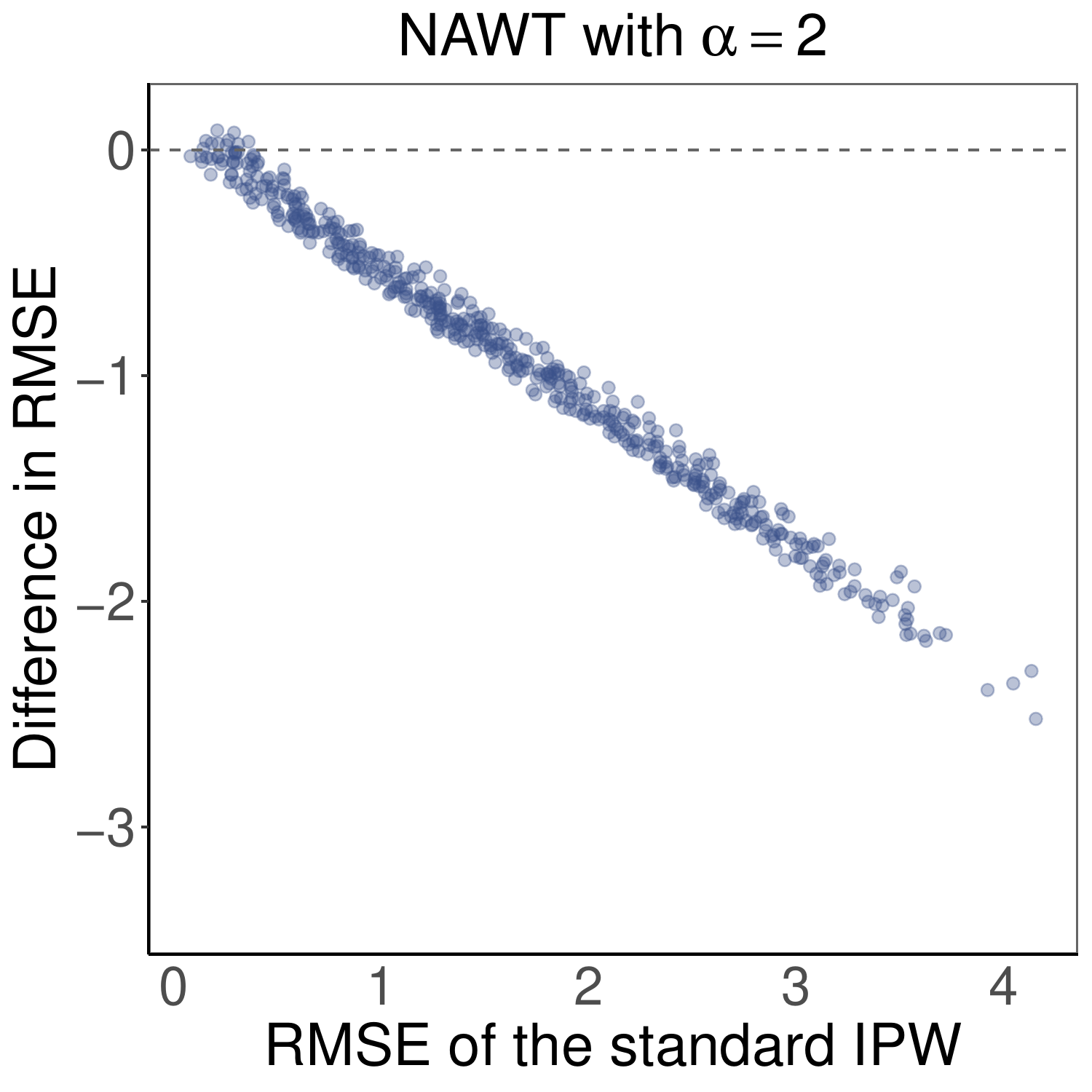}
\includegraphics[width=0.325\textwidth]{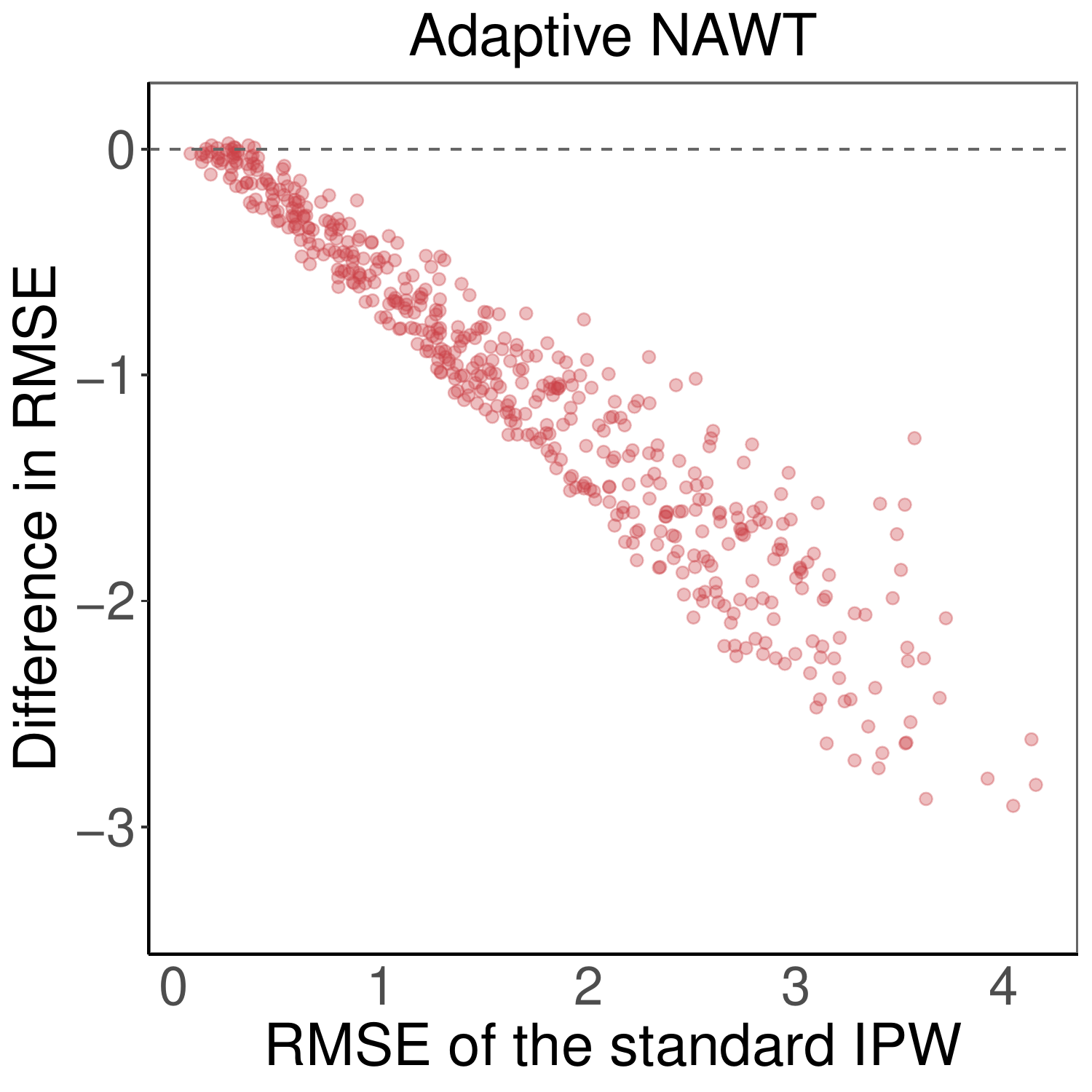}
\vspace{-8pt}
\caption{The difference in the RMSE between the NAWT and IPW for the ATE estimation against the RMSE of the standard IPW with the misspecified propensity score model (type~1).} \label{fig_atefindifmis1}
\end{figure}

\begin{figure}[h]
\centering
\includegraphics[width=0.325\textwidth]{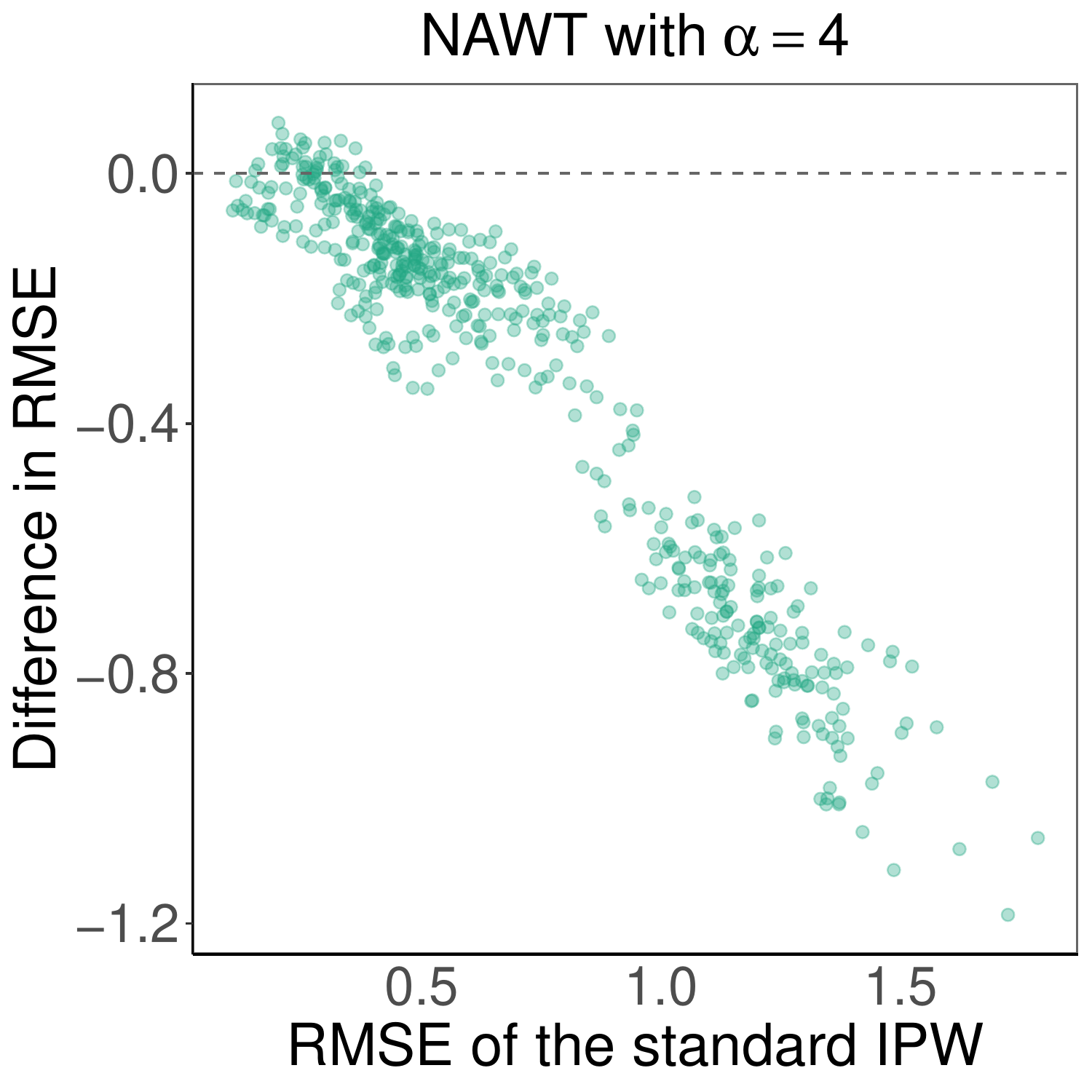}
\includegraphics[width=0.325\textwidth]{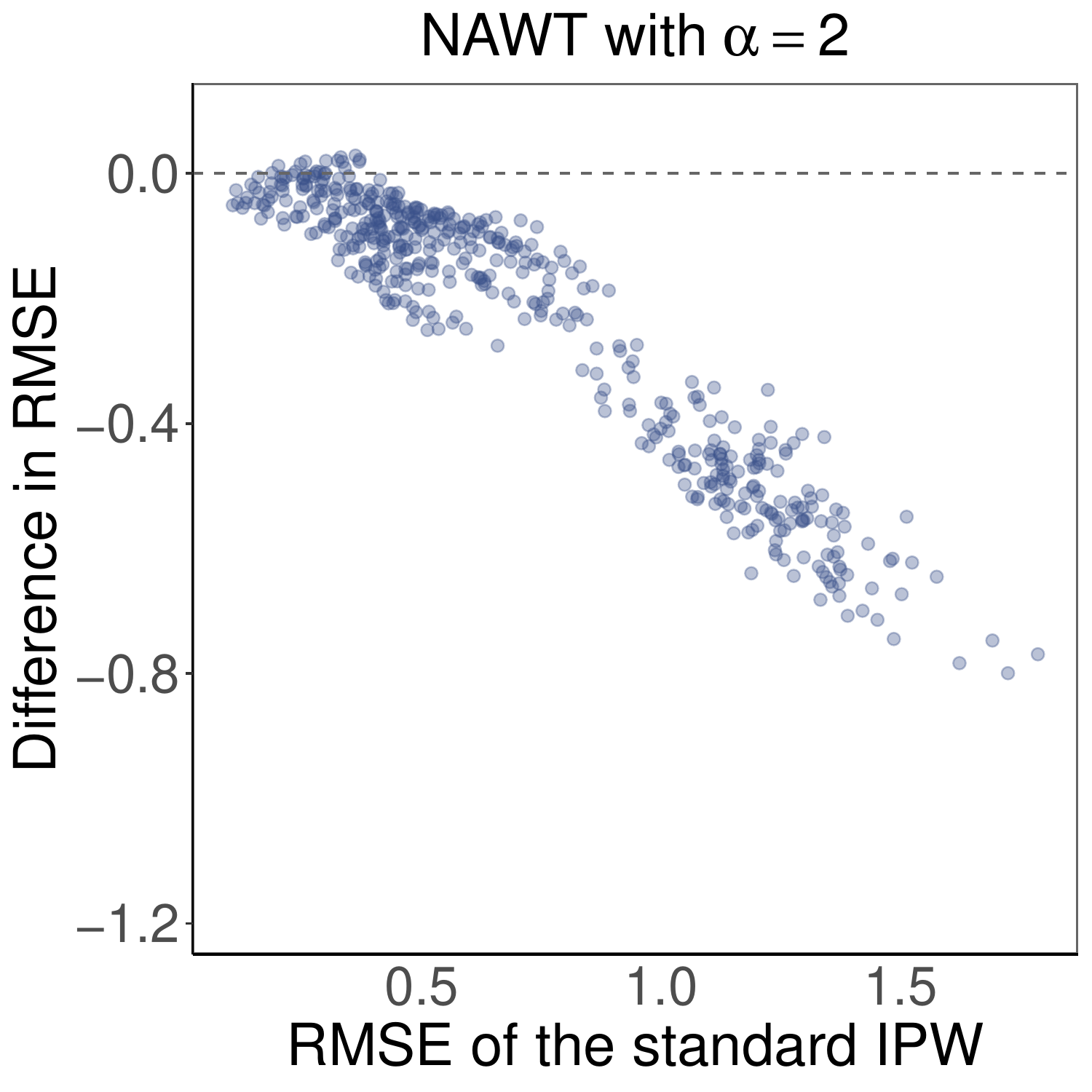}
\includegraphics[width=0.325\textwidth]{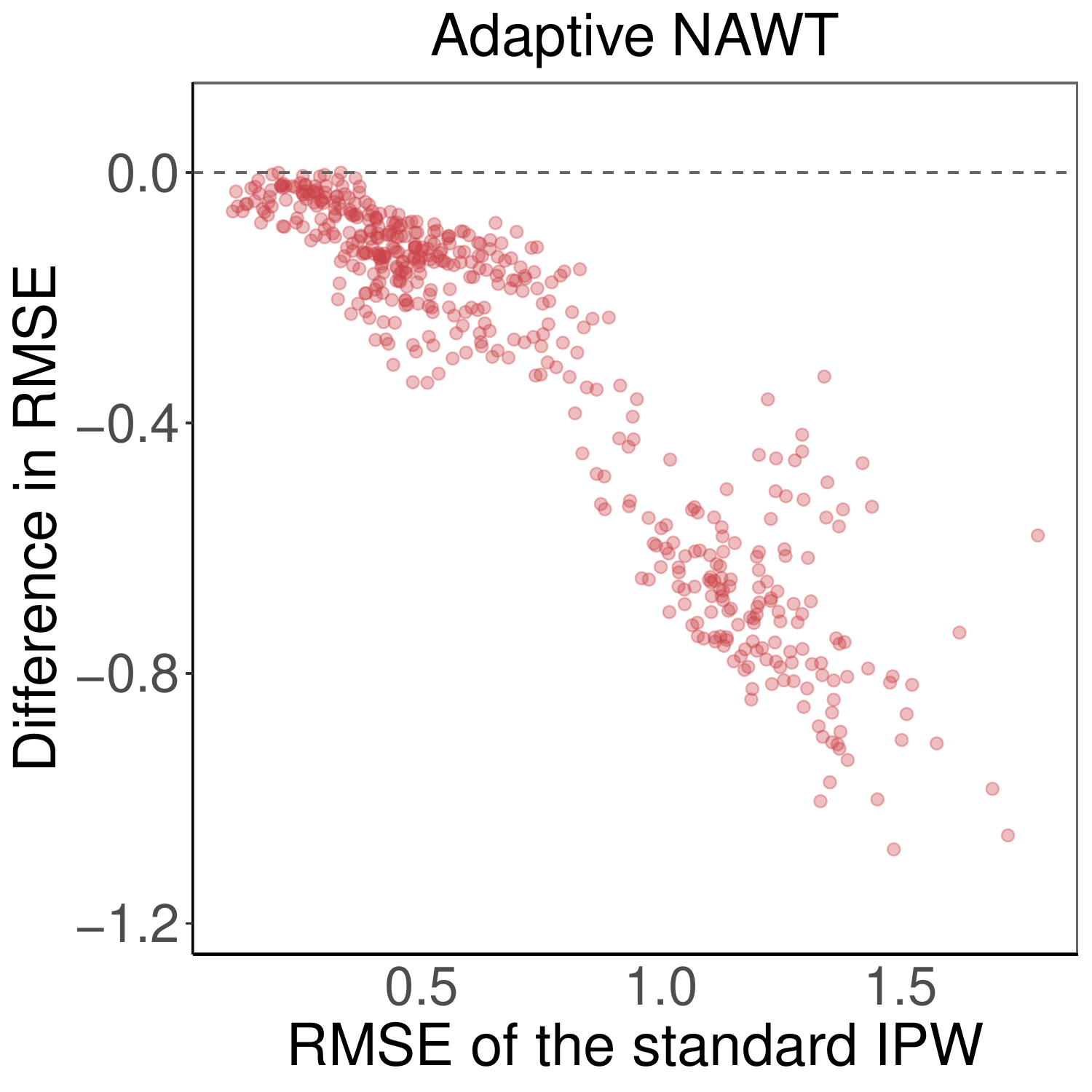}
\vspace{-8pt}
\caption{The difference in the RMSE between the NAWT and IPW for the ATE estimation against the RMSE of the standard IPW with the misspecified propensity score model (type~2).} \label{fig_atefindifmis2}
\end{figure}

\begin{figure}[hp]
\centering
\includegraphics[width=0.935\textwidth]{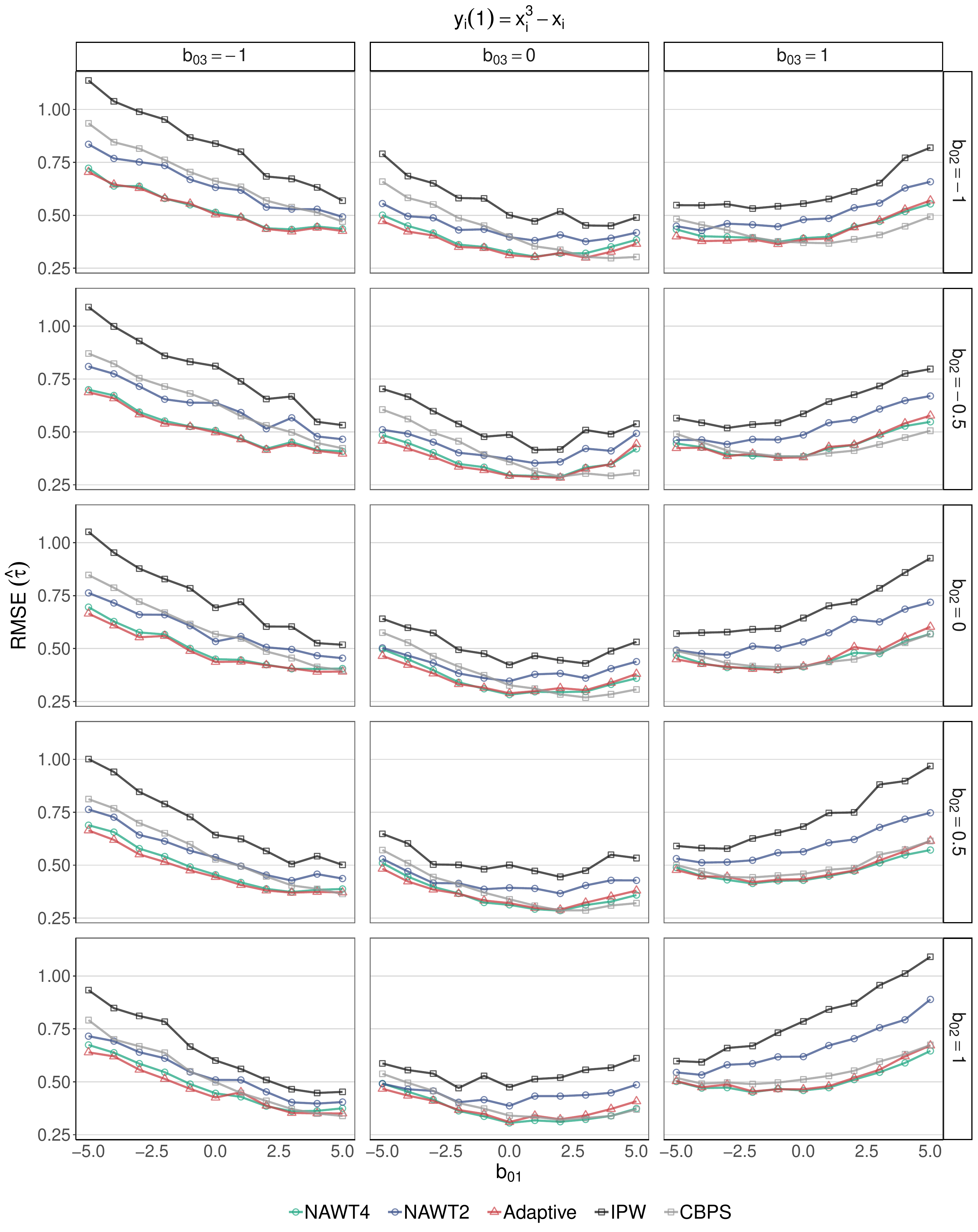}
\vspace{-12pt}
\caption{The finite-sample RMSE for the ATE estimation with the true propensity score model using the NAWT with $\alpha = 4$ (NAWT4) and $\alpha = 2$ (NAWT2), adaptive NAWT, standard IPW, and the CBPS under the various settings for the true outcome model for the controlled when the true outcome model for the treated is $y_i(1) = x_i^3 - x_i$.} \label{fig_atefincor3}
\end{figure}

The summary of the results for the ATE estimation are shown in \ref{fig_atefindifcor}--\ref{fig_atefindifmis2} and the details are shown in Figure~\ref{fig_atefincor3}--\ref{fig_atefinmis21}. The results are similar to the ATT estimation. Besides, like large-sample cases in Appendix~\ref{sec_simeff}, the true outcome model for the treated affects the performance of the NAWT, the NAWT performs much better than the standard IPW in the ATE estimation, especially with $\alpha = 4$, and the NAWT performs at least as well as and sometimes better than the CBPS.

In conclusion, these results imply that, in a wide variety of the data generating process and the propensity score model specification, the NAWT improves the standard IPW and the adaptive method works well.

\begin{figure}[hp]
\centering
\includegraphics[width=0.935\textwidth]{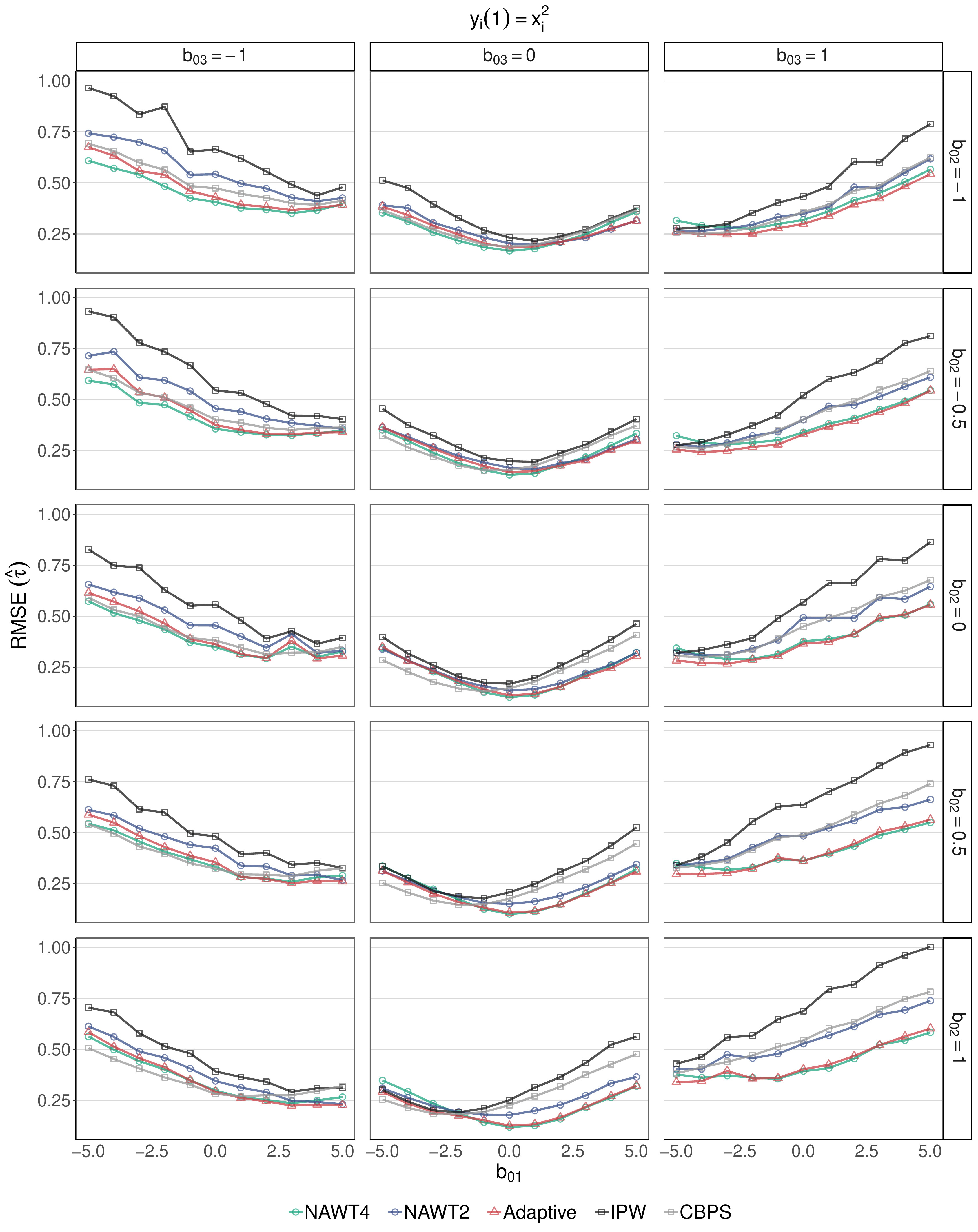}
\vspace{-12pt}
\caption{The finite-sample RMSE for the ATE estimation with the true propensity score model using the NAWT with $\alpha = 4$ (NAWT4) and $\alpha = 2$ (NAWT2), adaptive NAWT, standard IPW, and the CBPS under the various settings for the true outcome model for the controlled when the true outcome model for the treated is $y_i(1) = x_i^2$.} \label{fig_atefincor2}
\end{figure}

\begin{figure}[hp]
\centering
\includegraphics[width=0.935\textwidth]{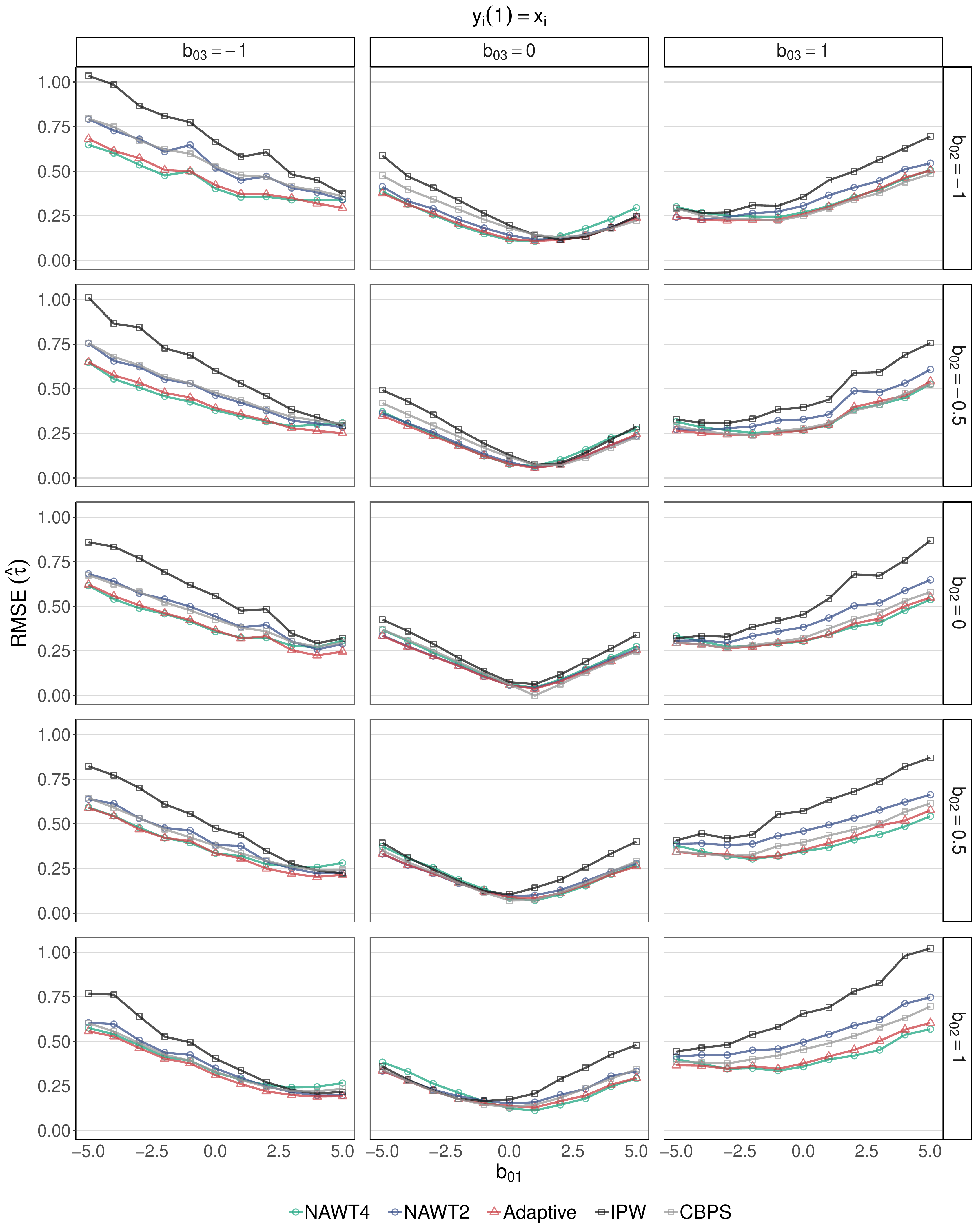}
\vspace{-12pt}
\caption{The finite-sample RMSE for the ATE estimation with the true propensity score model using the NAWT with $\alpha = 4$ (NAWT4) and $\alpha = 2$ (NAWT2), adaptive NAWT, standard IPW, and the CBPS under the various settings for the true outcome model for the controlled when the true outcome model for the treated is $y_i(1) = x_i$.} \label{fig_atefincor1}
\end{figure}

\begin{figure}[hp]
\centering
\includegraphics[width=0.935\textwidth]{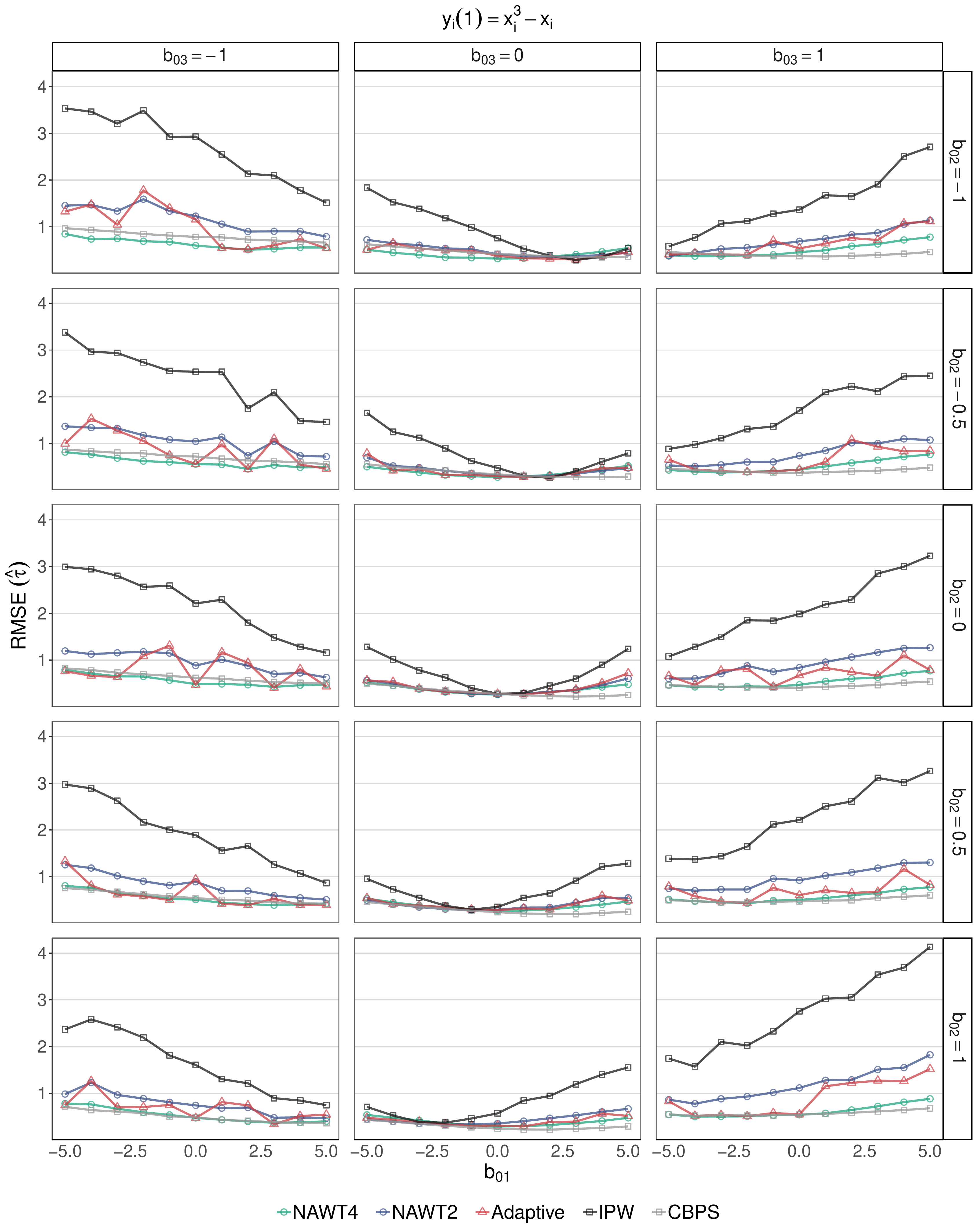}
\vspace{-12pt}
\caption{The finite-sample RMSE for the ATE estimation with the misspecified propensity score model (type~1) using the NAWT with $\alpha = 4$ (NAWT4) and $\alpha = 2$ (NAWT2), adaptive NAWT, standard IPW, and the CBPS under the various settings for the true outcome model for the controlled when the true outcome model for the treated is $y_i(1) = x_i^3 - x_i$.} \label{fig_atefinmis13}
\end{figure}

\begin{figure}[hp]
\centering
\includegraphics[width=0.935\textwidth]{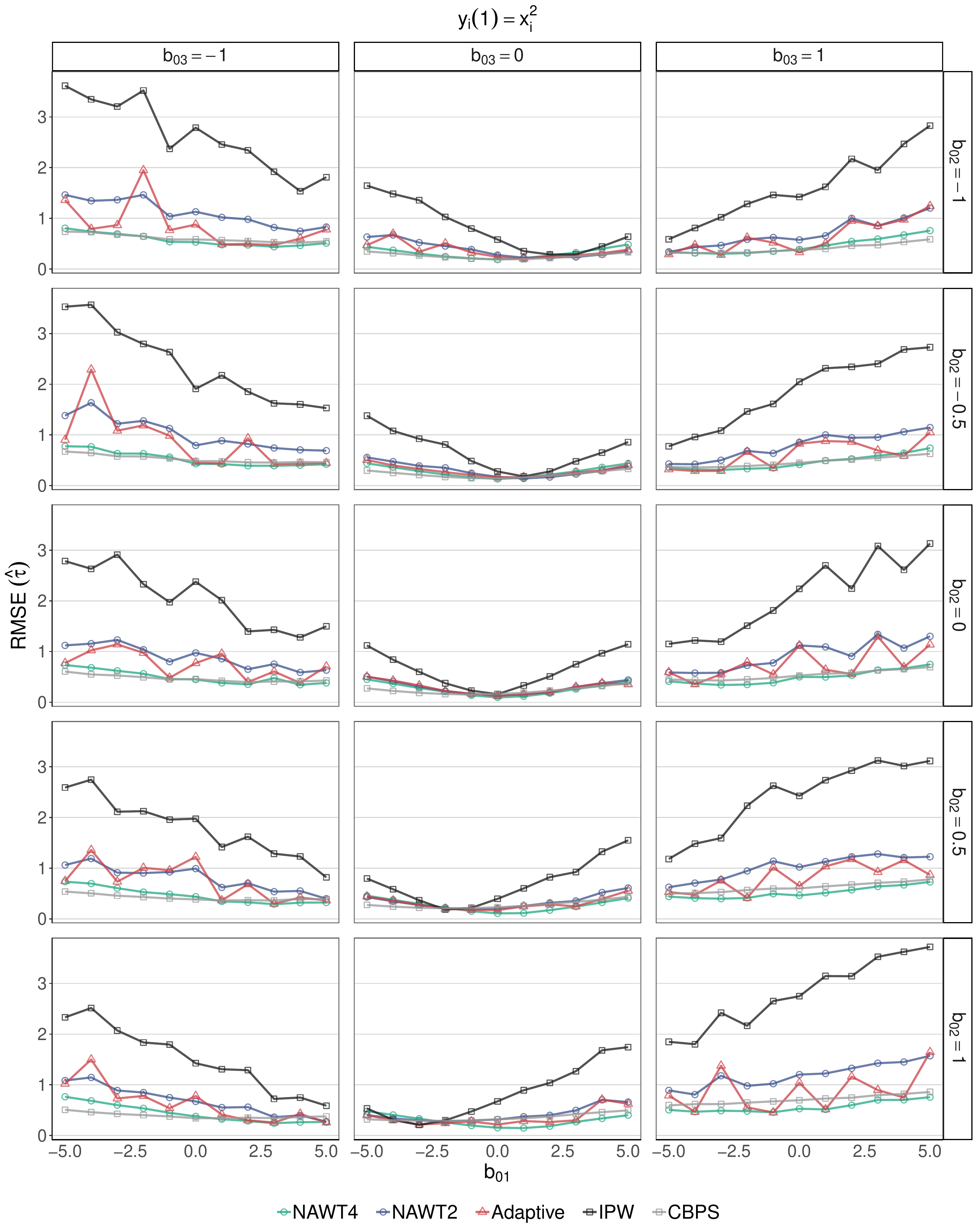}
\vspace{-12pt}
\caption{The finite-sample RMSE for the ATE estimation with the misspecified propensity score model (type~1) using the NAWT with $\alpha = 4$ (NAWT4) and $\alpha = 2$ (NAWT2), adaptive NAWT, standard IPW, and the CBPS under the various settings for the true outcome model for the controlled when the true outcome model for the treated is $y_i(1) = x_i^2$.} \label{fig_atefinmis12}
\end{figure}

\begin{figure}[hp]
\centering
\includegraphics[width=0.935\textwidth]{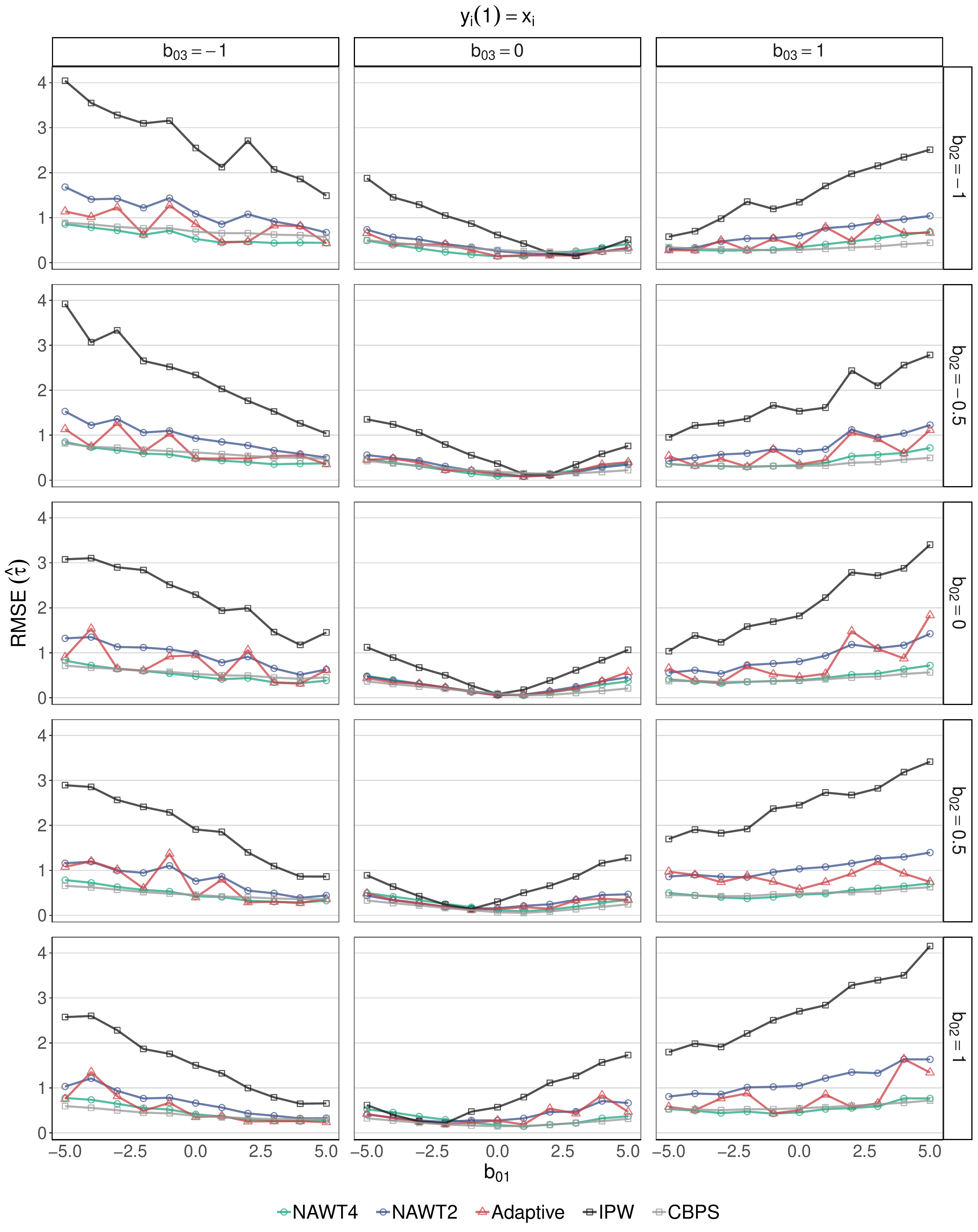}
\vspace{-12pt}
\caption{The finite-sample RMSE for the ATE estimation with the misspecified propensity score model (type~1) using the NAWT with $\alpha = 4$ (NAWT4) and $\alpha = 2$ (NAWT2), adaptive NAWT, standard IPW, and the CBPS under the various settings for the true outcome model for the controlled when the true outcome model for the treated is $y_i(1) = x_i$.} \label{fig_atefinmis11}
\end{figure}

\begin{figure}[hp]
\centering
\includegraphics[width=0.935\textwidth]{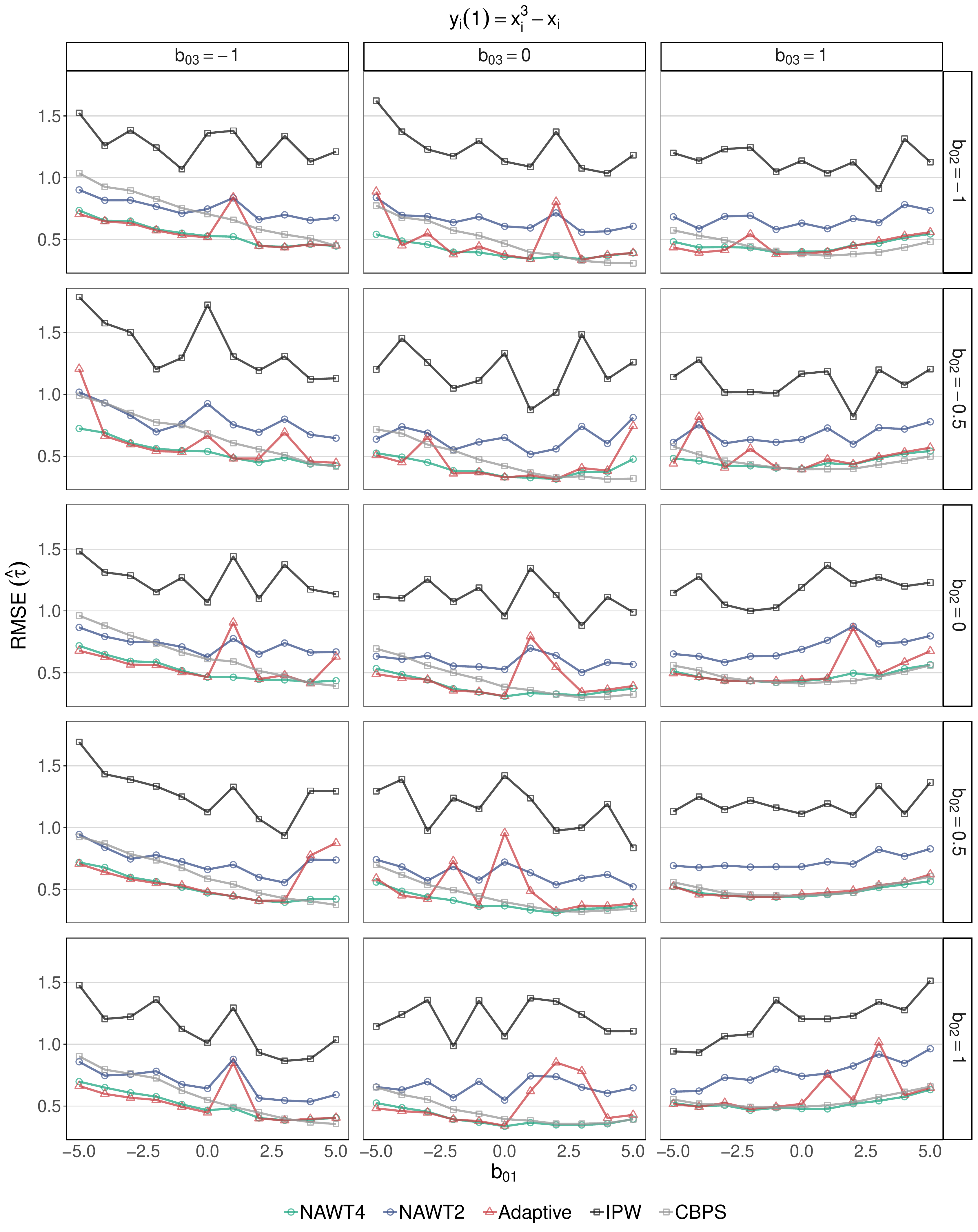}
\vspace{-12pt}
\caption{The finite-sample RMSE for the ATE estimation with the misspecified propensity score model (type~2) using the NAWT with $\alpha = 4$ (NAWT4) and $\alpha = 2$ (NAWT2), adaptive NAWT, standard IPW, and the CBPS under the various settings for the true outcome model for the controlled when the true outcome model for the treated is $y_i(1) = x_i^3 - x_i$.} \label{fig_atefinmis23}
\end{figure}

\begin{figure}[hp]
\centering
\includegraphics[width=0.935\textwidth]{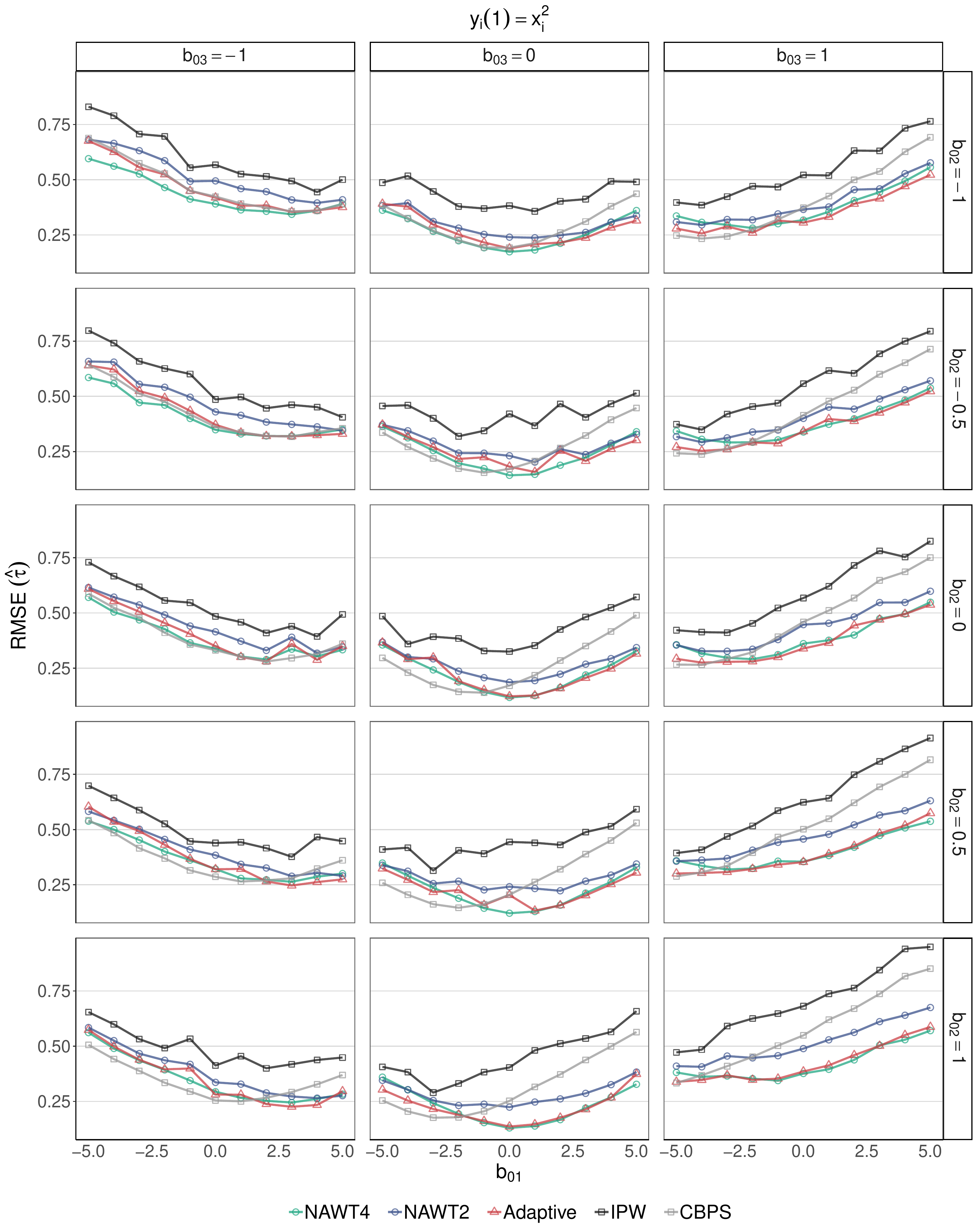}
\vspace{-12pt}
\caption{The finite-sample RMSE for the ATE estimation with the misspecified propensity score model (type~2) using the NAWT with $\alpha = 4$ (NAWT4) and $\alpha = 2$ (NAWT2), adaptive NAWT, standard IPW, and the CBPS under the various settings for the true outcome model for the controlled when the true outcome model for the treated is $y_i(1) = x_i^2$.} \label{fig_atefinmis22}
\end{figure}

\begin{figure}[hp]
\centering
\includegraphics[width=0.935\textwidth]{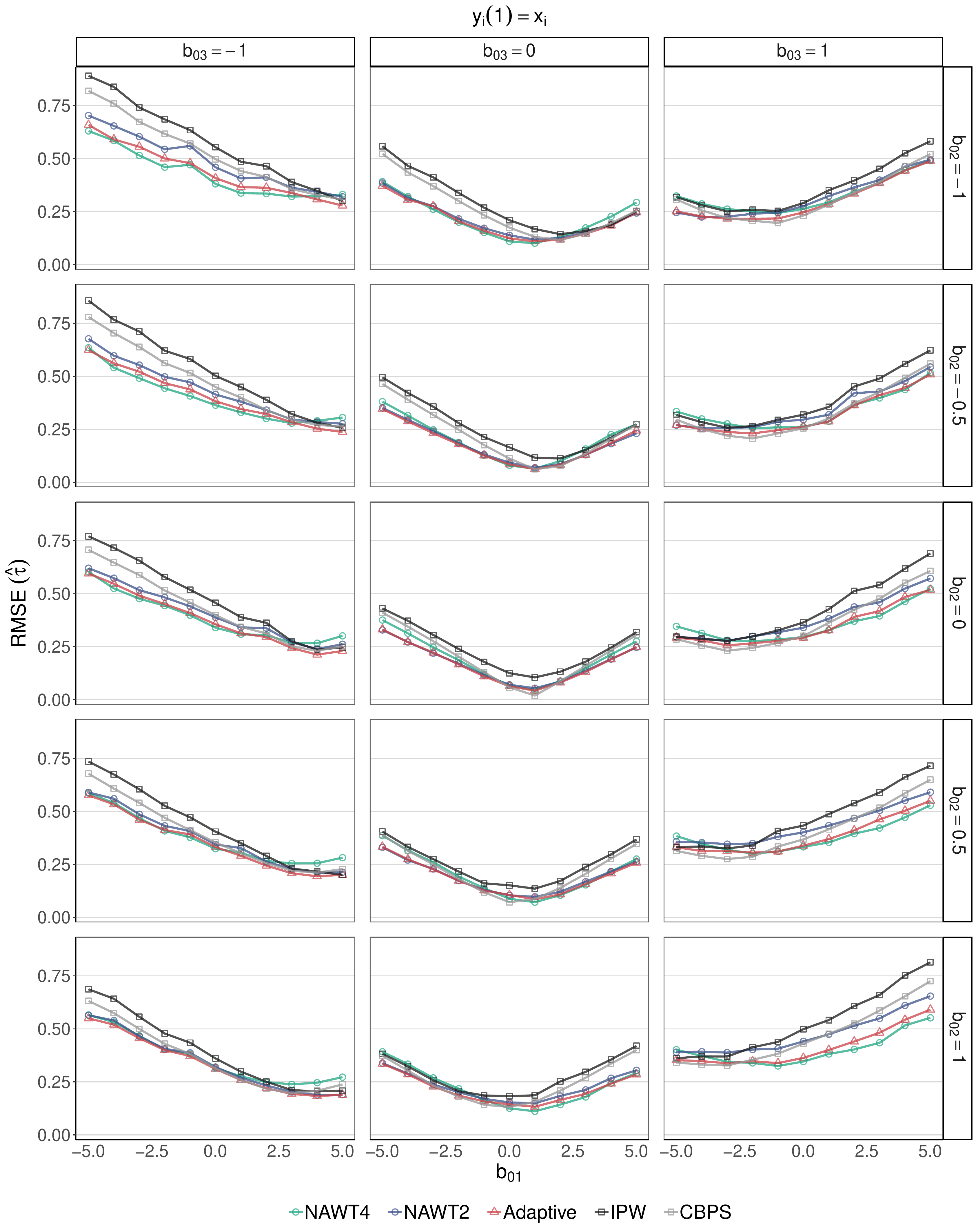}
\vspace{-12pt}
\caption{The finite-sample RMSE for the ATE estimation with the misspecified propensity score model (type~2) using the NAWT with $\alpha = 4$ (NAWT4) and $\alpha = 2$ (NAWT2), adaptive NAWT, standard IPW, and the CBPS under the various settings for the true outcome model for the controlled when the true outcome model for the treated is $y_i(1) = x_i$.} \label{fig_atefinmis21}
\end{figure}

\clearpage
\section{Details of the NAWT with covariate balance conditions} \label{sec_nawtcbps}

We can use the inverse of estimated covariance as the weighting matrix $\mathbf{A} = \hat{\mathbf{\Sigma}}(\beta, \mathbf{t}, \mathbf{X})^{-1}$ for efficiency. The continuously updating inverse covariance weight is
\begin{align}
\hat{\mathbf{\Sigma}}(\beta, \mathbf{t}, \mathbf{X}) &= \frac{1}{n} \sum \mathbb{E} [g(\beta, t_i, \mathbf{x}_i) g(\beta, t_i, \mathbf{x}_i)^\tr \mid \mathbf{x}_i] \\
&= \frac{1}{n} \sum \begin{pmatrix}
\omega(\pi_\beta(\mathbf{x}_i)) \pi_\beta(\mathbf{x}_i) (1 - \pi_\beta(\mathbf{x}_i)) \mathbf{x}_i \mathbf{x}_i^\tr & 
\omega(\pi_\beta(\mathbf{x}_i)) \pi_\beta(\mathbf{x}_i) \mathbf{x}_i \tilde{\mathbf{x}}_i^\tr \\
\omega(\pi_\beta(\mathbf{x}_i)) \pi_\beta(\mathbf{x}_i) \tilde{\mathbf{x}}_i \mathbf{x}_i^\tr & 
\pi_\beta(\mathbf{x}_i) / (1 - \pi_\beta(\mathbf{x}_i)) \tilde{\mathbf{x}}_i \tilde{\mathbf{x}}_i^\tr
\end{pmatrix}
\end{align}
where $t_i$ is integrated out conditional on the covariates $\mathbf{x}_i$, or we can utilize the two-step GMM \parencites{Imai2014}.

The asymptotic distribution of $\hat{\beta}_\mathrm{ATT}$ is:
\begin{equation}
\sqrt{n} (\hat{\beta}_\mathrm{ATT} - \beta) \xrightarrow[d]{} \mathcal{N}(0, (\mathbf{G}^\tr \mathbf{A} \mathbf{G})^{-1} \mathbf{G}^\tr \mathbf{A} \mathbf{\Sigma} \mathbf{A} \mathbf{G} (\mathbf{G}^\tr \mathbf{A} \mathbf{G})^{-1})
\end{equation}
where
\begin{align}
\mathbf{G} &\equiv \mathbb{E} \left[ \frac{\partial}{\partial \beta} \bar{g}(\beta, t_i, \mathbf{x}_i) \right] \\
&= \left( \begin{array}{ll}
\mathbb{E} \left[ - \omega(\pi_\beta(\mathbf{x}_i)) \pi_\beta(\mathbf{x}_i) (1 - \pi_\beta(\mathbf{x}_i)) \mathbf{x}_i \mathbf{x}_i^\tr \right] \\
\mathbb{E} \left[ \pi_\beta(\mathbf{x}_i) \tilde{\mathbf{x}}_i \tilde{\mathbf{x}}_i^\tr \right]
\end{array}
\right) \\
\mathbf{\Sigma} &\equiv \mathbb{E} [g(\beta, t_i, \mathbf{x}_i) g(\beta, t_i, \mathbf{x}_i)^\tr] \\
&= \left( \begin{array}{ll}
\mathbb{E} \left[ \omega(\pi_\beta(\mathbf{x}_i))^2 \pi_\beta(\mathbf{x}_i) (1 - \pi_\beta(\mathbf{x}_i))  \mathbf{x}_i \mathbf{x}_i^\tr \right] \\
\mathbb{E} \left[ \pi_\beta(\mathbf{x}_i) (1 - \pi_\beta(\mathbf{x}_i))^{-1} \tilde{\mathbf{x}}_i \tilde{\mathbf{x}}_i^\tr \right] 
\end{array}
\right) .
\end{align}

\newpage
\section{Distribution of estimates in the simulation studies} \label{sec_dist}

\begin{figure}[h]
\centering
\includegraphics[width=0.325\textwidth]{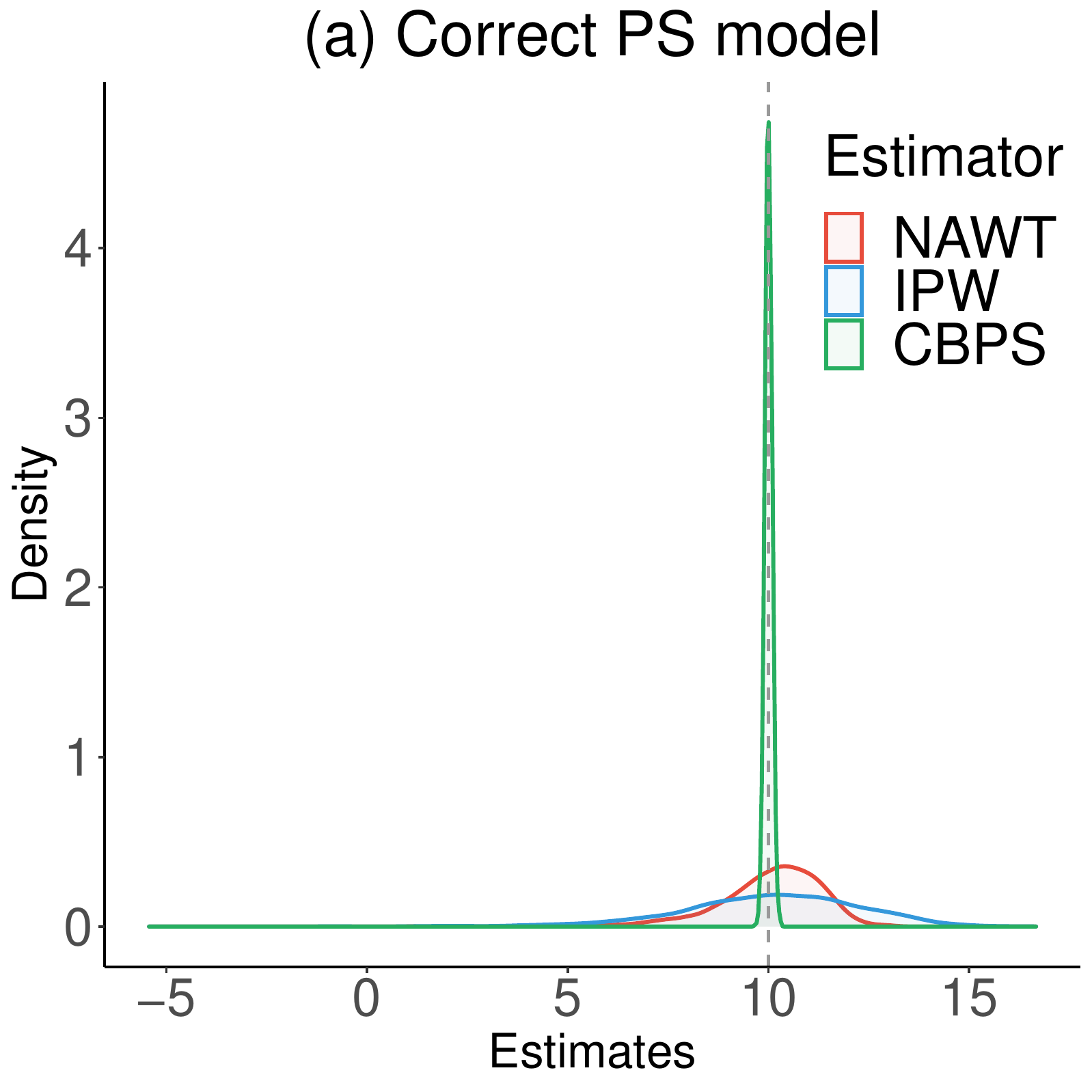}
\includegraphics[width=0.325\textwidth]{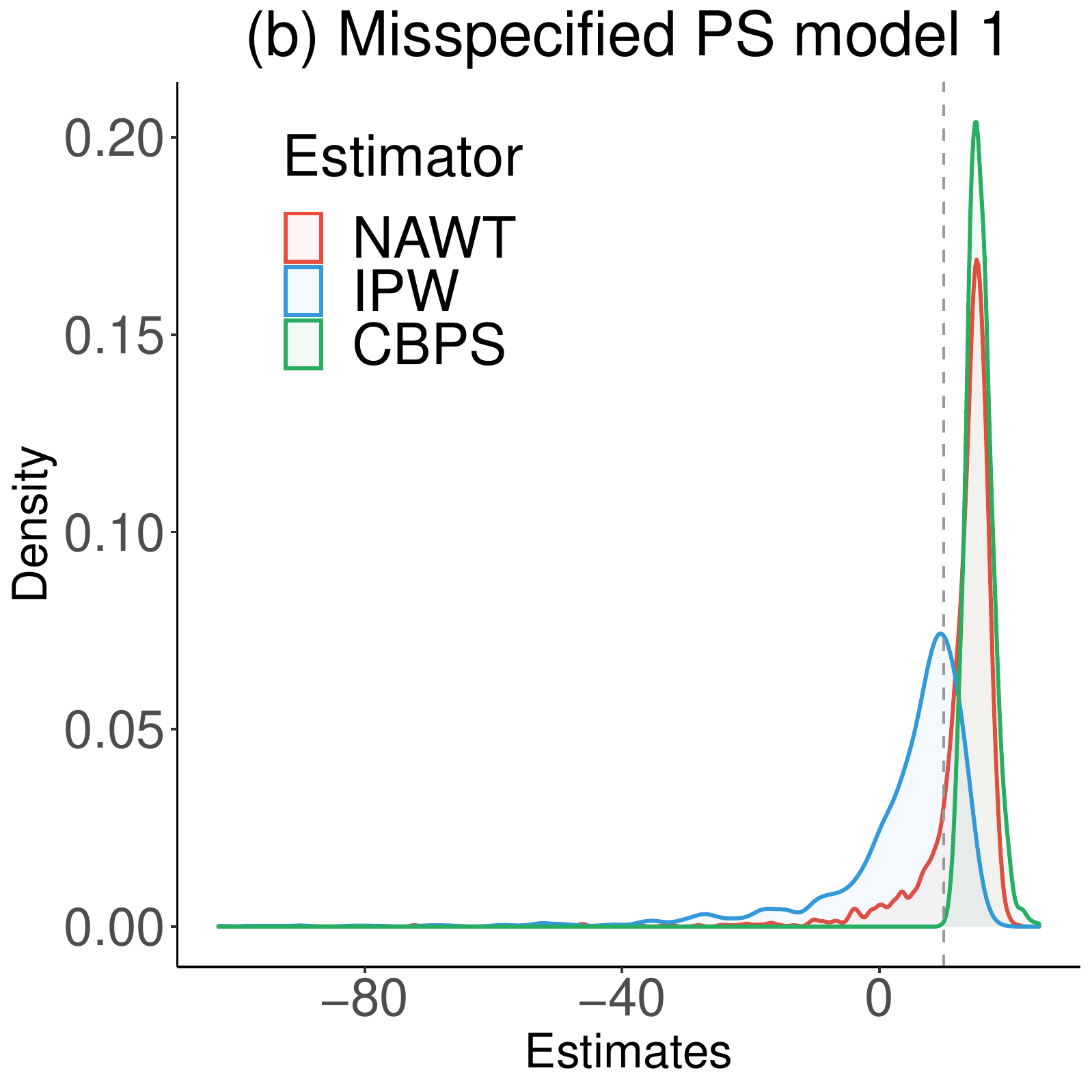}
\includegraphics[width=0.325\textwidth]{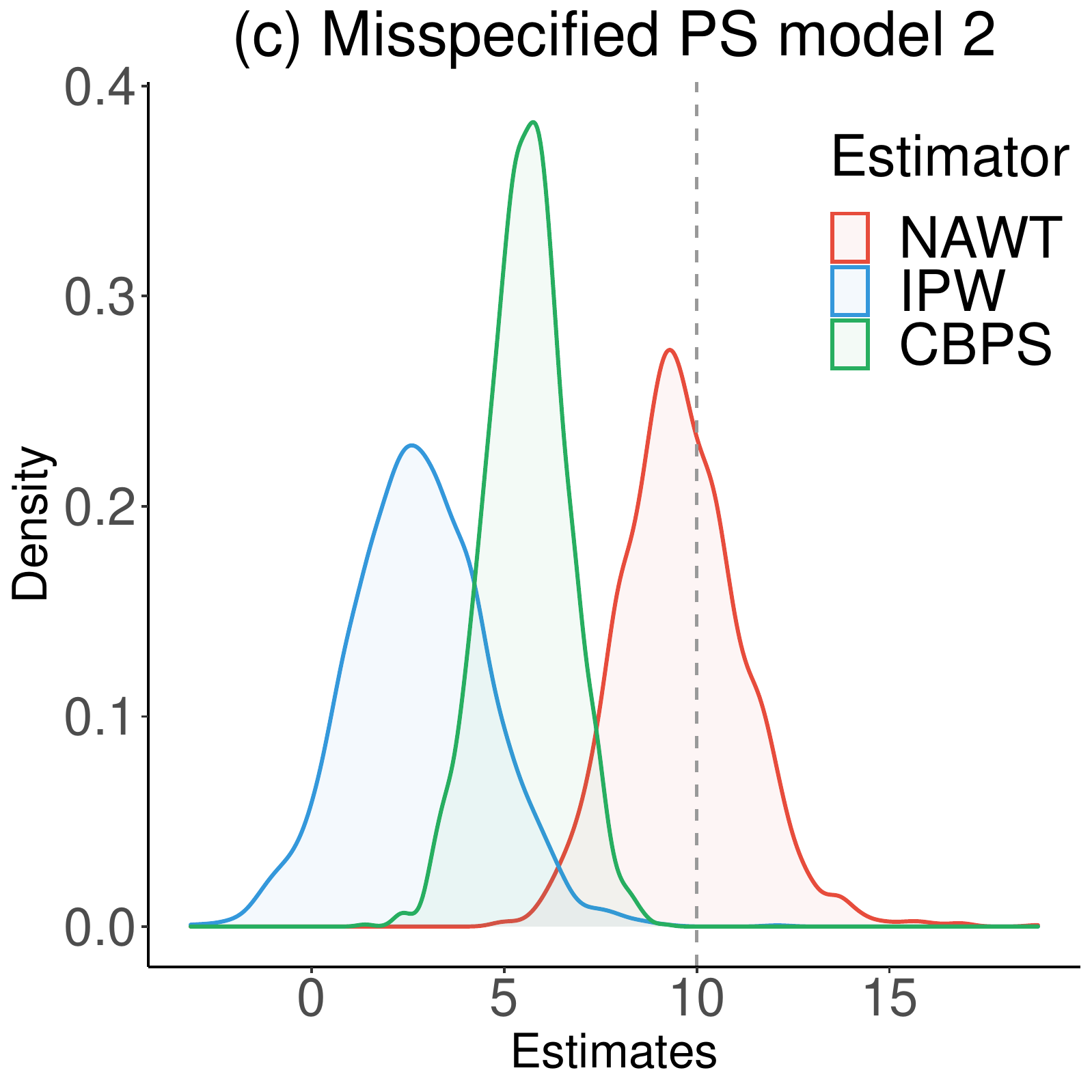}
\vspace{-8pt}
\caption{The distribution of estimates for the ATT estimation using the NAWT, standard IPW, and IPW with the CBPS under the following scenarios: (a) correct propensity score model and two types of propensity score model misspecification (b) and (c).} \label{fig_estatt}

\vspace{30pt}

\includegraphics[width=0.325\textwidth]{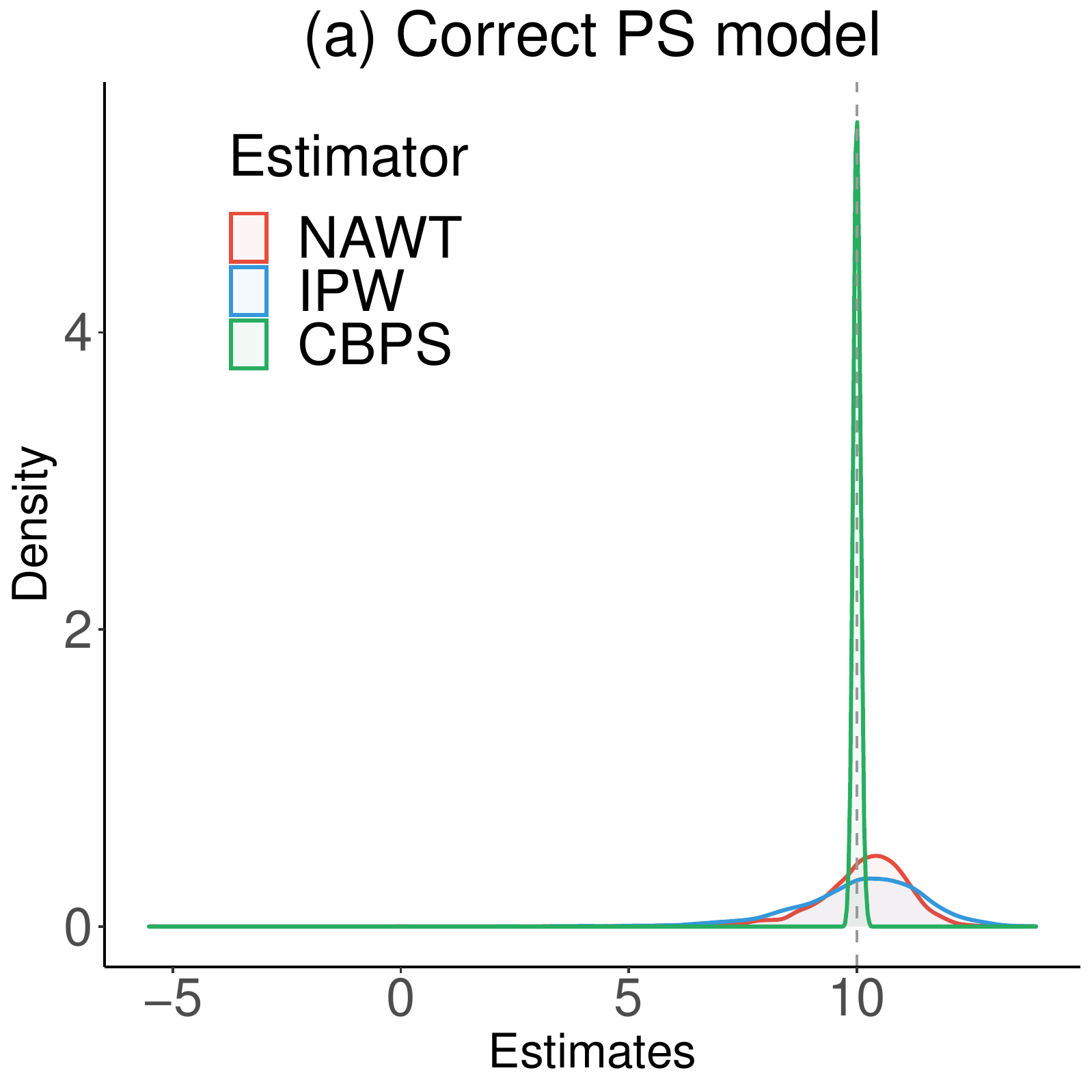}
\includegraphics[width=0.325\textwidth]{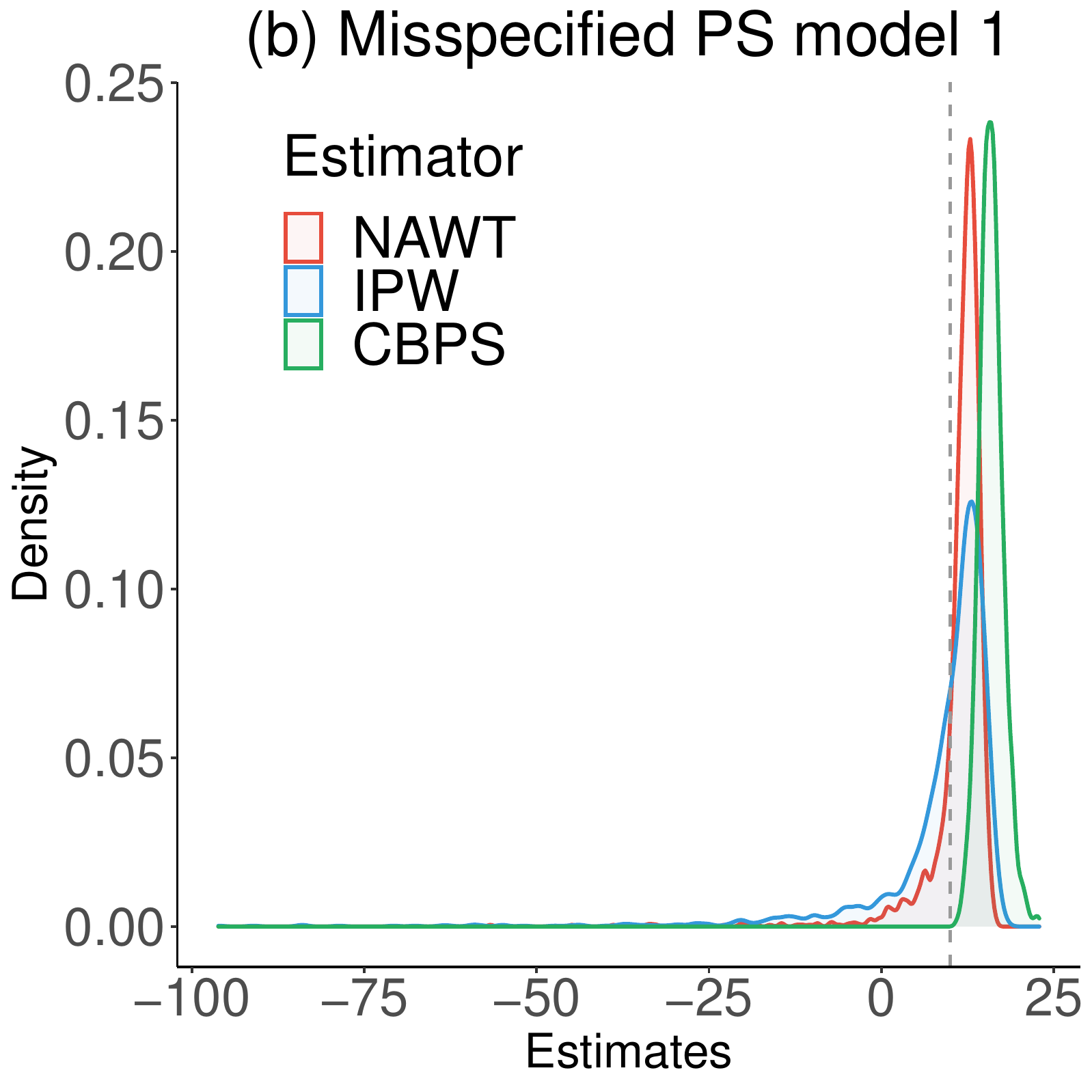}
\includegraphics[width=0.325\textwidth]{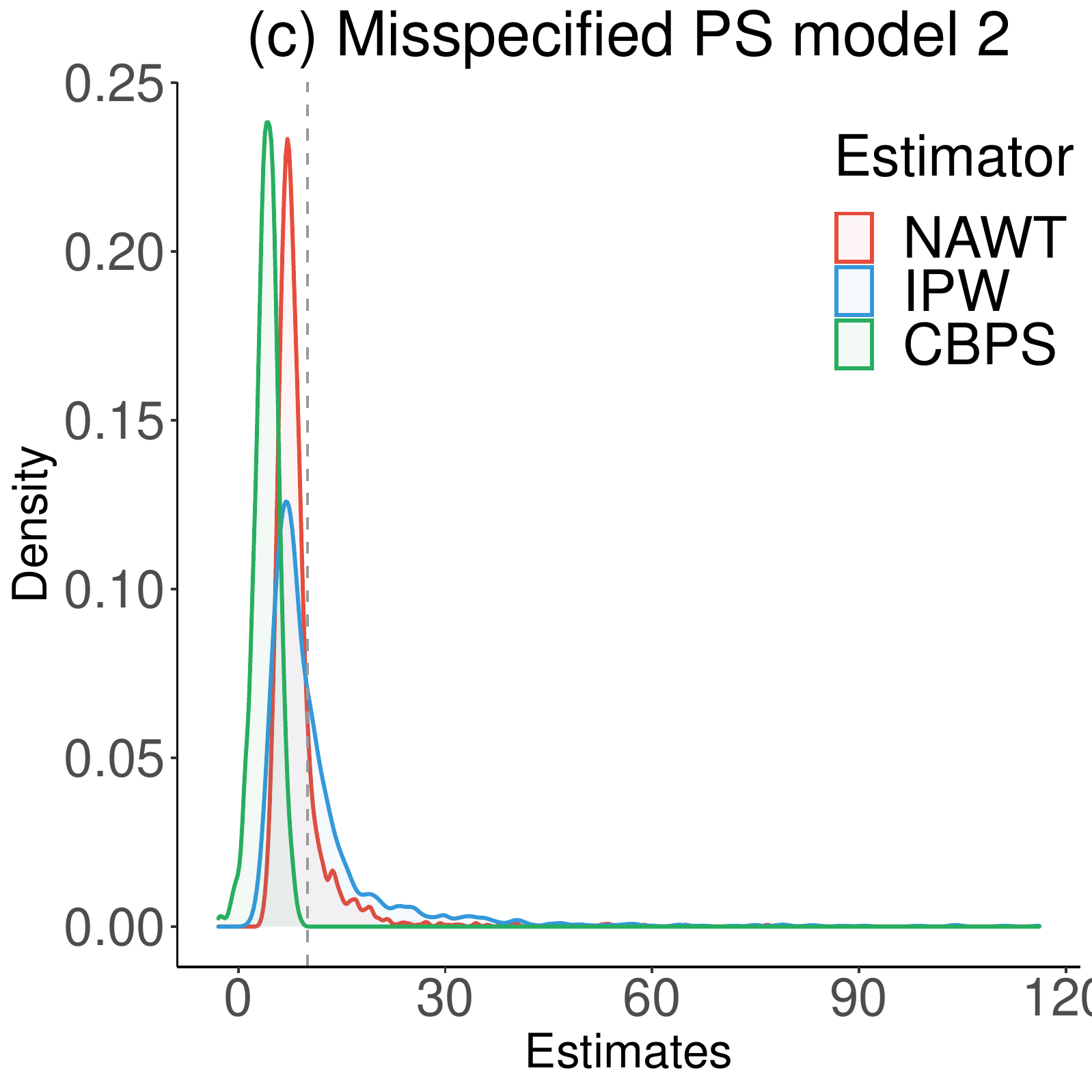}
\vspace{-8pt}
\caption{The distribution of estimates for the ATE estimation using the NAWT (the separate estimation), standard IPW, and IPW with the CBPS under the following scenarios: (a) correct propensity score model and two types of propensity score model misspecification (b) and (c).} \label{fig_estate}
\end{figure}

\clearpage

\section{The comparison of the NAWT for the ATE estimation} \label{sec_combined}

\vspace{15pt}

\begin{figure}[h]
\vspace{-15pt}
\centering
\includegraphics[width=0.3\textwidth, clip]{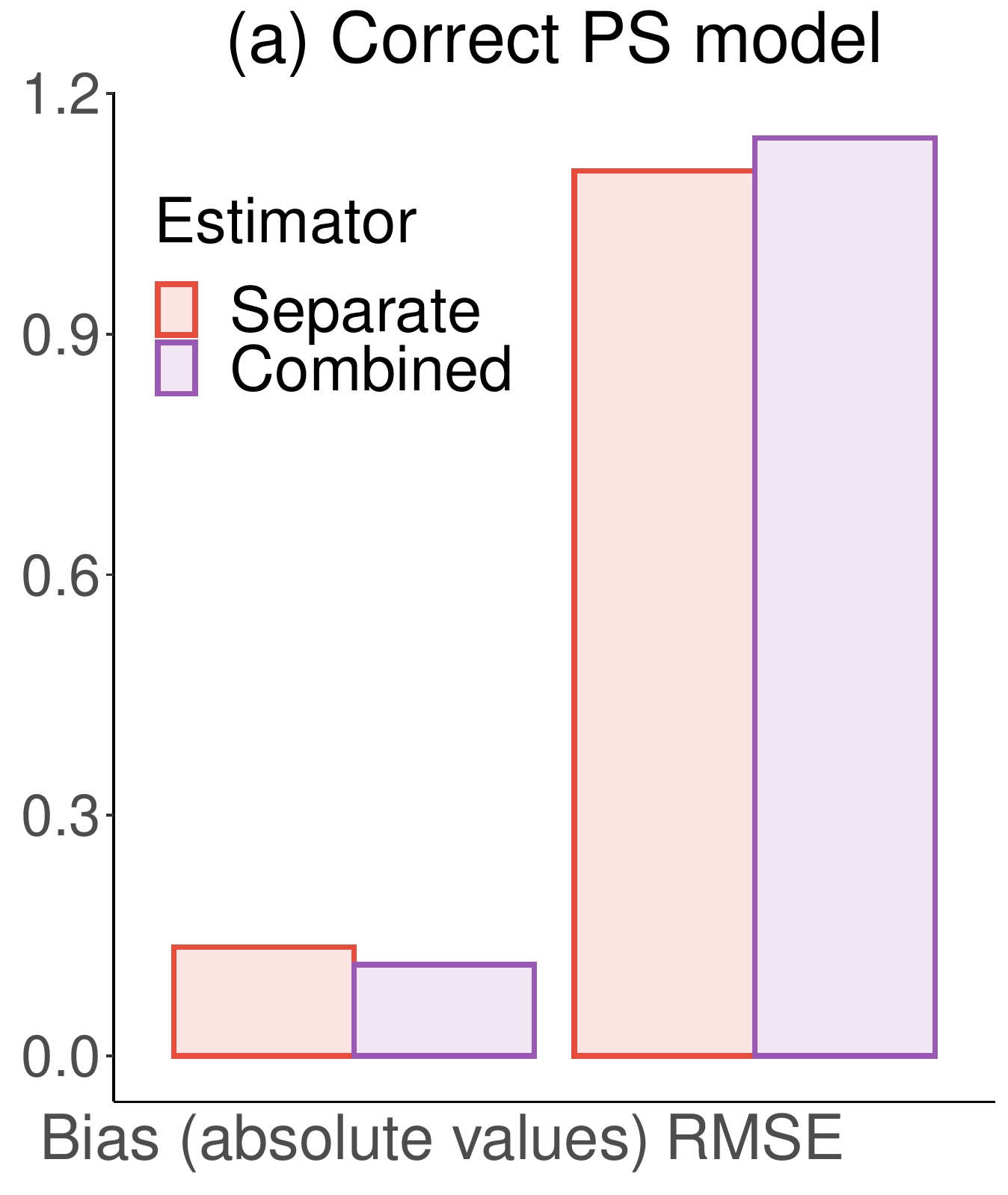}
\includegraphics[width=0.3\textwidth, clip]{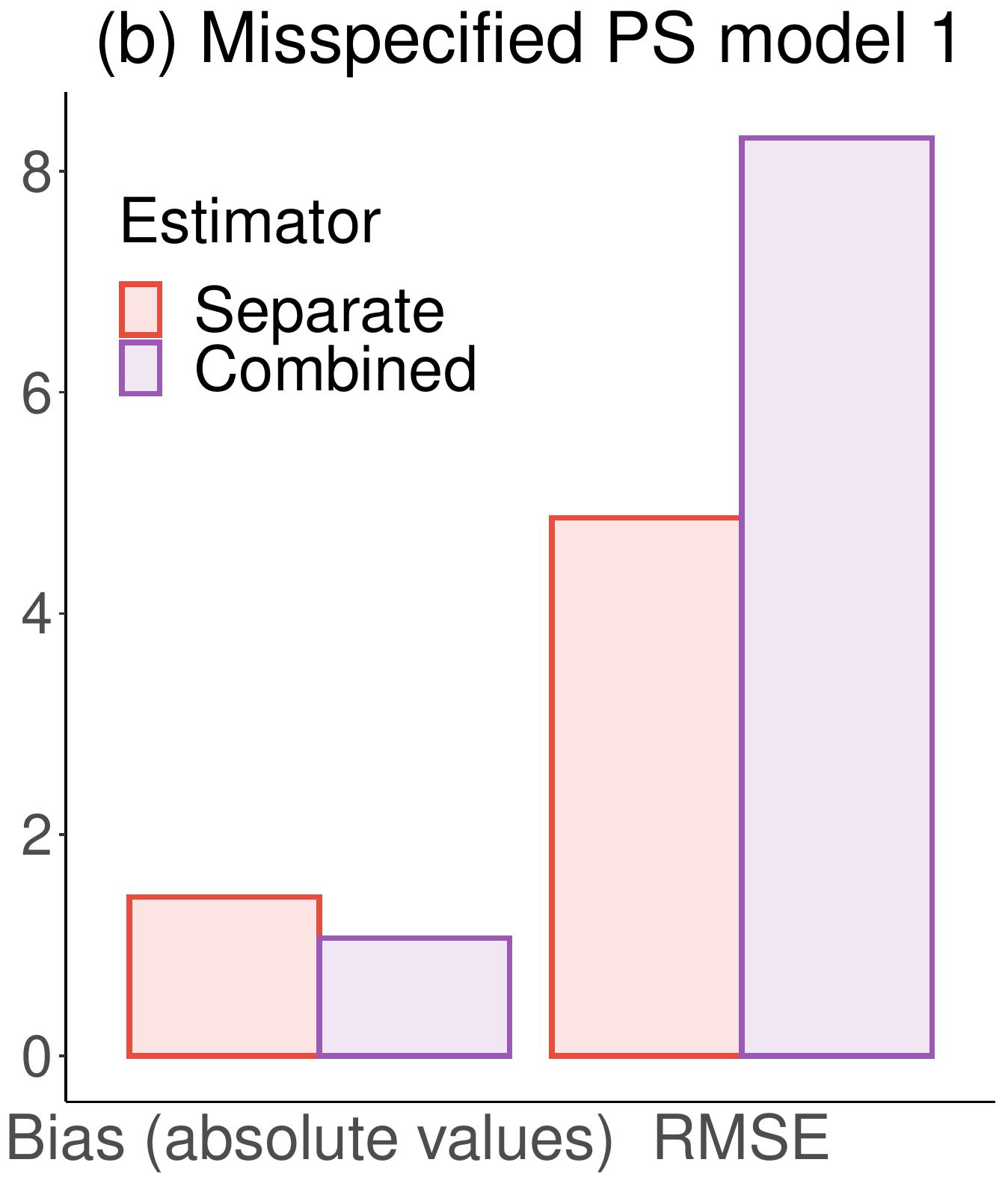}
\includegraphics[width=0.3\textwidth, clip]{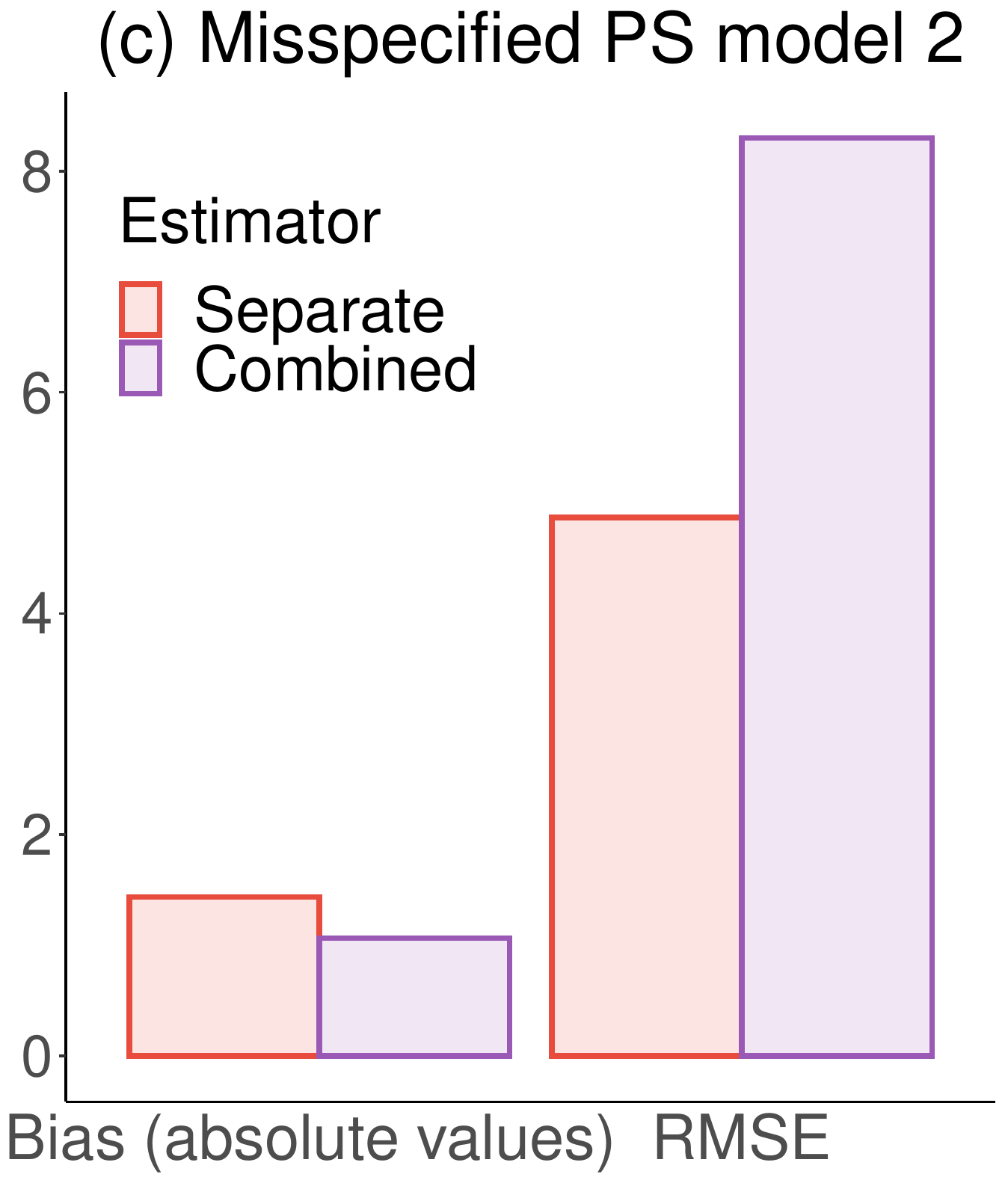}
\vspace{-10pt}
\caption{The bias in absolute values and the RMSE for the ATE estimation using the NAWT (the separated estimation) and NAWT (the combined estimation) under the following scenarios: (a) correct propensity score model and two types of propensity score model misspecification (b) and (c). The separated estimation outperforms the combined estimation in terms of the RMSE in all the scenarios, especially in the model misspecification cases.} \label{fig_biasrmseate2}

\vspace{20pt}

\includegraphics[width=0.3\textwidth, clip]{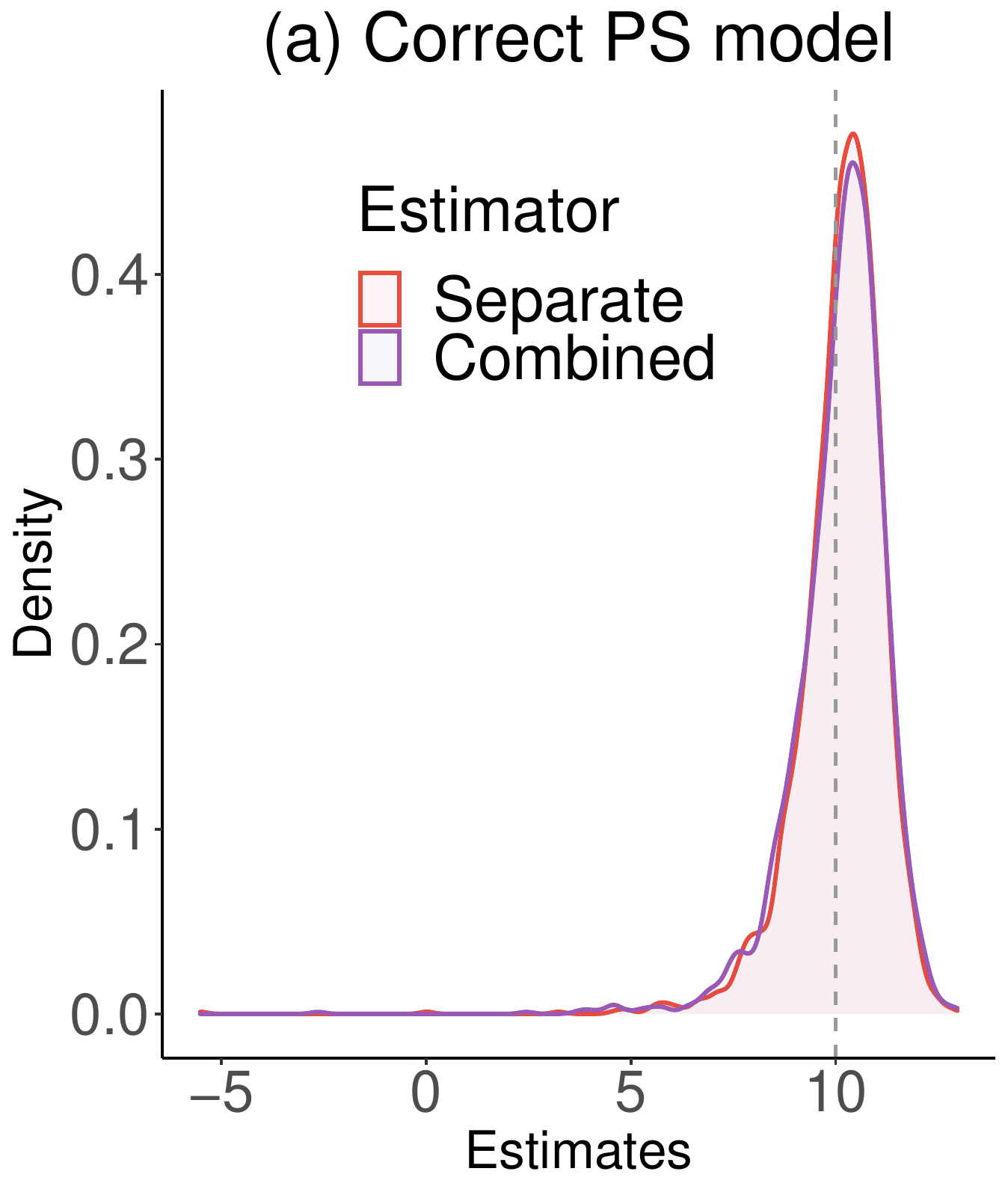}
\includegraphics[width=0.3\textwidth, clip]{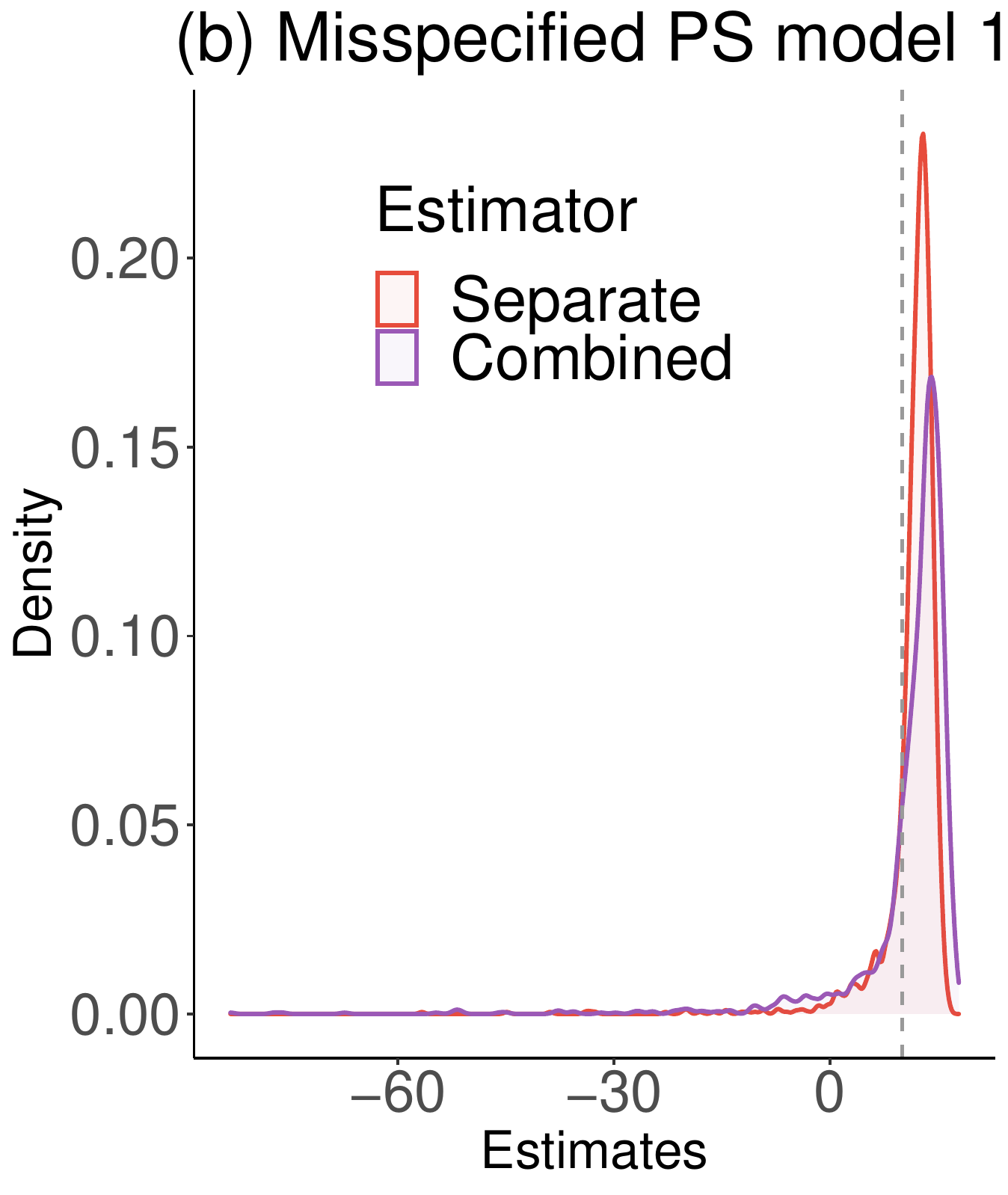}
\includegraphics[width=0.3\textwidth, clip]{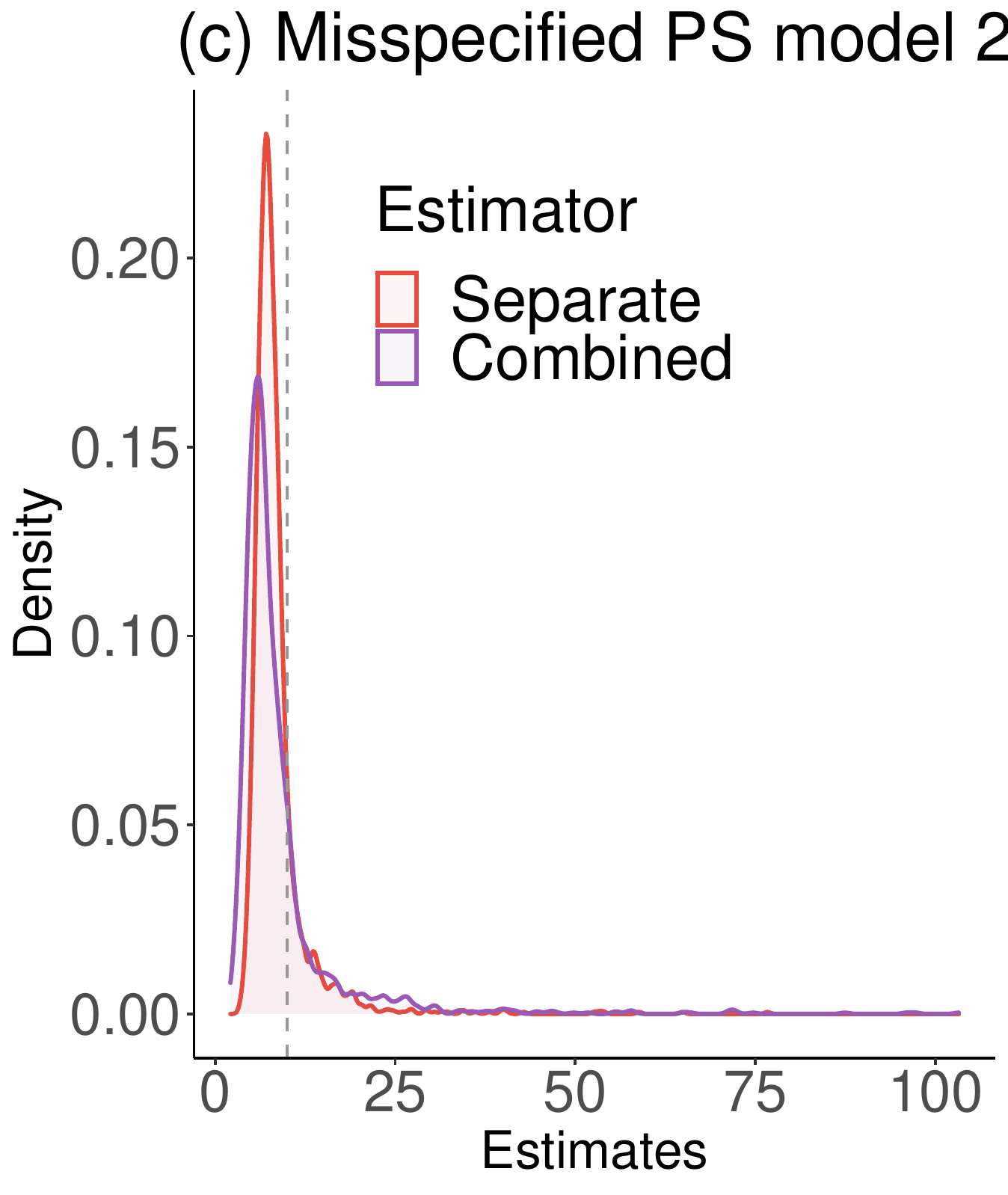}
\vspace{-10pt}
\caption{The distribution of estimates for the ATE estimation using the NAWT (the separated estimation) and NAWT (the combined estimation) under the following scenarios: (a) correct propensity score model and two types of propensity score model misspecification (b) and (c).} \label{fig_estate2}
\end{figure}

\clearpage

\section{Details of the simulation results} \label{sec_simtable}
\vspace{20pt}
\begin{table}[ht]
\renewcommand{\tabcolsep}{10pt}
\vspace{-20pt}
\centering
\caption{Details of the simulation results}\label{tb_simulation}
\begin{tabular}{lllrrc}
\hline
\multicolumn{1}{c}{Estimand}&\multicolumn{1}{c}{PS model}&\multicolumn{1}{c}{Estimator}&\multicolumn{1}{c}{Bias}&\multicolumn{1}{c}{RMSE}&\multicolumn{1}{c}{95\% CI Coverage}\tabularnewline
\hline
\multirow{11}{*}{ATT}&\multirow{3}{*}{(a) correct}&NAWT&$ 0.045$&$ 1.302$&$0.924$\tabularnewline
&&IPW&$ 0.034$&$ 2.269$&$0.909$\tabularnewline
&&CBPS&$ 0.003$&$ 0.086$&$0.941$\tabularnewline
&&&$$&$$&$$\tabularnewline
&\multirow{3}{*}{(b) misspecified 1}&NAWT&$ 2.743$&$ 7.191$&$0.354$\tabularnewline
&&IPW&$-6.479$&$14.292$&$0.618$\tabularnewline
&&CBPS&$ 5.550$&$ 5.912$&$0.077$\tabularnewline
&&&$$&$$&$$\tabularnewline
&\multirow{3}{*}{(c) misspecified 2}&NAWT&$-0.366$&$ 1.637$&$0.873$\tabularnewline
&&IPW&$-7.204$&$ 7.415$&$0.023$\tabularnewline
&&CBPS&$-4.435$&$ 4.558$&$0.016$\tabularnewline
&&&$$&$$&$$\tabularnewline
\hline&&&$$&$$&$$\tabularnewline
\multirow{14}{*}{ATE}&\multirow{4}{*}{(a) correct}&NAWT&$ 0.135$&$ 1.104$&$0.857$\tabularnewline
&&IPW&$ 0.046$&$ 1.472$&$0.910$\tabularnewline
&&CBPS&$ 0.003$&$ 0.077$&$0.944$\tabularnewline
&&Combined&$ 0.114$&$ 1.145$&$0.865$\tabularnewline
&&&$$&$$&$$\tabularnewline
&\multirow{4}{*}{(b) misspecified 1}&NAWT&$ 1.437$&$ 4.867$&$0.440$\tabularnewline
&&IPW&$-1.317$&$10.537$&$0.583$\tabularnewline
&&CBPS&$ 5.941$&$ 6.188$&$0.015$\tabularnewline
&&Combined&$ 1.064$&$ 8.299$&$0.410$\tabularnewline
&&&$$&$$&$$\tabularnewline
&\multirow{4}{*}{(c) misspecified 2}&NAWT&$-1.437$&$ 4.867$&$0.440$\tabularnewline
&&IPW&$ 1.317$&$10.537$&$0.583$\tabularnewline
&&CBPS&$-5.941$&$ 6.188$&$0.015$\tabularnewline
&&Combined&$-1.064$&$ 8.299$&$0.410$\tabularnewline
\hline
\multicolumn{6}{p{.95\textwidth}}{\small Note: The performance of the NAWT, standard IPW, and IPW with the CBPS is compared in terms of the bias and RMSE under the scenario of the correct propensity score model and two misspecified propensity score models for the ATT and ATE estimation. The NAWT outperforms the standard IPW in terms of the RMSE in all the scenarios, and it depends on the situation whether the NAWT works better than the IPW with CBPS in terms of the bias and RMSE.}
\end{tabular}
\end{table}


\clearpage

\section{The comparison among the NAWT with different weighting functions} \label{sec_alpha}

\begin{table}[ht]
\renewcommand{\tabcolsep}{15pt}
\centering
\caption{Comparison among weighting functions}\label{tb_alpha}
\begin{tabular}{llrrrc}
\hline
\multicolumn{1}{c}{N}&\multicolumn{1}{c}{PS model}&\multicolumn{1}{c}{$\alpha$}&\multicolumn{1}{c}{Bias}&\multicolumn{1}{c}{RMSE}&\multicolumn{1}{c}{95\% CI Coverage}\tabularnewline
\hline
\multirow{12}{*}{400}&\multirow{4}{*}{(a) correct}&$ 0$&$  0.145$&$ 3.706$&$0.901$\tabularnewline
&&$ 1$&$  0.229$&$ 2.684$&$0.862$\tabularnewline
&&$ \mathbf{2}$&$ \mathbf{-0.047}$&$ \mathbf{2.281}$&$0.932$\tabularnewline
&&$ 3$&$ -0.980$&$ 3.696$&$0.958$\tabularnewline
&\multirow{4}{*}{(b) misspecified 1}&$ 0$&$ -3.666$&$12.634$&$0.641$\tabularnewline
&&$ 1$&$  \mathbf{1.226}$&$ 8.545$&$0.570$\tabularnewline
&&$ \mathbf{2}$&$  4.472$&$ \mathbf{7.144}$&$0.387$\tabularnewline
&&$ 3$&$  6.677$&$ 7.744$&$0.308$\tabularnewline
&\multirow{4}{*}{(c) misspecified 2}&$ 0$&$ -7.232$&$ 7.759$&$0.208$\tabularnewline
&&$ 1$&$ -3.398$&$ 4.076$&$0.548$\tabularnewline
&&$ \mathbf{2}$&$ \mathbf{-0.308}$&$ \mathbf{2.692}$&$0.875$\tabularnewline
&&$ 3$&$  2.302$&$ 5.338$&$0.770$\tabularnewline
&&$$&$$&$$&$$\tabularnewline
\multirow{12}{*}{2000}&\multirow{4}{*}{(a) correct}&$ 0$&$  \mathbf{0.014}$&$ 1.696$&$0.925$\tabularnewline
&&$ 1$&$  0.025$&$ 1.341$&$0.889$\tabularnewline
&&$ \mathbf{2}$&$ -0.023$&$ \mathbf{1.176}$&$0.920$\tabularnewline
&&$ 3$&$ -0.135$&$ 1.270$&$0.950$\tabularnewline
&\multirow{4}{*}{(b) misspecified 1}&$ 0$&$ -8.351$&$15.752$&$0.551$\tabularnewline
&&$ 1$&$ -2.597$&$11.114$&$0.646$\tabularnewline
&&$ \mathbf{2}$&$  \mathbf{1.402}$&$ 8.777$&$0.380$\tabularnewline
&&$ 3$&$  4.431$&$ \mathbf{8.097}$&$0.179$\tabularnewline
&\multirow{4}{*}{(c) misspecified 2}&$ 0$&$ -7.152$&$ 7.253$&$0.001$\tabularnewline
&&$ 1$&$ -3.187$&$ 3.345$&$0.113$\tabularnewline
&&$ \mathbf{2}$&$ \mathbf{-0.274}$&$ \mathbf{1.215}$&$0.868$\tabularnewline
&&$ 3$&$  1.551$&$ 2.233$&$0.712$\tabularnewline
&&$$&$$&$$&$$\tabularnewline
\multirow{12}{*}{10000}&\multirow{4}{*}{(a) correct}&$ 0$&$ -0.012$&$ 0.751$&$0.940$\tabularnewline
&&$ 1$&$ \mathbf{-0.006}$&$ 0.556$&$0.922$\tabularnewline
&&$ \mathbf{2}$&$ -0.012$&$ \mathbf{0.469}$&$0.926$\tabularnewline
&&$ 3$&$ -0.030$&$ 0.521$&$0.944$\tabularnewline
&\multirow{4}{*}{(b) misspecified 1}&$ 0$&$-12.692$&$19.978$&$0.094$\tabularnewline
&&$ 1$&$ -5.707$&$13.820$&$0.692$\tabularnewline
&&$ \mathbf{2}$&$ \mathbf{-0.920}$&$10.283$&$0.464$\tabularnewline
&&$ 3$&$  2.668$&$ \mathbf{8.764}$&$0.204$\tabularnewline
&\multirow{4}{*}{(c) misspecified 2}&$ 0$&$ -7.154$&$ 7.174$&$0.000$\tabularnewline
&&$ 1$&$ -3.142$&$ 3.174$&$0.000$\tabularnewline
&&$ \mathbf{2}$&$ \mathbf{-0.247}$&$ \mathbf{0.599}$&$0.840$\tabularnewline
&&$ 3$&$  1.510$&$ 1.673$&$0.232$\tabularnewline
\hline
\multicolumn{6}{p{.93\textwidth}}{\small Note: The performance of the NAWT with different weighting functions is compared in terms of the bias and RMSE in nine different situations.}
\end{tabular}
\end{table}


Using the same procedure as in Section~\ref{sec_simulation}, the performance of the NAWT with different weighting functions is compared for the ATT estimation. Specifically, I compare the NAWT with $\omega(\pi_\beta(\mathbf{x}_i)) = \pi_\beta(\mathbf{x}_i)^\alpha$ with $\alpha =$ 0, 1, 2, and 3, where $\alpha = 0$ corresponds to the standard IPW. I compare nine situations in total, where combinations of three scenarios ((a) correct propensity score model and two types of propensity score model misspecification (b) and (c)) with $N =$ 400, $2,000$, and $10,000$ are utilized. I conduct $10,000$, $4,000$ and $2,000$ Monte Carlo simulations, respectively, for $N =$ 400, $2,000$, and $10,000$, respectively.

The results are shown in Table~\ref{tb_alpha}. In almost all the situations, the NAWT with $\omega(\pi_\beta(\mathbf{x}_i)) = \pi_\beta(\mathbf{x}_i)^2$ performs the best. The NAWT with $\omega(\pi_\beta(\mathbf{x}_i)) = \pi_\beta(\mathbf{x}_i)^2$ works well irrespective of the sample size and propensity score model (mis)specification.

\clearpage

\end{document}